\documentclass[11pt,a4paper]{article}
\pdfoutput=1
\usepackage{jheppub}

\usepackage{amsmath}
\usepackage{verbatim}
\usepackage{amssymb}
\usepackage{inputenc, array}
\usepackage{textcomp}
\usepackage{bm}
\usepackage{physics}
\usepackage{simpler-wick}
\usepackage{appendix}
\usepackage{xcolor}
\usepackage{comment}

\newcommand{\be}[1]{\begin{equation}\label{#1} }
\newcommand{\ee}{\end{equation}}
\newcommand{\bea}[1]{\begin{eqnarray}\label{#1} }
\newcommand{\eea}{\end{eqnarray}}
\newcommand{\refb}[1]{(\ref{#1})}

\newcommand{\z}{{\bar z}}

\newcommand{\D}{\Delta}

\newcommand{\s}{\sigma}

 \newcommand{\p}{\textbf{P}}
\newcommand{\x}{\textbf{X}}
\newcommand{\y}{\textbf{Y}}
\newcommand{\mo}{\mathcal{O}}
\usepackage{hyperref}
\usepackage{cleveref}
\usepackage{float}
\usepackage{subcaption}
\usepackage{physics}
\usepackage{slashed}
\usepackage{amsfonts}
\usepackage{caption}
\usepackage{mathrsfs}
\usepackage{bbm}

\title{Holography in Flat Spacetimes: the case for Carroll}

\author{Arjun Bagchi,}  \author{Prateksh Dhivakar,} \author{and Sudipta Dutta.}\author{\\}

\affiliation{Indian Institute of Technology Kanpur, Kanpur 208016, India.\\} 

\emailAdd{(abagchi, prateksh, dsudipta)@iitk.ac.in}

\preprint{}
\abstract{We compare and contrast the two approaches of holography in asymptotically flat spacetimes, viz. the co-dimension two Celestial approach based on the Mellin transformation and the co-dimension one Carrollian approach based on the modified Mellin and elucidate how some of the problems of the Celestial approach can be rectified by the Carrollian one. Considering flat holography as a limit from AdS/CFT makes a co-dimension one dual more plausible, and our previous construction of Carrollian correlations from AdS Witten diagrams is testimony to this. In this paper, we show how to generalize our earlier analysis for operators with spin. We work out a large number of explicit non-trivial examples (twelve) and show matching between the limit of AdS$_4$ Witten diagrams and 3d boundary symmetry considerations, thus making the case for the Carrollian dual even stronger.}

\begin{document}
\maketitle

\section{Introduction}

Holography has been our best tool to probe the mysteries of quantum gravity. While progress in understanding holography in the context of Anti de Sitter (AdS) spacetimes has been spectacularly successful through the celebrated AdS/CFT correspondence, contraction of the hologram for spacetimes different from ads has proven to be challenging. This paper aims to address the question of building holography for asymptotically flat spacetimes and build on recent progress in this direction. 

\subsection{Celestial and Carrollian dual theories} 

Following the lead of Strominger \cite{Strominger:2013jfa,He:2014laa,Cachazo:2014fwa,Kapec:2014opa,Strominger:2014pwa,He:2014cra,Strominger:2017zoo}, there has been a recent resurgence in activities in investigating aspects of flat space physics. There has been a renewed understanding of various aspects of infrared physics connecting asymptotic symmetries, soft theorems, and memory effects in a triangle of relations nowadays called the infrared triangle. A proposal of a hologram of flat space has emerged out of this and is called Celestial holography \footnote{For a review of this approach with a comprehensive literature guide, the reader is referred to \cite{Pasterski:2021rjz,Raclariu:2021zjz,Pasterski:2021raf}.}. This relates bulk physics in 4d asymptotically flat spacetimes to a 2d conformal field theory (CFT) living on the celestial sphere at the null boundary of flatspace.
The principle observation here is that the Lorentz group in 4d acts as the global conformal group on the sphere at the boundary, and through Mellin transformations, one can map between bulk S-matrix elements in flat spacetimes to $n$-point correlation functions in 2d CFT \cite{Pasterski:2016qvg,Pasterski:2017kqt,Pasterski:2017ylz}. The Celestial programme has had many successes discovering new results in asymptotic symmetries and scattering amplitudes.  

\medskip

There are attempts at holography in asymptotically flat spacetimes, which predates the Celestial programme \cite{deBoer:2003vf, Barnich:2010eb, Bagchi:2010zz}. We will especially be interested in the approach that is nowadays called the Carrollian approach and originates from the observation of the isomorphism between the asymptotic symmetries of flat spacetimes and non-Lorentzian conformal symmetries in one lower dimension \cite{Bagchi:2010zz, Bagchi:2012cy, Duval:2014uva, Bagchi:2016bcd}. While the initial thrust in this direction was understanding the case of 3d asymptotically flat spacetimes in terms of 2d field theories with Carrollian conformal or BMS$_3$ symmetry \cite{Bagchi:2012yk,Bagchi:2012xr,Afshar:2013vka,Gonzalez:2013oaa,Bagchi:2014iea,Hartong:2015usd,Bagchi:2015wna,Jiang:2017ecm,Hijano:2017eii,Apolo:2020bld}, of late there is an emerging body of interesting work which relates to 4d asymptotically flat spacetimes
\cite{Bagchi:2016bcd, Banerjee:2018gce, Bagchi:2022emh,Bagchi:2023fbj,Donnay:2022aba,Donnay:2022wvx,Nguyen:2023vfz}.

\medskip

In this paper, our aim is two-fold. We first begin by comparing and contrasting the Carrollian and Celestial approaches to flat holography. We will justify why we think the Carrollian co-dimension one approach, which explicitly keeps track of the null nature of the asymptotic boundary of Minkowski spacetimes, is more appropriate for constructing a hologram of asymptotically flat spacetimes. The modifed Mellin transformation, introduced in \cite{Banerjee:2018gce}, would be central to our arguments and analysis. 

\medskip

We will then focus on our recent construction of the flat limit of Witten diagrams in AdS$_4$, which yields Carroll conformal correlation functions on the boundary \cite{Bagchi:2023fbj}. In the earlier paper \cite{Bagchi:2023fbj}, we focussed on spinless two and three-point correlations. We show in this paper that our construction generalises to more involved correlation functions, e.g., four-point scalar correlators, correlators in split signature, and spinning correlators, and we explore a whole range of examples (a total of 12 explicit cases) and show matching between bulk results obtained in the limit from AdS and boundary results analyzed with the help symmetry structures. 

\medskip

Of these cases, we also examine a particular one that does not have a Celestial analogue. This is a non-MHV $(+++)/(---)$ three point amplitude. Such amplitudes arise from loop corrections or higher dimensional operators \cite{Dixon:1993xd,Dixon:2004za}. These amplitudes are UV divergent in the Mellin basis. Thus, it is natural that these amplitudes were not focused in the Celestial literature \cite{PhysRevD.96.085006,Schreiber:2017jsr}. We will show that such amplitudes are UV finite in the modified Mellin basis, and moreover, they can be captured by the finite holographic Carrollian CFT correlators. This enhances the validity of our analysis first put forward in \cite{Bagchi:2023fbj} and shows that the Carrollian picture emerges naturally out of a careful limit of AdS Witten diagrams.

\medskip

\subsection{Advantages of the Carrollian approach}

There are various reasons why we believe that the Carrollian perspective is more appropriate for Minkowskian holography. For the reader wanting to take away the main message without going through the whole of the paper, we provide a list below with brief descriptions. We will be elaborating on (most of) these points in the next section.  

\begin{itemize}
\item{\underline{\em{Symmetries:}}} The most rudimentary check of a holographic correspondence is the matching of symmetries between the bulk and the boundary theories. In AdS$_5$/CFT$_4$, this is the isomorphism between the isometry algebra $so(4,2)$ of the bulk and the conformal algebra, which is also $so(4,2)$ in $d=4$. Even without going into the intricacies of asymptotic symmetries, the isometry of Minkowski spacetime is the Poincare algebra $iso(3,1)$, and not just its Lorentz sub-algebra $so(3,1)$. We believe that there is no reason to treat the translation sub-algebra differently and disregard this from the global symmetry algebra of the boundary theory. The celestial CFT, which is based on the observation of the realisation of the Lorentz algebra as the global conformal algebra, unfortunately, does this. The Carrollian programme, on the other hand, is based on the isomorphism between the 4d Poincare algebra and the global symmetries of 3d Carroll CFTs, which is also $iso(3,1)$ and treats translations democratically. 

\item{\underline{\em{Changing of scaling dimensions under translation}}}: As a direct consequence of not considering translations in the symmetries of the dual field theory, Celestial CFTs inherit undesirable properties. The operators in the 2d dual Celestial CFT are labeled by weights under $L_0, \bar{L}_0$, which are $(h, \bar{h})$. If we perform spacetime translations on these operators, these weights shift. Assuming that there is a map between bulk fields and boundary operators, this means that if you translate the bulk field, it would map into a different operator on the boundary. This is very strange from the point of a dual field theory because translations are very natural transformations in the bulk, and this manifestation of shifting weights is a direct consequence of not taking these symmetries properly into account in the putative dual field theory. The Carrollian higher dimensional dual theory has fields that transform properly under spacetime translations, thus eradicating this problem.  

\item{\underline{\em{Mellin transformation and divergences}}}: As indicated above and will be detailed later, the Mellin transformation forms the basis of the relation between 2d Celestial CFTs and 4d asymptotically flat spacetimes by relating S-matrix elements to correlation functions of the 2d field theory through the integral transformation. However, Mellin transformations on, e.g., graviton amplitudes yield infinite answers. Although it is generally argued that this is not a problem since general relativity is not an ultra-violet complete theory, we find this argument unconvincing. Graviton amplitudes are connected to observables and, hence, in the age of gravitational wave detection, of prime experimental relevance. 

In \cite{Banerjee:2018gce}, it was shown that by considering covariance under the whole of the Poincare group and, in particular, spacetime translations, the Mellin transformation can be improved to a {\em Modified Mellin} transformation. It was later shown in \cite{Banerjee:2019prz} that this transformation leads to finite graviton amplitudes in the bulk. The map between the correlation function of Carrollian primaries and S-matrix elements utlilise the modified Mellin transformation instead of the usual Mellin transformation and hence also addresses the problem of the ultraviolet divergence of original Mellin transformations.

\item{\underline{\em{Limit from AdS}}}: Holography is, of course, best understood in its AdS avatar, where the dual theory is a co-dimension one conformal field theory living on the boundary of AdS. The infinite radius limit of AdS leads to flat spacetimes. The bulk theory does not reduce in dimension while the limit is taken. It is natural to think of flat holography as a singular limit of AdS holography. Perhaps all features of flat spacetimes would not appear in this limit, but given the wealth of information we have assembled in AdS/CFT, taking limits, albeit singular, is an attractive proposition. It is thus also natural that on the dual side, the infinite radius limit is perceived as a singular limit on the CFT but does not reduce the dimension. 

It was shown in \cite{Bagchi:2012cy} that this limit on the dual side actually sends the speed of light to zero on the field theory side, which is a natural consequence of a timelike boundary being boosted to a null boundary. This process contracts the CFT algebra in $(d-1)$ dimensions to a Carroll conformal algebra in $(d-1)$ and not just a $(d-2)$ dimensional relativistic conformal algebra. The $(d-2)$-dimensional CFT algebra forms a subalgebra of the $(d-1)$-dimensional Carroll conformal algebra. This sub-algebra maps to the Lorentz subalgebra of the $d$ dimensional Minkowski spacetime. But, crucially, the translation generators are missed. It is thus important to take the whole of the conformal Carroll algebra, which appears in the limit of the vanishing speed of light. 

\item{\underline{\em{The case of three bulk dimensions}}}: The asymptotic symmetry algebra for 3d asymptotically flat spacetimes is the BMS$_3$ algebra. For Einstein gravity, this involves a central term between the superrotations and supertranslations. The superrotations themselves form a Virasoro algebra, which should form the basis of a celestial construction. This Virasoro algebra does not contain a central term for Einstein gravity and hence cannot account for the entropy of Flat Space Cosmologies (FSCs), which, on the other hand, is captured perfectly by the BMS-Cardy analysis that follows from a Carrollian approach \cite{Bagchi:2012xr}{\footnote{See \cite{Bhattacharjee:2023sfd} for a recent attempt at an effective 1d field theory approach reproducing the FSC entropy, although this seems to be a restriction of the 2d dual theory to the $u=0$ slice, where $u$ is the null time direction on $\mathscr{I}^\pm$.}}. The co-dimension one Carrollian dual theory has also had many other successes, e.g., in reproducing stress-tensor correlators \cite{Bagchi:2015wna}, entanglement entropy \cite{Bagchi:2014iea, Jiang:2017ecm,Hijano:2017eii}. It seems unlikely that the Celestial approach, without the proper inclusion of the supertranslations and, hence, the crucial central term, would be able to reproduce bulk answers. 

\end{itemize}

Apart from all of these specific points listed above, the holographic principle itself was historically based on the observation that the black hole entropy is proportional to the area of the event horizon, and hence, from the extensivity of entropy, this leads to the proposal that a theory of quantum gravity which describes the black hole is equivalent to a quantum field theory in one lower dimension where this entropy becomes extensive. It is thus in the spirit of the original holographic principle to have a co-dimension one dual for flat spacetimes. 

\newpage

\subsection{What is in this paper?}
Our paper, as advertised above, is made of two parts. 

The first part, which is essentially Sec. \ref{sec:celestialvscarroll}, is a comparison between the Celestial and Carrollian approaches to flat holography. We give the reader a quick overview of Celestial holography, based principally on the observation that S-matrix elements can be related to 2d Celestial CFT correlation functions by the Mellin transformation. We then show that by taking care of translations properly, we arrive at a three-dimensional boundary theory whose correlators are also related to S-matrix elements, now by the {\em modified Mellin transformation}. 
We also comment on a different approach for Carrollian holography advocated in \cite{Donnay:2022aba,Donnay:2022wvx}. 

\medskip

The second part of the paper addresses the flat limit of AdS/CFT and specifically how to obtain Carrollian correlation functions from Witten diagrams, generalizing our construction \cite{Bagchi:2023fbj} to operators with spin. The time-dependent correlation functions of Carrollian CFTs emerge naturally as the leading order term under a large $R$ series expansion ($R$ being the AdS radius) for the scattering amplitudes of Witten diagrams \cite{Bagchi:2023fbj}. This result, though generic, was just established for massless scalar particles. In this paper, we extend it to massless spinning particles.

\medskip

Sec. \ref{sec:carrollfieldtheory} and Sec. \ref{sec:carrollcorrelators} are devoted to the field theory analysis of Carrollian CFTs where various correlation functions are computed from symmetries. Sec. \ref{sec:flatlimitwitten} contains a review of the construction of \cite{Bagchi:2023fbj} of scalar Witten diagrams and the flat limit, and we then set up the formulation for more complicated operators with non-zero spin in the latter half of this section. 

\medskip

In Sec. \ref{sec:wittendiagex}, we provide explicit examples of our construction and show matching between the limiting construction of Sec. \ref{sec:flatlimitwitten} and the symmetry analysis of Sec. \ref{sec:carrollcorrelators}. We have a total of {\em twelve} examples. The reader is pointed to Sec. \ref{sec:summarywitten} for a quick summary of all the computations. We end with some discussions in Sec. \ref{sec:conclusions}.   




\newpage

\part{Carrollian Holography: An overview}

\bigskip \bigskip

In this part, which essentially consists of the following section, we give an extensive overview of Carrollian holography and highlight the advantages over its Celestial counterpart. 

\medskip
The version of Carrollian holography that we will discuss is the one which is intimately linked to the {\em{Modified Mellin transformation}}. The principal references for this are \cite{Banerjee:2018gce,Bagchi:2022emh, Bagchi:2023fbj}. There exists another school of research on Carrollian holography advocated in \cite{Donnay:2022aba,Donnay:2022wvx}. (See also \cite{Salzer:2023jqv,Nguyen:2023vfz,Saha:2023hsl,Saha:2023abr}.) We will briefly compare and contrast this to our approach at the end of Sec. \ref{sec:celestialvscarroll}. 

\bigskip \bigskip

\section{Flat holography and integral transformations}
\label{sec:celestialvscarroll}

The main observable for quantum gravity in asymptotically flat spacetimes are the S-matrix elements. Thus, it is mandatory that the holographic dual should be able to capture these S-matrix elements in terms of its correlation functions. Below, we detail how two different integral transformations on the S-matrix elements lead to two different dual pictures of holography in asymptotically flat spacetime.

\subsection{Mellin transformations and 2d Celestial CFT}

The Celestial approach to Minkowskian holography proposes  2d CFT on the celestial sphere as a dual to 4d gravity in asymptotically flat spacetimes. The essence of this proposal emerges from the observation that the double cover of the Lorentz group in four spacetime dimensions is SL$(2,\mathbb{C})$, which is the group of global conformal transformation in $d=2$. Furthermore, the existence of the subleading soft graviton theorems suggests that this global group should be extended to include the infinite local conformal transformations as well. The enhancement of these symmetries indicates a 2d CFT dual. Let us elaborate on these statements by considering the scattering of massless particles. 
The momenta of a massless particle $p^\mu$ can be parametrized as 
\begin{equation} \label{parameter}
	p^\mu=\omega (1-z\bar{z},z+\bar{z},-i(z-\bar{z}),1+z\bar{z})
\end{equation}
Here $\omega$ denotes the energy of the particle and $z,\bar{z}$ are coordinates of the celestial sphere on null infinity. A generic Lorentz transformation or SL$(2,\mathbb{C})$ transforms the massless momenta as 
\begin{align} \label{lorentz}
	\mathbb{P}  &\to \Lambda \mathbb{P} \Lambda^{-1},  \quad  \mathbb{P}=p_\mu\sigma^\mu \\  \nonumber
	i.e. \quad	\omega&\to \omega |cz+d|, \quad z\to \frac{az+b}{cz+d}, \quad \bar{z} \to \frac{\bar{a}\bar{z}+\bar{b}}{\bar{c}\bar{z}+\bar{d}}
\end{align}
Here $\sigma^\mu=(1, \s^i)$ are the Pauli matrices and $a,b,c,d$ parametrise Lorentz transformations:
\begin{equation}
	\Lambda= \begin{bmatrix}
		a  \quad b  \\
		c   \quad d
	\end{bmatrix}  \quad \text{such that} \quad  ad-bc=1.
\end{equation}
From \eqref{lorentz}, it is evident that the Lorentz group acts as global conformal transformations on the celestial sphere. Now let us consider the particle states with definite momentum, which can be created or annihilated by the standard momentum space operators $a(\omega,z,\bar{z},\sigma)$ and $a^{\dagger}(\omega,z,\bar{z},\sigma)$. The additional level $\sigma$ of $(a,a^\dagger)$ denotes the helicity of the corresponding massless particle. We shall use a compact notation $a(\epsilon\omega,z,\bar{z},\sigma)$, where $\epsilon=\pm 1$ would decide if the operator is a creation or an annihilation operator.

\medskip

Using these momentum space operators, one can define  local operators on the celestial sphere via a Mellin transformation as
\begin{align} \label{mellin}
	\mathcal{O}^{\epsilon}_{h,\bar{h}}(z,\bar{z})=\int_{0}^{\infty} d\omega \, \omega^{\Delta-1}a(\epsilon\omega,z,\bar{z},\sigma), \quad  h,\bar{h}=\frac{\Delta \pm \sigma}{2}
\end{align}
Essentially, this Mellin transformation amounts to a change of basis in which the action of the boost generators is diagonalized. The transformation trades the energy $\omega$ with the boost weights $\Delta$.  One advantage of choosing this basis is a manifestation of the conformal properties in the wave functions defined in \eqref{mellin}. As suggested by the notation, the local operator on celestial sphere $\mathcal{O}_{h,\bar{h}}(z,\bar{z})$ now transforms like a primary field of weights $h$ and $\bar{h}$ with respect to the SL$(2,\mathbb{C})$.
\begin{align}\label{eq:celpritrans}
	\mathcal{O}^{\epsilon}_{h,\bar{h}}(z,\bar{z}) \to \mathcal{O}^{\epsilon}_{h,\bar{h}}(z',\bar{z'}) =\frac{1}{|cz+d|^{2h}}\frac{1}{|\bar{c}\bar{z}+\bar{d}|^{2\bar{h}}}	\mathcal{O}^{\epsilon}_{h,\bar{h}}(z,\bar{z})
\end{align}

\medskip

Having this new basis at hand, the scattering amplitude now readily translates into the conformal correlators on the celestial sphere,
\begin{align}
	\langle \mathcal{O}_1(z_1,\bar{z}_1)\mathcal{O}_2(z_2,\bar{z}_2)....\mathcal{O}_n(z_n,\bar{z}_n)\rangle = \prod_{i=1}^n \int_{0}^{\infty} d\omega_i \omega_i^{\Delta_1-1}\mathcal{S}_n(\omega_i,z_i,\bar{z}_i,\sigma_i)
\end{align}
In the above, $\mathcal{O}_i(z_i,\bar{z}_i)$ is the shorthand for $\mathcal{O}^{\epsilon}_{h_i,\bar{h}_i}(z_i,\bar{z}_i)$. While doing this transformation, we go from the basis of plane waves to the basis of conformal primary wavefunctions. These wavefunctions are given by
\begin{equation}\label{eq:confpri}
	\Phi(x|z,\bar{z}) = \int d\omega \, \omega^{\Delta-1} \, e^{\pm i \tilde{q} \cdot x} e^{-\epsilon \omega} = \dfrac{(\mp i)^{\Delta} \Gamma(\Delta)}{(-\tilde{q} \cdot x \mp i \epsilon)^{\Delta}}.
\end{equation}
$\Phi(x|z\bar{z})$ satisfies the massless Klein-Gordon equation and it transforms like \eqref{eq:celpritrans} under an SL$(2,\mathbb{C})$ transformation. 

\medskip

Although the conformal properties are explicit in this basis, the Mellin transformation runs into some peculiarities while dealing with the translations. For example, under the translation by $P^3$, the the boundary operator $\mathcal{O}_{h,\bar{h}}(z,\bar{z})$ changes as 
\begin{align}
	\delta_{P^3}\mathcal{O}_{h,\bar{h}}(z,\bar{z})&=\int_{0}^{\infty}d\omega \omega^{\Delta-1}[P^3,a(\epsilon\omega,z,\bar{z},\sigma)]   \\ \nonumber
	&=(1-z\bar{z})\int_{0}^{\infty}d\omega \omega \omega^{\Delta-1}a(\epsilon\omega,z,\bar{z},\sigma)  \\ \nonumber
	&=(1-z\bar{z})\mathcal{O}_{h+\frac{1}{2},\bar{h}+\frac{1}{2}}(z,\bar{z})
\end{align}
The above equation shows that the bulk translations shift the conformal dimensions of the boundary fields by $\frac{1}{2}$. This is a feature we find undesirable from the point of view of a boundary dual theory. For us, this is an indication that the symmetry algebra of the boundary theory was not taken into account appropriately, and there is something amiss with this 2d Celestial picture of holography in flat spacetimes. 

\medskip

As discussed in the introduction, the Mellin transformation comes with the inherent problem of being UV divergent. This, e.g., shows up in the transformations of flat space graviton amplitudes. We will see below that there is a single solution to both problems above when we take translations into account properly.  

\subsection{Covariance under Poincare and Modified Mellin Transformations}
Our aim now is to take translations into account properly so that the weights of the boundary fields do not shift under this subgroup of the Poincare group. In a sense, this is demanding covariance of the theory under the whole Poincare group and not its Lorentz subgroup. To do this, we do the following time translation:
\be{}
\mathcal{O}^{\epsilon}_{h,\bar{h}}(z,\bar{z}) \mapsto e^{-iHU}\mathcal{O}^{\epsilon}_{h,\bar{h}}(z,\bar{z}) e^{iHU}.
\ee
Following the Mellin transformation \eqref{mellin}, this leads to some three dimensional fields: 
\begin{align}\label{modmellin}
	\Phi^{\epsilon}_{h,\bar{h}}(u,z,\bar{z})&=e^{-iHU}\mathcal{O}^{\epsilon}_{h,\bar{h}}(z,\bar{z}) e^{iHU} =\int_{0}^{\infty}d\omega \omega^{\Delta-1}e^{-iHU}a(\epsilon\omega,z,\bar{z},\sigma) e^{iHU}  \nonumber \\ 
	\Rightarrow \Phi^{\epsilon}_{h,\bar{h}}(u,z,\bar{z})&=\int_{0}^{\infty}d\omega \omega^{\Delta-1}e^{i\epsilon\omega u}a(\epsilon\omega,z,\bar{z},\sigma).
\end{align}
In the above equation, we define $u=U(1+z\bar{z})$, where $U$ and $u$ are the time in the bulk and the boundary, which are related by a conformal factor $(1+z\bar{z})$. This modification now gives rise to a field that lives on the whole of null infinity ($\mathscr{I}^\pm$) instead of just the celestial sphere, and the relation \refb{modmellin} was called the {\em Modified Mellin transformation} in \cite{Banerjee:2018gce}, where it was first introduced. 

\medskip

The construction above gives rise to a field on $\mathscr{I}^\pm$, and we briefly study the transformation properties of $\Phi^{\epsilon}_{h,\bar{h}}(u,z,\bar{z})$ under the global Poincar\'e group. These transformation rules directly follow from the above integral transformation of creation and annihilation operators \cite{Banerjee:2018gce}. For simplicity, let us consider scalar particles in the bulk (i.e, $h- \bar{h}=0$). $\Phi^{\epsilon}_{h,\bar{h}}(u,z,\bar{z})$ then transforms under the Lorentz group as 
\begin{align}
	\Phi^{\epsilon}_{h,\bar{h}}(u,z,\bar{z}) \to \Phi^{\epsilon}_{h,\bar{h}}(u',z',\bar{z'})&=\int_{0}^{\infty}d\omega \omega^{\Delta-1}e^{i\epsilon\omega u}\Lambda^{-1} a(\epsilon\omega,z,\bar{z})\Lambda    \\ \nonumber
	&=\int_{0}^{\infty}d\omega \omega^{\Delta-1}e^{i\epsilon\omega u}a\Big(\epsilon\omega|cz+d|^2,\frac{az+b}{cz+d},\frac{\bar{a}\bar{z}+\bar{b}}{\bar{c}\bar{z}+\bar{d}}\Big)   \\ \nonumber
	&=\frac{1}{|cz+d|^{2\Delta}}\Phi^{\epsilon}_{h,\bar{h}}\Big(\frac{u}{|cz+d|^2},\frac{az+b}{cz+d},\frac{a\bar{z}+\bar{b}}{\bar{c}\bar{z}+\bar{d}}\Big) 		
\end{align}

We performed the last step by redefining the integration variable $\omega$. Following similar arguments, it can be shown that the  generic transformation rules for the spinning primaries under the Lorentz group are given by
\begin{align} \label{suprot}
	\Phi^{\epsilon}_{h,\bar{h}}(u,z,\bar{z}) \to \Phi^{\epsilon}_{h,\bar{h}}(u',z',\bar{z'})=\frac{1}{|cz+d|^{2h}}\frac{1}{|\bar{c}\bar{z}+\bar{d}|^{2\bar{h}}}	\Phi^{\epsilon}_{h,\bar{h}}\Big(\frac{u}{|cz+d|^2},\frac{az+b}{cz+d},\frac{a\bar{z}+\bar{b}}{\bar{c}\bar{z}+\bar{d}}\Big) 
\end{align}

\medskip

Further, the action of bulk translations can also be derived following similar prescriptions. For example, the action of the generator $P^3$ is given by
\begin{align}
	\delta_{P^3}\Phi^{\epsilon}_{h,\bar{h}}(u,z,\bar{z})&=\int_{0}^{\infty}d\omega \omega^{\Delta-1}e^{i\epsilon\omega u}[P^3,a(\epsilon\omega,z,\bar{z})]  \\ \nonumber
	&= \int_{0}^{\infty}d\omega \omega^{\Delta-1}e^{i\epsilon\omega u} \omega(1-z\bar{z})a(\epsilon\omega,z,\bar{z})  \\  \nonumber
	&=(1-z\bar{z})\partial_u\Phi^{\epsilon}_{h,\bar{h}}(u,z,\bar{z})
\end{align}
Generically $\Phi^{\epsilon}_{h,\bar{h}}(u,z,\bar{z})$  transforms under translations as
\begin{align} \label{suptra}
	\Phi^{\epsilon}_{h,\bar{h}}(u,z,\bar{z})\to\Phi^{\epsilon}_{h,\bar{h}}(u',z',\bar{z'})=		\Phi^{\epsilon}_{h,\bar{h}}(u+p+qz+r\bar{z}+sz\bar{z},z,\bar{z})
\end{align}
where $p,q,r$ and $s$ parametrises the four spacetime translations. 

\medskip

Now, the basis of plane waves becomes modified under the modified Mellin transformation \eqref{modmellin}. They take up the form
\begin{equation}\label{eq:carrollwave}
	\psi(x|u,z,\bar{z}) =  \int d \omega \omega^{\Delta-1} e^{\pm i \omega \, u} e^{\pm i \tilde{q} \cdot x} e^{-\epsilon \omega} = \dfrac{(\mp i)^{\Delta} \Gamma(\Delta)}{(-u-\tilde{q} \cdot x \mp i \epsilon )^{\Delta}} \, .
\end{equation}
This \textit{Carrollian primary wavefunction} is a modified avatar of \eqref{eq:confpri} \cite{Banerjee:2018gce}.  \eqref{eq:carrollwave} satisfies the massless Klein-Gordon equation for a fixed value of $u$ i.e.,
\begin{equation}
	\pdv{x_{\mu}} \pdv{x^{\mu}} \psi(x|u,z,\bar{z}) = 0 ~~~ \text{for a fixed $u$} \, .
\end{equation}
They also have the crucial feature that they transform covariantly under the complete set of bulk isometries: bulk Lorentz transformations and bulk translations i.e., they transform like \eqref{suprot} and \eqref{suptra}. In doing the modified Mellin transform, we go from the momentum eigenstate (plane wave) basis to the boost eigenstate basis.

\medskip

\subsection{BMS and Carrollian holography} \label{primary}

Interestingly, the asymptotic symmetries of asymptotically flat spacetimes are not only the Poincare group but an infinite dimensional group called the BMS$_4$ group, which includes supertranslations and superrotations extending usual translations and rotations of the Poincare group. Historically, Bondi, van der Burg, Metzner \cite{Bondi:1962px} and Sachs \cite{Sachs:1962wk} discovered the supertranslations while studying the asymptotic structures of flat spacetimes. Later on, it was realised that the asymptotic symmetry group could further be extended to include the superrotations \cite{Barnich:2010eb}. Gravitational scattering in the bulk respects not only global Poincare symmetries but these infinite dimensional supertranslations and superrotations as well. The leading and subleading soft graviton theorems are direct consequences of the existence of these infinite dimensional asymptotic symmetries.

\medskip

The most celebrated example of asymptotic symmetry is the case of AdS$_3$ by Brown and Henneaux \cite{Brown:1986nw}, who discovered two copies of the Virasoro algebra arising from their canonical analysis extending the finite dimensional $so(2,2)$ isometry algebra of AdS$_3$. This was, of course, also realised as the symmetries of a 2d relativistic CFT and is seen as one of the principle precursors of the AdS/CFT correspondence. 

\medskip

BMS symmetries can also be realised in a similar holographic manner in a lower dimensional theory \cite{Bagchi:2010zz, Bagchi:2012cy, Duval:2014uva, Bagchi:2016bcd}. These symmetries are the same as Carrollian conformal symmetries in one lower dimension \cite{Duval:2014uva}. The proposal is thus: 
$${\boxed{\textit{Gravity in D-dim asymptotically flat spacetime} \equiv \textit{Carrollian CFT in D-1 dim}.}}$$
The dual theory resides on the null boundary of the asymptotically flat spacetime. This is the foundation of the Carrollian approach to flat holography. The most rudimentary check of holography, the matching of symmetries between the bulk and boundary theory, is inbuilt into this approach, unlike Celestial holography where translations and, as a consequence, supertranslations are not included as symmetries of the dual field theory. 

\medskip

We will focus on the relation between 4d asymptotically flat spacetimes and 3d Carrollian CFTs. The algebra in question is thus the BMS$_4$ algebra: 
\begin{align} \label{BMS}
	[L_n,L_m]&=(n-m)L_{n+m}, \quad [\bar{L}_n,\bar{L}_m]=(n-m)\bar{L}_{n+m}  \\ \nonumber
	[L_n,M_{r,s}]&=\left(\frac{n+1}{2}-r \right)M_{n+r,s}, \quad [\bar{L}_n,M_{r,s}]=\left(\frac{n+1}{2}-s \right)M_{r,n+s}    \\   \nonumber
	[M_{r,s},M_{p,q}]&=0.
\end{align}
This is the asymptotic symmetry algebra at null infinity ($\mathscr{I}^\pm$), which is topologically $\mathbb{R}_u \times$ S$^2$. In the above, $L_n$ and $\bar{L}_n$ are the generators of superrotations and form the Virasoro sub-algebra. $M_{r,s}$ are the supertranslations, the angle dependent translations of the null direction. $\{L_a, \bar{L}_a; M_{i,j}\}$ where $a=0, \pm 1$ and $i,j = 0, 1$ forms the 10-parameter Poincare subalgebra with  $\{L_a, \bar{L}_a\}$ forming the Lorentz part. The vector field representation of the generators of this algebra, which also can be obtained as asymptotic Killing vectors projected on null infinity, is given by 
\begin{subequations}\label{eq:BMSgenerators}
	\begin{align}
		&L_n= z^{n+1}\partial_z+\frac{1}{2}(n+1)z^nu\partial_u, \quad
		\bar{L}_n=\bar{z}^{n+1}\partial_{\bar{z}}+\frac{1}{2}(n+1)\bar{z}^nu\partial_u,  \\ 
		&M_{r,s}=z^r\bar{z}^s\partial_u  \qquad  \forall \quad n,r,s \in Z
	\end{align}
\end{subequations}
Here $z, \z$ are the stereographic coordinates on S$^2$, and $u$ is the null direction on $\mathscr{I}^\pm$. 

\medskip

We define our putative dual 3d Carrollian CFT to live on $\mathscr{I}^\pm$ and the generators \refb{eq:BMSgenerators} reflect the $\mathbb{R}_u \times$ S$^2$ topology of the background manifold. We will now build the quantum field theory living on $\mathscr{I}^\pm$ with \refb{BMS} as its underlying symmetry. We label states $\Phi(0,0,0)$ by their weights under $L_0, \bar{L}_0$: 
\begin{align}
	[L_0,\Phi(0,0,0)]=h\Phi(0,0,0), \quad [\bar{L}_0, \Phi(0,0,0)]=\bar{h}\Phi(0,0,0)   
\end{align}
Taking a cue from the usual relativistic CFT, we propose a class of Carrollian primary states $\Phi_p(0,0,0)$, which are given by: 
\begin{subequations}
	\begin{align}
		[L_n,\Phi_p(0,0,0)]&=[\bar{L}_n,\Phi_p(0,0,0)] =0, \quad \forall n > 0 \label{pr-a}\\ 
		[M_{r,s},\Phi_p(0,0,0)]&=0, \quad  \forall r,s > 0 \label{pr-b}
	\end{align}
\end{subequations}
Notice that on top of the relations \refb{pr-a} that are identical to a 2d relativistic CFT, we also impose \refb{pr-b}, which is crucial to a 3d Carroll CFT. These conditions induce the following transformation rules at arbitrary spacetime point
\begin{subequations}\label{primary}
	\begin{align} 
		[L_n,\Phi_p(u,z,\bar{z})]&=z^{n+1}\partial_z \Phi_p(u,z,\bar{z})+\Big(h+\frac{u}{2}\partial_u \Big)z^n\Phi_p(u,z,\bar{z}) \\ 
		[\bar{L}_n,\Phi_p(u,z,\bar{z})]&=\bar{z}^{n+1}\partial_{\bar{z}} \Phi_p(u,z,\bar{z})+\Big(\bar{h}+\frac{u}{2}\partial_u \Big)\bar{z}^n\Phi_p(u,z,\bar{z}) \\ 
		[M_{r,s},\Phi_p(u,z,\bar{z})]&=z^r\bar{z}^s\partial_u\Phi_p(u,z,\bar{z})
	\end{align}
\end{subequations}
These transformation rules define conformal Carroll primaries. The Carrollian operators that obey the primary conditions \refb{primary} only for the global Poincare subgroup would be called quasi-primary operators. 

\medskip

\subsection{Modified Mellin and 3d Carroll CFT}

From our discussions above, it is now straight-forward to see that the transformation rules \refb{primary} of the Carrollian quasi-primary operators are the infinitesimal versions of the ones we derived from the modified Mellin transformations earlier in \refb{suprot} and \refb{suptra}. So we see that the fields earlier defined from demanding covariance of the Mellin transformed operators under spacetime translation are identical to 3d Carrollian quasi-primary operators \cite{Bagchi:2022emh}. This confirms that the modified Mellin transformations of the momentum space operators in the bulk indeed give rise to Carrollian conformal fields that live on the null boundary.

\medskip

Having this framework at hand, one can relate the scattering amplitudes in the bulk with the correlation functions of Carrollian primaries defined by \eqref{primary}. This is given by \cite{Bagchi:2022emh}: 
\begin{align} \label{scattering}
	\langle \prod_{i=1}^{n}	\Phi^{\epsilon_i}_{h_i,\bar{h}_i}(u_i,z_i,\bar{z}_i)\rangle = \prod_i \int_{0}^{\infty} d\omega_i ~ \omega_i^{\Delta_i-1}e^{i\omega_iu_i}\mathcal{S}_n (\omega_i,z_i,\bar{z}_i,\sigma_i)
\end{align}
Here $\mathcal{S}_n$ are the $n$-point massless scattering amplitudes in the bulk. This can equivalently be thought of as doing a basis change from plane waves to Carrollian primary wavefunctions of \eqref{eq:carrollwave}.

\medskip

In \cite{Bagchi:2022emh}, after laying out the basic construction, the two point function was verified from the free propagation amplitude of a single massless particle. The three point scattering amplitudes of massless particles in Minkowski spacetime are zero apart from a few specific scenarios. These cases were also shown to be consistent with \eqref{scattering} in \cite{Bagchi:2023fbj}. Later in this paper, we will put this formula to further tests by considering various scattering amplitudes involving spinning particles and in spacetimes of signature (2,2). 

\medskip

\paragraph{Bulk energy conservation from Carroll.} The modified Mellin transformation has several very nice features. Here, we highlight an important connection between bulk energy conservation and boundary Carroll time translations. Acting $M_{00}$ of \eqref{eq:BMSgenerators} on both sides of \eqref{mt}, we get
\begin{equation}\label{eq:wardm00}
	-i\sum_{i=1}^{n} \partial_{u_i} \langle \prod_{i=1}^{n} \phi^{\epsilon_i}_{h,\bar{h}}(u_i,z_i,\bar{z}_i) \rangle = \prod_{i=1}^{n} \int  d\omega_i \, \omega^{\Delta_i-1}_i \,  e^{-i\epsilon_i \omega_i u_i} \, \left( - i \sum_{j=1}^{n} \epsilon_j \omega_j \right) S(\{\omega_i,z_i,\bar{z}_i,\sigma_i\})
\end{equation}
From the Ward identities of Carroll time translations, we know that the LHS of \eqref{eq:wardm00} is zero. Thus, we must have
\begin{equation}
	\sum_{j=1}^{n} \epsilon_j \omega_j = 0 \, .
\end{equation}
This is just the energy conservation between the incoming and outgoing particles ($\epsilon_i = \pm  1$ denotes the incoming or outgoing particles).

\medskip

\paragraph{Limit from AdS and Witten diagrams.} We will also show that the correlation functions associated with these amplitudes can also be obtained by implementing a suitable flat limit of AdS-Witten diagrams. An essential component of the Witten diagrams is the bulk to boundary propagator $\textbf{K}_\Delta(\p,\x)$. In the embedding space representation, it is given by \cite{Penedones:2010ue}:
\begin{equation}
	\textbf{K}_\Delta(\p,\x)=\frac{C_\Delta^d}{(-P \cdot X +i\epsilon)^\Delta} \, .
\end{equation}
Our flat limit, which contracts the symmetries of both the bulk and the boundary, ensures that $\textbf{K}_\Delta(\p,\x)$ reduces to the Carrollian primary wavefunction \eqref{eq:carrollwave}:
\begin{equation}
	\lim_{R \to \infty} \textbf{K}_\Delta(\p,\x)  \propto \dfrac{(\mp i)^{\Delta} \Gamma(\Delta)}{(-u-\tilde{q} \cdot x \mp i \epsilon )^{\Delta}}
\end{equation}
Here, $R$ is the AdS radius. We have implemented a careful large AdS radius limit that keeps track of the emergent null direction. This is what one should expect from the point of view of flat holography \cite{Bagchi:2022emh}. In our limit, flat space emerges in the center of AdS. We expect the S-Matrix to arise from CFT data \cite{Penedones:2010ue,Fitzpatrick:2011hu,Hijano:2019qmi} \footnote{See Section 3.1 of \cite{Bagchi:2023fbj} for a comprehensive review of the literature utilizing this approach.}. Thus, it is satisfying to see that Carrollian primary wavefunctions (the basis functions encoding Carrollian primary fields) emerge in a carefully constructed limit. We give the details of this construction in Section \ref{sec:flatlimitwitten}.

\subsection{Modified Mellin as the superstructure}
We begin by claiming that the above version of the holographic principle for asymptotically flat spacetimes is the superstructure from which other formulations, including Celestial holography and other approaches to Carrollian holography, \cite{Donnay:2022aba,Donnay:2022wvx} appear as subsectors. To substantiate our claims, we consider again the modified Mellin transformation \cite{Banerjee:2018gce} 
\begin{equation}\label{mmt}
	\langle \Phi_1 (u_1, {z}_1) \ldots \Phi_n (u_n, z_n) \rangle = \prod_i \int_0^\infty d \omega_i \ \omega_i^{\Delta_i-1} e^{-i \epsilon_i \omega_i u_i} S(\{\epsilon_i \omega_i, z_i, \sigma_i\}).
\end{equation}
As we have reviewed in \cite{Bagchi:2023fbj}, this relates the 4d flat space scattering amplitudes to 3d Carrollian primary ($\Phi$) correlators which depend on all three boundary directions $(u, z, \bar{z})$ of $\mathscr{I}^+$. (We have suppressed the $\bar{z}$ dependence in the above and also will do so below.) 

\medskip

\paragraph{Carroll and Celestial duals.} 
It is clear that the $u\to0$ limit of the modified Mellin transformation lands one up in the usual Mellin transformation:
\begin{equation}\label{mt}
	\langle \phi_1 ({z}_1) \ldots \phi_n (z_n) \rangle = \prod_i \int_0^\infty d \omega_i \ \omega_i^{\Delta_i-1} S(\{\epsilon_i \omega_i, z_i, \sigma_i\})
\end{equation}
Here, we are relating 4d flat space scattering amplitudes to 2d Celestial primary ($\phi$) correlators. The usual Mellin transformation has the problem that this leads to non-finite amplitudes in the bulk. E.g., four point graviton amplitudes in the bulk 4d asymptotically flat spacetime diverge. 

\medskip

The modified Mellin transformation, on the other hand, helps eradicate this problem, and it has been shown in \cite{Banerjee:2019prz} that plugging in the extra $u$ dependent term renders the graviton amplitudes finite by acting like a regulator.

\medskip

It was shown in \cite{Banerjee:2018gce} that $u\to0$ also makes sense on the field theory side. From the point of view of the Celestial CFT, $u$ acts as a UV regulator, which can be taken to zero {\em at the end of the calculation}. It is plausible that one would be able to understand a 2d Celestial CFT as the reduction of a 3d Carroll CFT, which lives on a particular $u=0$ slice of the 3d Carroll manifold. 

\medskip

\paragraph{$\D=1$ Carrollian CFTs.} In \cite{Donnay:2022aba,Donnay:2022wvx}, a different proposal for Carrollian holography was put forward\footnote{See also \cite{Salzer:2023jqv,Nguyen:2023vfz}} from asymptotic expansions of fields in the bulk spacetime and IR physics in asymptotically flat spacetimes was related to sourced Ward identities on the Carroll side. While this is an interesting approach and seems to yield results consistent with some 4d bulk analysis, below we point out some possible issues with this in light of our formulation. 

\smallskip

We begin by observing that if one takes $\Delta_i=1$ in the modified Mellin transformation, one reduces this to a Fourier transform:
\begin{equation}\label{ft}
	\langle \chi_1 (u_1, {z}_1) \ldots \chi_n (u_n, z_n) \rangle = \prod_i \int_0^\infty d \omega_i \  e^{-i \epsilon_i \omega_i u_i} S(\{\epsilon_i \omega_i, z_i, \sigma_i\})
\end{equation}
This relation maps 3d Carroll primaries $\chi_i$ with $\Delta_i=1$ to 4d scattering in flatspace. The two transformations Eq.(\ref{mt}) and Eq.(\ref{ft}) are related by the map considered in \cite{Donnay:2022aba,Donnay:2022wvx}. Both of these are clearly sub-structures to the super-structure Eq.(\ref{mmt}). 

\medskip
The approach advocated in \cite{Donnay:2022aba,Donnay:2022wvx} purports that the 3d Carrollian CFT dual to 4d asymptotically flat spacetime should only have Carroll primaries with weights $\Delta_i=1$. We believe that there are reasons why this subsector of $\Delta_i=1$ may not be consistent by itself. Below, we list a couple of these reasons. 

\medskip

\noindent {\underline {\em {IR divergences:}}} The way to arrive at the Carrollian CFTs \cite{Donnay:2022aba,Donnay:2022wvx} is to restrict the more generic Carroll CFTs with arbitrary $\D_i$ to $\D_i = 1$. However, taking the limit of $\D_i \to 1$ in the modified Mellin transformation is fraught with danger as there are poles at $\D_i=1$. These are connected to the leading soft graviton theorems, as we elaborate below. The problem with divergences at $\D_i = 1$ manifests itself explicitly in the Ward identities of Carrollian CFTs as well, as the correlation functions blow up \cite{Salzer:2023jqv,Nguyen:2023miw}. A complimentary way to see this divergences is by taking the bulk correlation functions to the null boundary. This approach was taken recently in \cite{Nguyen:2023miw}. 

\medskip

As is well documented, S-matrices in asymptotically flat spacetimes are meddled with IR divergences. These divergences are readily manifest in the expressions of the soft expansions of scattering amplitudes. The implications of this expansion on three-dimensional fields have been studied in \cite{Banerjee:2020kaa}. Let us review the basic arguments here. The map that relates bulk and boundary operators is the modified Mellin transformation \eqref{modmellin}. As this is an integral transformation on the energy variable $\omega$, one cannot directly take $\omega \to 0$ limit. It turns out that, analogous to celestial holography, the poles in the energy variable of the bulk operators map to the poles in the scaling dimensions in the boundary theory. Consider the following soft expansion of a momentum space annihilation operator $a(\omega,z,\bar{z},\sigma)$ 
\begin{align} 
	a(\omega,z,\bar{z},\sigma)=\sum_{n}\frac{S_{1-n}(z,\bar{z},\sigma)}{\omega^n}.
\end{align}
Here the coefficients $S_0(z,\bar{z},\sigma)$, $S_1(z,\bar{z},\sigma)...$ are respectively leading and subleading soft contributions and so on. Now, if we take this formula and plug it in \eqref{modmellin}, then we shall have an expansion in $\Delta$, i.e.
\begin{align}
	\Phi_{h,\bar{h}}(u,z,\bar{z})= \sum_{n}\frac{\Gamma(2-\Delta-n)}{2-\Delta-n}S_{1-n}(z,\bar{z},\sigma).
\end{align}
The presence of these Gamma functions in conformal dimension $\Delta$ is the manifestation of IR divergences. For example, the coefficient of the leading soft contribution is divergent at $\Delta=1$. Thus in order to extract the leading soft operator, one needs to take the residue of $\Phi_{\Delta,\sigma}(u,z,\bar{z})$ around $\Delta=1$, i.e.
\begin{align}
	S_0(z,\bar{z},\sigma)=-\lim_{\Delta \to 1}(\Delta-1)[(iu)^{\Delta-1}\Phi_{\Delta,\sigma}(u,z,\bar{z})]
\end{align}
In a similar fashion, the subleading soft contribution can be expressed as
\begin{align}
	S_1(z,\bar{z},\sigma)=\lim_{\Delta \to 0}\Delta(1-u\partial_u)[(iu)^{\Delta}\Phi_{\Delta,\sigma}(u,z,\bar{z})]
\end{align}
The insertions of these soft operators in the correlation functions lead to the standard Ward identities of the boundary theory. The reader is directed to \cite{Banerjee:2020kaa} for further details. However, the main point we would like to emphasize here is that in order to implement the soft expansion of massless particles in boundary theory, it is crucial that we should keep the scaling dimension of the boundary operators arbitrary. 

\bigskip

\noindent {\underline {\em {Carroll Stress Tensor:}}} Any translationally invariant quantum field theory contains an energy-momentum tensor. Carrollian CFTs are, of course translationally invariant by construction, and hence, any Carroll CFT must have an EM tensor. Below, we argue that the stress tensor in a 3d Carrollian CFT is a primary of weight $\D=3$ and hence, a Carroll CFT with $\D_i=1$ would not have an EM tensor and be inconsistent. 

\medskip

A generic way of obtaining Carroll CFTs is taking the Carroll limit of a 3d relativistic CFT. Now, consider a $d=3$ Carroll CFT that is obtained as a $c\to0$ limit of a 3d relativistic CFT. This limit does not change the conformal weight $\D$ of an operator since the Dilatation operator $D= t\partial_t + x^i \partial_i$ does not change under the limit. 

\medskip

The EM tensor of a Carroll CFT can be systematically obtained in the $c\to0$ limit of a relativistic CFT. In a $d=3$ relativistic CFT, the EM tensor is a primary of weight $\D=3$. In the 3d Carroll CFT, the Carrollian EM tensor is also a weight $\D=3$ primary. The above statement can also be made from intrinsic Carrollian symmetry arguments without invoking any limiting procedure and can be checked in explicit examples of Carroll CFTs, e.g., in the Carrollian scalar theory \cite{Dutta:2022vkg}. Clearly, the EM tensor is a $\D\neq 1$ primary that has to exist in every Carrollian CFT, and hence, it is implausible that the $\D=1$ subsector is consistent on its own. 

\bigskip

\noindent {\underline{\em {Possible resolutions:}}} We now speculate about possible resolutions to the above points. Regarding the EM tensor, the problem may be circumvented if the BMS$_4$ algebra generates Virasoro central terms, and the EM tensor no longer remains a primary. Also, there have been recent investigations of logarithmic structures in 2d Celestial CFTs \cite{Fiorucci:2023lpb}. If these have Carrollian counterparts, perhaps one might be able to find a way around the arguments we presented above. 

The $\D_i \to 1$ subsector, although possibly not a self-consistent Carrollian CFT, is expected to play a significantly important role in understanding structures related to the soft sector. We find indications of this in the explicit amplitudes we compute later in the paper.

\medskip

\paragraph{Conclusions.} We conclude this rather important section with a summary of our arguments above as to why (our version of) Carrollian holography provides a natural hologram of asymptotically flat spacetimes. 

\medskip

If we believe that we would be able to recover (at least some aspects of) flat holography as an infinite radius limit of AdS/CFT, it is natural that the boundary theory does not suddenly reduce in dimension in the limit, nor is it expected to get restricted to just a particular set of operators which have a single scaling dimension. It is thus satisfying that the Carroll wavefunctions, which are directly related to the modified Mellin transformation, are the ones we get by looking carefully at the AdS Witten diagrams in our work. This modified Mellin basis is at the heart of the connection of Carrollian theories and holography in flat spacetimes. We will illustrate this further in the next part of our paper. 

\medskip

We think we have convincingly argued that the co-dimension one holography of asymptotically flat spacetimes provided by Carrollian holography is perhaps the better way of constructing flat holography. This approach seems more natural than the Celestial one, given that the translations (and supertranslations) are naturally encoded on the dual field theory, and the null direction of the boundary of the gravitational theory is taken into account. The Carrollian picture is also what comes about when we take the limit from AdS/CFT, making sure we do not neglect the null direction. 

\medskip

The mapping between Carrollian and Celestial holography described in \cite{Donnay:2022wvx} is a mapping between a particular subsector of the Carroll theory and a Celestial CFT. One can only ``trade" the null direction $u$ and the conformal dimension $\Delta$ in this case as $\Delta$ is fixed to a particular value $\Delta=1$ in this sector of the Carroll theory. A Carroll CFT has operators with $\Delta\neq 1$. So, in general, this seems to suggest that the Carrollian and Celestial pictures of flat holography are not identical and not just a change of basis as has been claimed elsewhere. It seems that the Carrollian version may contain more information as it keeps track of the null direction of the boundary theory{\footnote{In the lower dimensional 3d-2d duality, entanglement entropy \cite{Bagchi:2014iea} is such an observable of the theory which depends explicitly on the null direction and is zero otherwise in Einstein gravity.}}. 

\medskip

We now present a summary of our claims so that the reader can take away the main points from this subsection. 
\begin{itemize}
	\item[$\square$] The integral transformations indicate that 2d Celestial CFTs may arise as a certain $u\to0$ limit of 3d Carroll CFTs, where $u$ is the null direction of $\mathscr{I}^+$.  
	
	\item[$\square$] The Carroll CFTs described in \cite{Donnay:2022aba,Donnay:2022wvx} are the $\Delta=1$ (where $\Delta$ is the weight of a Carroll primary) sub-sector of a generic Carroll CFT. 
	
	\item[$\square$] There exists a map between the $\Delta=1$ subsector of a Carroll CFT and a Celestial CFT \cite{Donnay:2022aba} {\footnote{This has been explicitly shown to work for the $\Delta=1$ subsector of the Celestial CFT and conjectured for arbitrary $\D$.}}, and not between a generic Carroll CFT and a Celestial CFT. 
	
	\item[$\square$] A Carroll CFT generically contains primary operators with dimensions $\Delta\neq1$ in its spectrum. If the mapping between $\D=1$ Carroll CFTs and Celestial CFTs constructed in \cite{Donnay:2022wvx} holds, this points to the fact that a Carroll CFT would contain more operators (the ones with $\Delta\neq1$) than a Celestial CFT. The mapping between the two would not be one-to-one. So, in general, it seems that the Carrollian and Celestial pictures of flat holography are not equivalent. 
	
	\item[$\square$] It is likely that the truncation of Carroll CFTs to just the $\Delta=1$ subsector is not consistent since, e.g., the EM tensor of a 3d Carroll CFT is not contained in this sector. A Carroll CFT must always contain an EM tensor, and hence, it is likely that just the $\Delta=1$ subsector cannot describe a consistent Carroll CFT.  
	
\end{itemize}

While it seems that the $\Delta=1$ Carroll CFTs may not be consistent by themselves from what we have discussed above, the approach taken in \cite{Donnay:2022aba,Donnay:2022wvx} of building an extrapolate dictionary seems intriguing and a potentially correct way of approaching holography in asymptotically flat spacetimes from a co-dimension one viewpoint. It would be very interesting to reconsider the asymptotic expansions to see how to allow for generic values of $\Delta$ that, on general grounds, should be allowed for holographic Carroll CFTs. Some of these aspects are investigations in progress. For the rest of the paper, we will focus on the flat limit of AdS/CFT and, in particular, recovering explicit 3d Carrollian correlators (for arbitrary $\D$) from AdS$_4$ Witten diagrams.

\bigskip

\newpage

\part{Spinning Witten Diagrams and Carroll correlators}

\bigskip \bigskip

In this part, we focus on the limit of AdS, which lands us up in flat spacetimes. We show how to recover 3d Carrollian correlations from Witten diagrams in AdS$_4$. This part is a direct follow up of the formulation outlined earlier in \cite{Bagchi:2023fbj}. We will generalise our earlier construction to include operators with non-zero spin. We start with a boundary analysis including spin in the Carrollian field theoretic set-up in Sec. \ref{sec:carrollfieldtheory}. The highest weight representations would translate to Ward identities constraining Carrollian correlators in the null boundary. These correlators will be worked out in Sec. \ref{sec:carrollcorrelators}. We then move to the bulk in Sec. \ref{sec:flatlimitwitten} and review the construction of the limit of scalar Witten diagrams before adding spin to the game and refining our previous formulation to include these more general structures. We then move to explicit examples in Sec. \ref{sec:wittendiagex} and work out a total of twelve examples, which are matched with the boundary analysis of Sec. \ref{sec:carrollcorrelators}.

\bigskip \bigskip

\section{Carrollian symmetries and field theory analysis}
\label{sec:carrollfieldtheory}

In this section, we return to the field theoretic discussion of the previous section and have a closer look at the primary operators. We have outlined the primary transformation rules in \eqref{primary}. Here, we shall elaborate on this formula emphasizing the spinning, i.e., vector and tensor primaries, and derive their correlation function using symmetry arguments. For this purpose, we first need to define the Carrollian quasi primaries based on the global subgroup and consider its finite dimensional representations induced from the little group. 

\subsection{Conformal Carrollian symmetries} 

The Carroll group \cite{LevyLeblond,NDS} in arbitrary dimensions comprises the following generators:
\begin{align}
	&P_i=-i\partial_i \, , \quad  H=-i\partial_u \, ,   \\  \nonumber
	&B_i=-ix_i\partial_u\, ,  \quad   J_{ij}=i(x_i\partial_j-x_j\partial_i) \, .
\end{align}
Here, we have denoted $u$ as the Carroll time and $x_i$ as the spatial coordinates. The generators $J_{ij}$, $P_i$, $B_i$, and $H$ are spatial rotations, translations, Carroll boosts, and the Hamiltonian, respectively. They can be obtained from the relativistic generators in the same dimensions via the $c\to 0$ contractions. The non-trivial commutation relations of the Carroll algebra are  given by
\begin{align}
	[J_{ij},B_k]=-i\delta_{k[i}B_{j]}, \quad [J_{ij},P_k]=-i\delta_{k[i}P_{j]}, \quad
	[B_i,P_j]=i\delta_{ij}H
\end{align}
One noteworthy difference from the Poincare group is that the Carroll boosts, unlike the Lorentz boosts, commute among themselves. Also, notice that the Hamiltonian only enters the algebra as a central element. The conformal extension of the Carroll group in arbitrary spacetime dimensions can be obtained by further adding the dilatation $D$ and the Carrollian special conformal generators $K$ and $K_i$. These generators would also follow similarly by contractions from the relativistic conformal generators. The vector fields associated with the conformal generators are given by 
\begin{align}
	D=-i(u\partial_u+x^i\partial_i), \quad K=-ix^2\partial_u, \quad K_i=-i(x^2\partial_i-2x_ix^j\partial_j-2x_iu\partial_u) \, ,
\end{align}
and the extension of the algebra is as follows
\begin{align}
	&[B_i,K_j]=i\delta_{ij}K\, , \quad [D,K]=-iK\, , \quad [K,P_i]=2iB_i \, , \\ \nonumber
	&[K_i,P_j]=-2i(D\delta_{ij}-J_{ij})\, , \quad [H,K_i]=2iB_i\, , \quad [D,H]=iH \, , \\ \nonumber
	&[D,P_i]=iP_i\, , \quad [D,K_i]=-iK_i \, .
\end{align}

\medskip

Although the contraction from the relativistic conformal algebra yields only a finite dimensional algebra, in the Carrollian regime, this group of conformal transformations also admits an infinite extension, just like the 2d relativistic CFTs. This is a peculiar feature of Carrollian conformal field theories that continues to hold for higher dimensional cases as well.
Particularly in $d=3$, this infinite extension is isomorphic with the superrotation extended BMS$_4$ algebra \eqref{BMS} generated by the vector fields \eqref{eq:BMSgenerators}.    

The global part of the algebra, spanned by $n=0,\pm1$ and $r,s=0,1$, forms the Poincare subgroup of the infinite dimensional BMS group. These generators can be identified in terms of finite conformal Carroll generators as 
\begin{align} \label{Map}
	&M_{00}=H, \quad M_{10}=B_x+iB_y, \quad  M_{01}=B_x-iB_y, \quad M_{11}=K   \\  \nonumber
	&L_{-1}= \frac{1}{2}(P_x-iP_y), \quad L_0=\frac{1}{2}(D+iJ), \quad L_{1}=-\frac{1}{2}(K_x+iK_y)   \\ \nonumber
	&\bar{L}_{-1}=\frac{1}{2}(P_x+iP_y), \quad \bar{L}_{0}=\frac{1}{2}(D-iJ),\quad \bar{L}_{1}=-\frac{1}{2}(K_x-iK_y)
\end{align}
In order to match these generators, we have redefined the spatial coordinates as  $z=x+iy$ and $\bar{z}=x-iy$. Field theories respecting the Carroll symmetries and its above mentioned conformal extension have been previously addressed in \cite{Bagchi:2019xfx,Bagchi:2019clu,Henneaux:2021yzg,deBoer:2021jej,Barnich:2022bni,deBoer:2023fnj,Chen:2021xkw}.

\subsection{Carrollian quasi-primary fields}
\label{sec:primaryfields}

The representation of the Carrollian primary fields can be obtained by implementing similar methods to the relativistic conformal case. We first identify the little group that leaves the origin, $\textbf{x}=0$ invariant. These transformations are generated by the rotations, Carroll boosts, dilatations, and Carroll SCTs.
Thus, at the origin, the transformation rules are given by 
\begin{align} \label{little}
	&[J_{ij},\Phi(0)]=\mathcal{S}_{ij}\Phi(0), \quad [B_i,\Phi(0)]=\mathcal{B}_i \Phi(0), \quad [D,\Phi(0)]=-i\mathbf{{\Delta}}\Phi(0)  \\ \nonumber
	&[K,\Phi(0)]=\mathit{k}\Phi (0), \quad [K_i,\Phi(0)]=\mathit{k}_i \Phi(0).
\end{align}
The generators of the little group at the origin should form a matrix representation of the corresponding algebra. Now, using Schur's Lemma, it is possible to diagonalize the action of $\mathbf{\Delta}$, as they commute with the $\mathcal{S}_{ij}$ and $\mathcal{B}_i$s. Further, $\mathit{k}$ and $\mathit{k}_i$ can be set to zero as a consequence of the algebra.
The transformation rules at arbitrary points can be figured out by translating these relations at the origin  (\ref{little}). These are \cite{Bagchi:2016bcd}{\footnote{Reader is referred to eq (5.8) in this paper.}}:
\begin{subequations}\label{eq:carrollianprimaries}
	\begin{align}
		&[H,\Phi(u,x^i]=-i\partial_u \Phi (u,x^i)  \\ 
		&[P_i,\Phi(u,x^i)]=-i\partial_i \Phi (u,x^i)  \\ 
		&[J_{ij},\Phi(u,x^i)]=-i(i\mathcal{S}_{ij}-x_i\partial_j+x_j\partial_i) \Phi (u,x^i)  \\ 
		&[D,\Phi(u,x^i)]=-i(\Delta+u\partial_u+x^i\partial_i) \Phi (u,x^i)  \\ 
		&[K,\Phi(u,x^i)]=(-ix^2\partial_u+2x^i\mathcal{B}_i) \Phi (u,x^i)  \\ 
		&[B_i,\Phi(u,x^i)]=(-ix_i\partial_u+\mathcal{B}_i) \Phi (u,x^i)  \\
		&[K_i,\Phi(u,x^i)]=-i(-2x_i \Delta +2ix^j\mathcal{S}_{ij}-2iu\mathcal{B}_i-2ux_i\partial_u-2x_ix^j\partial_j+x^2\partial_j) \Phi (u,x^i)  
	\end{align}
\end{subequations}
These transformation rules define the Carroll primary in arbitrary dimensions. However, we shall be interested in only 3d Carroll CFTs that concern holography of 4 dimensional flat spacetimes. Specifically, in $d=3$, the Carroll group is generated by one rotation, two Carroll boosts, and the spatial translations. The algebra of the spin matrices associated to rotations ($\mathcal{J}$) and Carroll boosts ($\mathcal{B}_x, \mathcal{B}_y$)  in this case is given by
\begin{align}  \label{spin algebra}
	[\mathcal{J},\mathcal{B}_x]=-i\mathcal{B}_y, \quad [\mathcal{J},\mathcal{B}_y]=i\mathcal{B}_x, \quad [\mathcal{B}_x,\mathcal{B}_y]=0
\end{align}

\subsection*{Spin 0}

The spin 0 or the scalar representation of the conformal Carroll fields can be obtained by trivially setting 
\begin{equation}
	\mathcal{J}=\mathcal{B}_x=\mathcal{B}_y=0
\end{equation}
With this input, the transformation  of the quasi primaries defined in \eqref{eq:carrollianprimaries} can be cast into the standard form
\begin{align}
	&	[L_n,\Phi(u,z,\bar{z})]=-i[z^{n+1}\partial_z+(\frac{\Delta}{2}+\frac{u}{2}\partial_u)z^n]\Phi(u,z,\bar{z})  \\ \nonumber
	&	[\bar{L}_n,\Phi(u,z,\bar{z})]=-i[\bar{z}^{n+1}\partial_{\bar{z}}+(\frac{\Delta}{2}+\frac{u}{2}\partial_u)\bar{z}^n]\Phi(u,z,\bar{z})     \quad    \forall n  \in 0, \pm 1  \\  \nonumber
	&\text{and} \quad
	[M_{r,s},\Phi(u,z,\bar{z})]=-iz^r\bar{z}^s\partial_u \Phi(u,z,\bar{z}) \quad   \forall r,s \in 0,1
\end{align} 
These transformation properties expectedly match up with the previously defined Carroll primaries in \eqref{primary}  with $h=\bar{h}=\frac{\Delta}{2}$.

\subsection*{Spin 1}

The spin 1 representation of the rotation generator is given by
\begin{align} \label{eq:spin1rot}
	\mathcal{J}_{s=1}=
	\begin{bmatrix}
		0 & 0 & 0 \\
		0 & 0 & -i  \\
		0 & i & 0
	\end{bmatrix}
\end{align}
This is just the fundamental representation of $SO(2)$. We should choose the boost matrices in a way that is consistent with the commutation relations given by (\ref{spin algebra}). There are two possible scenarios. One choice would be to set these matrices equal to zero, i.e., simply
\begin{equation} \label{choice}
	\mathcal{B}_x=\mathcal{B}_y=0 \, .
\end{equation}
For this choice, the required commutators are trivially satisfied. However, there is another non-trivial choice consistent with \eqref{spin algebra}. We shall stick to \eqref{choice} for our purposes as they induce the transformation rules that agree with the bulk analysis. We discuss this choice and another possible choice below around equation \eqref{bnonzero}. Once this choice is assumed, we can readily derive the transformations of the vector primaries. They are given by
\begin{itemize}
	\item $\mathbf{\Phi^u}(u,z,\bar{z})$
	\begin{align}\label{eq:phiu}
		&	[L_n,\Phi^u(u,z,\bar{z})]=-i[z^{n+1}\partial_z+(\frac{\Delta}{2}+\frac{u}{2}\partial_u)z^n]\Phi^u(u,z,\bar{z})   \\ \nonumber
		&	[\bar{L}_n,\Phi^u(u,z,\bar{z})]=-i[\bar{z}^{n+1}\partial_{\bar{z}}+(\frac{\Delta}{2}+\frac{u}{2}\partial_u)\bar{z}^n]\Phi^u(u,z,\bar{z})     \quad    \forall n  \in 0, \pm 1  \\  \nonumber
		&[M_{r,s},\Phi^u(u,z,\bar{z})]=-iz^r\bar{z}^s\partial_u \Phi^u(u,z,\bar{z})
	\end{align}
	\item $\mathbf{\Phi^z}(u,z,\bar{z})$
	\begin{align}
		&	[L_n,\Phi^z(u,z,\bar{z})]=-i[z^{n+1}\partial_z+(\frac{\Delta-1}{2}+\frac{u}{2}\partial_u)z^n]\Phi^z(u,z,\bar{z})   \\ \nonumber
		&	[\bar{L}_n,\Phi^z(u,z,\bar{z})]=-i[\bar{z}^{n+1}\partial_{\bar{z}}+(\frac{\Delta+1}{2}+\frac{u}{2}\partial_u)\bar{z}^n]\Phi^z(u,z,\bar{z})     \quad    \forall n  \in 0, \pm 1  \\  \nonumber
		&[M_{r,s},\Phi^z(u,z,\bar{z})]=-iz^r\bar{z}^s\partial_u \Phi^z(u,z,\bar{z})
	\end{align}
	\item  $\mathbf{\Phi^{\bar{z}}}(u,z,\bar{z})$
	\begin{align}
		&	[L_n,\Phi^{\bar{z}}(u,z,\bar{z})]=-i[z^{n+1}\partial_z+(\frac{\Delta+1}{2}+\frac{u}{2}\partial_u)z^n]\Phi^{\bar{z}}(u,z,\bar{z})   \\ \nonumber
		&	[\bar{L}_n,\Phi^{\bar{z}}(u,z,\bar{z})]=-i[\bar{z}^{n+1}\partial_{\bar{z}}+(\frac{\Delta-1}{2}+\frac{u}{2}\partial_u)\bar{z}^n]\Phi^{\bar{z}}(u,z,\bar{z})     \quad    \forall n  \in 0, \pm 1  \\  \nonumber
		&[M_{r,s},\Phi^{\bar{z}}(u,z,\bar{z})]=-iz^r\bar{z}^s\partial_u \Phi^{\bar{z}}(u,z,\bar{z})
	\end{align}
\end{itemize}
For these vector primaries, we have redefined the components as  $\Phi^z=\Phi^x-i\Phi^y$ and $\Phi^{\bar{z}}=\Phi^x+i\Phi^y$. Looking at the transformation rules of these components, it can be immediately figured out that each component of the vector primary again transforms like \eqref{primary}. 

\medskip

As a consequence of the trivial choice of the Carroll boost matrices, these components do not transform into each other. However, the information about the spin is still manifested in the holomorphic and anti-holomorphic weights ($h$ and $\bar{h}$) of these components. Here we should remind the reader that $\Delta$ is the three dimensional scaling eigenvalue (as evident from the Killing vector fields), as opposed to celestial CFT. However, because of the trivial choice of these boost matrices, the spin $\sigma$ effectively reduces to a 2d spin.
The weights of the components are organised as in Table \ref{tab:weightsspin1upper}. 
\begin{table}[h!]
	\centering
	\begin{tabular}{||c | c c c||} 
		\hline
		weights &  $\Phi^u$ & $\Phi^z$ & $\Phi^{\bar{z}}$ \\ [0.75ex] 
		\hline
		$h$ &  ${\Delta}/{2}$ & $\frac{\Delta-1}{2}$ & $\frac{\Delta+1}{2}$ \\ 
		\hline
		$\bar{h}$ & ${\Delta}/{2}$ & $\frac{\Delta+1}{2}$ & $\frac{\Delta-1}{2}$ \\
		\hline
	\end{tabular}
	\caption{Weights of vector primaries with upper indices}
	\label{tab:weightsspin1upper}
\end{table}

\medskip

We have considered Carrollian vector primaries, which are naturally defined with the upper indices. However, we need vector fields with lower indices for several practical purposes, specifically for calculating flat space scattering amplitudes later in the paper. We will elaborate on why it is important to have lower indices in our bulk analysis. If we lower these indices using the degenerate metric, we shall get
\begin{equation}\label{eq:spin1lowerindex}
	\Phi_u = g_{uu} \Phi^u = 0 \, , ~~~~~ \Phi_{z} = g_{z\bar{z}} \Phi^{\z} \, , ~~~~~ \Phi_\z = g_{\z z} \Phi^z \, .
\end{equation}
We see that the $\Phi_u$ component is identically zero if we work with the lower index instead of the upper one
\begin{equation}\label{eq:spin1carru}
	\Phi_u = 0 \, .
\end{equation}
The holomorphic and anti-holomorphic weights for the non-zero $\Phi_z$ and $\Phi_\z$ are given by Table \ref{tab:weightsspin1lower}.
\begin{table}[h!]
	\centering
	\begin{tabular}{||c | c c||} 
		\hline
		weights &   $\Phi_z$ & $\Phi_{\bar{z}}$ \\ [0.75ex] 
		\hline
		$h$ &    $\frac{\Delta+1}{2}$ & $\frac{\Delta-1}{2}$ \\ 
		\hline
		$\bar{h}$  & $\frac{\Delta-1}{2}$ & $\frac{\Delta+1}{2}$ \\
		\hline
	\end{tabular}
	\caption{Weights of vector primaries with lower indices}
	\label{tab:weightsspin1lower}
\end{table}
\\

The weights for these spatial components swap simply because of the off diagonal nature of the metric component $g_{z\bar{z}}$. These results also agree with that of \cite{Salzer:2023jqv,Nguyen:2023vfz}.

\medskip

We mentioned above that in our analysis, the boost matrices were trivially set to zero. There is another set of matrices forming a finite dimensional irrep of \eqref{spin algebra}. These matrices are given by 
\begin{align}\label{bnonzero}
	\mathcal{B}_x=	\begin{bmatrix}
		0 & 0 & 0 \\
		1 & 0 & 0 \\
		0 & 0 & 0
	\end{bmatrix}, \quad
	\mathcal{B}_y=	\begin{bmatrix}
		0 & 0 & 0 \\
		0 & 0 & 0  \\
		1 & 0 & 0
	\end{bmatrix}
\end{align}
These non-trivial boost matrices are realised on various Carrollian models \cite{Bagchi:2016bcd}. As these matrices described above are non-diagonalisable, the components of spinning primaries would mix under boost transformations when they are turned on. However, in the holographic context, the presence of the boost matrices is undesirable because the Carrollian boosts are isomorphic to bulk translations. Thus, when turned on, they would lead to non-trivial transformations of these primaries.
For example,using the mapping of generators in \eqref{Map}, one can show that for the vector primary,
\begin{equation}
	[M_{10},\Phi_z (u,z,\z)] = -i z \partial_u \Phi_z + 2 \Phi_u \,.
\end{equation}
Thus, the fields will not transform covariantly under bulk translations anymore. Hence, if we consider holographic Carrollian CFTs, we should choose the representation with the trivial boost matrices. We leave a careful analysis of non-trivial boost matrices and their possible implications on flat holography for future work.

\subsection*{Spin 2}

The spin 2 representation of the rotation generator is given by
\begin{align} \label{eq:spin2rot}
	\mathcal{J}_{s=2}=
	\begin{bmatrix}
		0 & 0 & 0 & 0 & 0 \\
		0 & 0 & -i & 0 & 0  \\
		0 & i & 0 & 0 & 0 \\
		0 & 0 & 0 & 0 & -2i \\
		0 & 0 & 0 & 2i & 0
	\end{bmatrix}
\end{align}
The representation of \eqref{eq:spin2rot} for the 5 dimensional traceless and symmetric spin 2 of $SO(2)$ can be built out of the three dimensional spin 1 representation \eqref{eq:spin1rot} of $SO(2)$ through the direct product as follows:
\begin{equation}\label{eq:spin2repbuild}
	\mathbf{3} \otimes \mathbf{3} = \mathbf{1} \oplus \mathbf{5} \oplus \mathbf{3} \, .
\end{equation}
Here $\mathbf{1}$ denotes the one dimensional trace, $\mathbf{3}$ denotes the three dimensional anti-symmetric representation and $\mathbf{5}$ denotes the required five dimensional symmetric and traceless representation. Thus, to build the required representation, we exponentiate the spin 1 generator in\eqref{eq:spin1rot} and then take the direct product ($\theta$ is a real parameter):
\begin{equation}
	e^{-i \theta \mathcal{J}_{s=1}} \otimes e^{-i \theta \mathcal{J}_{s=1}} \, .
\end{equation}
One can block diagonalize this $9 \times 9$ matrix and extract the five dimensional spin 2 representation. This is same as \eqref{eq:spin2rot} because
\begin{equation}
	\text{Diag}_{5\times5} \left[ e^{-i \theta \mathcal{J}_{s=1}} \otimes e^{-i \theta \mathcal{J}_{s=1}} \right] = e^{-i\theta \mathcal{J}_{s=2}} \, .
\end{equation}
This five dimensional representation \eqref{eq:spin2rot} is for the traceless and symmetric spin 2 primary field with the following vector representation:
\begin{align}\label{eq:spin2carrfield}
	\Phi_{s=2}(u,z,\bar{z})=
	\begin{bmatrix}
		\Phi^{uu}(u,z,\bar{z}) \\ \Phi^{ux}(u,z,\bar{z}) \\ \Phi^{uy}(u,z,\bar{z}) \\
		\Phi^{xx}(u,z,\bar{z}) \\ \Phi^{xy}(u,z,\bar{z})  
	\end{bmatrix}
\end{align}
We have chosen a symmetric and traceless representation for the spin 2 field since we want to holographically describe the bulk graviton field. Using the tracelessness condition we have eliminated $\Phi^{yy}$ as
\begin{equation}
	g_{\mu\nu} \Phi^{\mu\nu} = \Phi^{xx} + \Phi^{yy} = 0 \, ,
\end{equation}
where we used $g_{uu} = 0$ and $g_{ij} = \delta_{ij} $. Clearly, to impose this tracelessness, which selects only the spatial components, we had to work with tensor primaries with upper indices. This also justifies why we worked with upper indices for the vector primaries, which ultimately resulted in the crucial \eqref{eq:spin1carru}. From \eqref{eq:spin2rot}, we have
\begin{align}
	\mathcal{J}\Phi_{s=2}(u,z,\bar{z})=
	\begin{bmatrix}
		0 \\
		-i \Phi^{uy}(u,z,\bar{z}) \\
		i \Phi^{ux}(u,z,\bar{z}) \\
		-2i  \Phi^{xy}(u,z,\bar{z}) \\
		2i 	\Phi^{xx}(u,z,\bar{z})
	\end{bmatrix}
\end{align}
To build the transformation rules for the spin 2 Carrollian primary field from \eqref{eq:carrollianprimaries}, we use the following combination to work out an example of $L_0$
\begin{align}
	\Delta \Phi_{s=2}(u,z,\bar{z}) -  \mathcal{J} \Phi_{s=2}(u,z,\bar{z})=
	\begin{bmatrix}
		\Delta \Phi^{uu} \\
		\Delta \Phi^{ux} +i \Phi^{uy} \\
		\Delta \Phi^{uy}-i \Phi^{ux} \\
		\Delta \Phi^{xx} + 2i  \Phi^{xy} \\
		\Delta \Phi^{xy} -2i \Phi^{xx}
	\end{bmatrix}
\end{align}
We now define the fields in terms of the $z$ and $\bar{z}$ components as
\begin{equation}\label{eq:zzbarspin2}
	\begin{split}
		\Phi^{zz} &= 2\Phi^{xx} - 2i \Phi^{xy} \, , \\
		\Phi^{\bar{z}\bar{z}} &= 2\Phi^{xx} + 2i \Phi^{xy} \, , \\
		\Phi^{uz} &= \Phi^{ux} - i \Phi^{uy} \, , \\
		\Phi^{u \bar{z}} &= \Phi^{ux} + i \Phi^{uy} \, .
	\end{split}
\end{equation}
Once we have the \eqref{eq:zzbarspin2}, we can straightforwardly work out the transformation properties of the various components of \eqref{eq:spin2carrfield}. We set the Boost matrices of  \eqref{spin algebra} zero. The primary transformation rules thus become
\begin{itemize}
	\item $\mathbf{\Phi^{uu}}(u,z,\bar{z})$
	\begin{align}
		&	[L_n,\Phi^{uu}(u,z,\bar{z})]=-i[z^{n+1}\partial_z+(\frac{\Delta}{2}+\frac{u}{2}\partial_u)z^n]\Phi^{uu}(u,z,\bar{z})   \\ \nonumber
		&	[\bar{L}_n,\Phi^{uu}(u,z,\bar{z})]=-i[\bar{z}^{n+1}\partial_{\bar{z}}+(\frac{\Delta}{2}+\frac{u}{2}\partial_u)\bar{z}^n]\Phi^{uu}(u,z,\bar{z})     \quad    \forall n  \in 0, \pm 1  \\  \nonumber
		&[M_{r,s},\Phi^{uu}(u,z,\bar{z})]=-iz^r\bar{z}^s\partial_u \Phi^{uu}(u,z,\bar{z})
	\end{align}
	\item $\mathbf{\Phi^{uz}}(u,z,\bar{z})$
	\begin{align}
		&	[L_n,\Phi^{uz}(u,z,\bar{z})]=-i[z^{n+1}\partial_z+(\frac{\Delta-1}{2}+\frac{u}{2}\partial_u)z^n]\Phi^{uz}(u,z,\bar{z})   \\ \nonumber
		&	[\bar{L}_n,\Phi^{uz}(u,z,\bar{z})]=-i[\bar{z}^{n+1}\partial_{\bar{z}}+(\frac{\Delta+1}{2}+\frac{u}{2}\partial_u)\bar{z}^n]\Phi^{uz}(u,z,\bar{z})     \quad    \forall n  \in 0, \pm 1  \\  \nonumber
		&[M_{r,s},\Phi^{uz}(u,z,\bar{z})]=-iz^r\bar{z}^s\partial_u \Phi^{uz}(u,z,\bar{z})
	\end{align}
	\item  $\mathbf{\Phi^{u\bar{z}}}(u,z,\bar{z})$
	\begin{align}
		&	[L_n,\Phi^{u\bar{z}}(u,z,\bar{z})]=-i[z^{n+1}\partial_z+(\frac{\Delta+1}{2}+\frac{u}{2}\partial_u)z^n]\Phi^{u\bar{z}}(u,z,\bar{z})   \\ \nonumber
		&	[\bar{L}_n,\Phi^{u\bar{z}}(u,z,\bar{z})]=-i[\bar{z}^{n+1}\partial_{\bar{z}}+(\frac{\Delta-1}{2}+\frac{u}{2}\partial_u)\bar{z}^n]\Phi^{u\bar{z}}(u,z,\bar{z})     \quad    \forall n  \in 0, \pm 1  \\  \nonumber
		&[M_{r,s},\Phi^{u\bar{z}}(u,z,\bar{z})]=-iz^r\bar{z}^s\partial_u \Phi^{u\bar{z}}(u,z,\bar{z})
	\end{align}
	\item $\mathbf{\Phi^{zz}}(u,z,\bar{z})$
	\begin{align}
		&	[L_n,\Phi^{zz}(u,z,\bar{z})]=-i[z^{n+1}\partial_z+(\frac{\Delta-2}{2}+\frac{u}{2}\partial_u)z^n]\Phi^{zz}(u,z,\bar{z})   \\ \nonumber
		&	[\bar{L}_n,\Phi^{zz}(u,z,\bar{z})]=-i[\bar{z}^{n+1}\partial_{\bar{z}}+(\frac{\Delta+2}{2}+\frac{u}{2}\partial_u)\bar{z}^n]\Phi^{zz}(u,z,\bar{z})     \quad    \forall n  \in 0, \pm 1  \\  \nonumber
		&[M_{r,s},\Phi^{zz}(u,z,\bar{z})]=-iz^r\bar{z}^s\partial_u \Phi^{zz}(u,z,\bar{z})
	\end{align}
	\item  $\mathbf{\Phi^{\bar{z}\bar{z}}}(u,z,\bar{z})$
	\begin{align}
		&	[L_n,\Phi^{\bar{z}\bar{z}}(u,z,\bar{z})]=-i[z^{n+1}\partial_z+(\frac{\Delta+2}{2}+\frac{u}{2}\partial_u)z^n]\Phi^{\bar{z}\bar{z}}(u,z,\bar{z})   \\ \nonumber
		&	[\bar{L}_n,\Phi^{\bar{z}\bar{z}}(u,z,\bar{z})]=-i[\bar{z}^{n+1}\partial_{\bar{z}}+(\frac{\Delta-2}{2}+\frac{u}{2}\partial_u)\bar{z}^n]\Phi^{\bar{z}\bar{z}}(u,z,\bar{z})     \quad    \forall n  \in 0, \pm 1  \\  \nonumber
		&[M_{r,s},\Phi^{\bar{z}\bar{z}}(u,z,\bar{z})]=-iz^r\bar{z}^s\partial_u \Phi^{\bar{z}\bar{z}}(u,z,\bar{z})
	\end{align}
	
\end{itemize}

The weights of the various components are summarized in Table \ref{tab:weightsspin2upper}. The tracelessness condition on the spin 2 primaries is manifested as
\begin{equation}\label{eq:spin2carrfieldtrace}
	\Phi^{z\bar{z}} = 0 \, .
\end{equation}

\begin{table}[h!]
	\centering
	\begin{tabular}{||c | c c c c c ||} 
		\hline
		weights &  $\Phi^{uu}$ & $\Phi^{uz}$ & $\Phi^{u\bar{z}}$ & $\Phi^{zz}$ & $\Phi^{\bar{z}\bar{z}}$   \\ [0.75ex] 
		\hline
		$h$ &  ${\Delta}/{2}$ & $\frac{\Delta-1}{2}$ & $\frac{\Delta+1}{2}$ & $\frac{\Delta-2}{2}$ & $\frac{\Delta+2}{2}$  \\ 
		\hline
		$\bar{h}$ & ${\Delta}/{2}$ & $\frac{\Delta+1}{2}$ & $\frac{\Delta-1}{2}$ & $\frac{\Delta+2}{2}$ & $\frac{\Delta-2}{2}$  \\
		\hline
	\end{tabular}
	\caption{Weights of tensor primaries with upper indices}
	\label{tab:weightsspin2upper}
\end{table}
We can now lower the indices similar to \eqref{eq:spin1lowerindex} to get
\begin{equation}
	\begin{split}
		\Phi_{uu} = g_{uu} g_{uu} \Phi^{uu} = 0 \, , ~~~~~ \Phi_{uz} = g_{uu} g_{z\z} \Phi^{u\z} = 0 \, , ~~~~~ \Phi_{u\z} = g_{uu} g_{\z z} \Phi^{u z} = 0 \, , \\
		\Phi_{zz} = g_{z\z} g_{z\z} \Phi^{\z\z} \, , ~~~~~ \Phi_{\z\z} = g_{\z z} g_{\z z} \Phi^{zz} \, , ~~~~~ \Phi_{z\z} = g_{\z z} g_{z \z} \Phi^{z\z} = 0 \, . 
	\end{split}
\end{equation}
Thus, the $u$ components of the field are similar to the spin 1 case \eqref{eq:spin1carru}:
\begin{equation}\label{eq:spin2carru}
	\Phi_{uu} = \Phi_{uz} = \Phi_{u \z} = 0 \, .
\end{equation}
The weights of the non-trivial tensor primaries with lower indices $\Phi_{zz}$ and $\Phi_{\z \z}$ are given by Table \ref{tab:weightsspin2lower}.
\begin{table}[h!]
	\centering
	\begin{tabular}{||c | c c||} 
		\hline
		weights &   $\Phi_{zz}$ & $\Phi_{\z\bar{z}}$ \\ [0.75ex] 
		\hline
		$h$ &    $\frac{\Delta+2}{2}$ & $\frac{\Delta-2}{2}$ \\ 
		\hline
		$\bar{h}$  & $\frac{\Delta-2}{2}$ & $\frac{\Delta+2}{2}$ \\
		\hline
	\end{tabular}
	\caption{Weights of vector primaries with lower indices}
	\label{tab:weightsspin2lower}
\end{table}

If we were to work with the non-trivial choice of boost matrices of \eqref{spin algebra}, this would mix up various components of \eqref{eq:spin2carrfield}, and the analysis of the correlations functions would thus change. This and consequences for holography in flat spacetimes will be addressed elsewhere.

\section{Carrollian correlation functions}
\label{sec:carrollcorrelators}

Now, we derive the correlation functions of the Carroll primary operators whose transformation rules we derived in the previous section. The two and three point correlators of these primaries can be exactly determined by solving the Ward identities associated with the global generators. {\footnote{Spinning Carrollian correlators have been previously addressed in \cite{Salzer:2023jqv} and more recently in \cite{Nguyen:2023miw} when this paper was being readied for submission. While \cite{Salzer:2023jqv} computes some specific cases, \cite{Nguyen:2023miw} focusses on more generic three-point functions. Their answers match with our previous construction in  \cite{Bagchi:2023fbj} and the answers presented later. We also have more general three-point functions on the celestial torus and four point functions in this paper.}} In order to determine these correlators, let us first study the correlation function of Carroll primaries of arbitrary spin in \eqref{primary}. For the reader's convenience, we repeat these expressions below
\begin{align} \label{Bms pri}
	&	\delta_{L_{n}}\Phi(u,z,\bar{z})=-i[z^{n+1}\partial_z+(n+1)(h+\frac{u}{2}\partial_u)z^n]\Phi(u,z,\bar{z}) \, ,   \\  \nonumber
	&	\delta_{\bar{L}_n}\Phi(u,z,\bar{z})=-i[\bar{z}^{n+1}\partial_{\bar{z}}+(n+1)(\bar{h}+\frac{u}{2}\partial_u)\bar{z}^n]\Phi(u,z,\bar{z}) \, ,    \\  \nonumber
	&
	\delta_{M_{r,s},\Phi(u,z,\bar{z}}=-iz^r\bar{z}^s\partial_u \Phi(u,z,\bar{z}) \, .
\end{align}


\subsection{Two point function}

The two point correlation function of these objects was determined, and its relationship with scattering amplitudes was established in \cite{Bagchi:2022emh}. Here, we review the basic arguments. It turns out that the Ward identities associated with these global supertranslations have two classes of solutions depending on whether the solutions have any $u$ coordinate dependence. For e.g. 
\begin{align} \label{supertranslations}
	\delta_{M_{10}}G^{(2)}(u,u',z,z',\bar{z},\bar{z}')=(z\partial_u+z'\partial_{u'})G^{(2)}(u,u',z,z',\bar{z},\bar{z}')=0 \, .
\end{align}
One way to solve the equation is to drop the $u$ dependence. The rest of the supertranslation equations would also be automatically satisfied. Further, the global superrotations $L_{0,\pm 1},\bar{L}_{0,\pm 1}$ would fix it to be 2d CFT primary correlator \cite{Bagchi:2016bcd}:
\begin{align}
	G^{(2)}(u,z,\bar{z})=\frac{\delta_{h,h'}\delta_{\bar{h},\bar{h}'}}{(z-z')^{2h}(\bar{z}-\bar{z}')^{2\bar{h}}} \, .
\end{align}
On the other hand, \eqref{supertranslations} can also be solved by a $u$ dependent branch. This is possible only when the spatial part is a contact term. In this case, the supertranslation equations force the ansatz to be of the form,
\begin{align}
	G^{(2)}(u,u',z,z',\bar{z},\bar{z}')=f(u-u')\delta^2(z-z') \, .
\end{align}

Invariance under $L_{0,\pm1}$ and $\bar{L}_{0,\pm1}$ would further fix it to be
\begin{align}
	G^{(2)}(u,z,\bar{z},u',z',\bar{z}')=\delta_{\sigma+\sigma', 0}\frac{\delta^{(2)}(z-z',\bar{z}-\bar{z}')}{(u-u')^{\Delta+\Delta'-2}} \, .
\end{align}
Notice that unlike the CFT branch, this branch of the correlator is non-trivial even if the primaries have unequal scaling dimensions. The constraint restricts only the spins. We would now use this construction to compute the correlators of different components of spinning Carroll primaries discussed above. For vectors, the ansatz associated with the delta-branch for the two point correlator can be taken as 
\begin{align} 
	G^{(2) \,IJ}(u,u',z,z',\bar{z},\bar{z}')=f^{IJ}(u-u')\delta^2(z-z') \, .
\end{align}
However now the invarince under the spatial rotations ($L_0-\bar{L}_0$)  would imply 
\begin{equation}
	\sigma^I+\sigma^J=0 \, .
\end{equation}
From the weights of these spinning primaries in Table \ref{tab:weightsspin1upper}, we have $	\sigma^u=0$, $ \sigma^z=-1$  and $\sigma^{\bar{z}}=1$.  It is immediately obvious from the above equation that only non-trivial $f^{IJ}$ s are $f^{uu}$ and $f^{z\bar{z}}$.
\begin{equation}\label{eq:carrollspin1upper2pt}
	f^{uu}=f^{z\bar{z}}=\frac{1}{(u-u')^{\Delta+\Delta'-2}} \, .
\end{equation}
These are the only non-trivial correlators for spinning primaries when the boost matrices are zero. However, we shall see in the next section that we need to work with primaries of lower indices in order to match up with the bulk analysis. It is evident from our discussion above that if we work with lower indices, the $u$ component vanishes, and as a consequence, any correlation function involving $u$-component fields also vanishes. Thus, we should have 
\begin{equation}\label{eq:spin1carrfieldu}
	f_{uu} = 0 \, .
\end{equation}
However, $	f_{z\z}$  remains non zero and is given by
\begin{equation}\label{eq:carrollspin12pt}
	f_{z\z} = \big(g_{z\z}(z) g_{z\z}(z') \big)\frac{1}{(u-u')^{\Delta+\Delta'-2}} \, .
\end{equation}
A similar analysis can be done for the spin 2 primaries using the weights of Table \ref{tab:weightsspin2upper}. The invariance under spatial rotations $(L_0 - \bar{L}_0)$ would imply that the only non-trivial correlators are
\begin{equation}\label{eq:carrollspin2upper2pt}
	f^{uu,uu} = f^{uz,u\bar{z}} = f^{zz,\bar{z}\bar{z}} = \frac{1}{(u-u')^{\Delta+\Delta'-2}} \, .
\end{equation}
Lowering the indices (analogous to how we arrived at  \eqref{eq:spin1carrfieldu}) and using \eqref{eq:spin2carru}, we have{\footnote{We will see later in our bulk analysis that \eqref{eq:spin2carrfieldu} is consistent with the fact that the $uu$ component of the bulk to boundary propagator becomes pure diffeomorphism in \eqref{eq:uupurediffeo}. }}
\begin{equation}\label{eq:spin2carrfieldu}
	f_{uu,uu} = f_{uz,u\bar{z}}  = 0 \, .
\end{equation}
Thus, the only non-trivial correlator is that of $f_{zz,\bar{z}\bar{z}}$ in \eqref{eq:carrollspin2upper2pt}:
\begin{equation}\label{eq:carrollspin22pt}
	f_{zz,\bar{z}\bar{z}} = \big(g_{z\z}(z)^2 g_{z\z}(z')^2 \big)\frac{1}{(u-u')^{\Delta+\Delta'-2}} \, .
\end{equation}

\subsection{Three point function}

We now focus on three point functions. The three point function also admits two different branches of solutions. The time independent branch again coincides with the 2d CFT three point correlators, i.e.
\begin{align}
G^{(3)}(u,z,\bar{z})=\frac{1}{(z_{12})^{h_1+h_2-h_3}(z_{23})^{h_2+h_3-h_1}(z_{13})^{h_1+h_3-h_2}}
\end{align}
However the more relevant delta function branch has intricate substructures, which computes different three point scattering amplitudes in the bulk. For generic momenta, the three point function is trivially zero and this can be thought of as a consequence of momentum conservation applied to massless particles in 4d Minkowski spacetimes \cite{Banerjee:2018gce,Bagchi:2022emh}.
 We addressed this earlier in \cite{Bagchi:2023fbj}, where we also considered a non-trivial three point solution. This solutions corresponds to a case where all three massless momenta are collinear. This subclass of solutions is given by
\begin{align}
G^{(3)}(u,z,\bar{z})=u_{12}^au_{23}^bu_{31}^c\delta^{(2)}(z_{12})\delta^{(2)}(z_{23}),  \\  \nonumber
 \text{subjected to} \quad  a+b+c=\sum_{i}\Delta_i, \quad \sum_{i}\sigma_i=0
\end{align}

Below we consider other  non-trivial cases. 

\medskip


\subsubsection{Non collinear case}

For this case, we assume only one delta function, which imposes collinearity in two particle momenta, unlike the previous case considered in \cite{Bagchi:2023fbj}, where all three particles were collinear. The ansatz we start off with is
\begin{align}
	G^{(3)}(u_i,z_i,\bar{z}_i)=F(u_i) \, {z_{12}}^p \,{\bar{z}_{12}}^q\delta^2(z_{13}) \, .
\end{align}
The lowest supertranslation generator fixes $F(u_i)$ as $F(u_{12},u_{23},u_{31})$ as usual. However, higher supertranslation modes have non-trivial consequences on this type of correlation function. For example, consider the variation with respect to the generator $M_{10}$
\begin{align}
	\delta_{M_{10}}G^{(3)}(u_i,z_i,\bar{z}_i)=0 \implies [z_1(\partial_{u_1}+\partial_{u_3})+z_2\partial_{u_2}]F(u_i)=0 \, .
\end{align}
This equation only allows terms which are independent of $u_2$. Hence 
\begin{align}
	F(u_i) \equiv F(u_{13})=\sum_{c} f_c \, {u_{13}}^c \, .
\end{align}
SL$(2,\mathbb{C})$ or Lorentz symmetries in the bulk would constrain the exponents. Again, the ansatz is already invariant under the translations along $z$ and $\bar{z}$. The scale transformation generators $L_0$ and $\bar{L}_0$ constraint the exponents in the following way
\begin{align}
	\sum_{i} h_i=1-p-\frac{c}{2} \, , \qquad \sum_{i}\bar{h}_i=1-q-\frac{c}{2} \, .
\end{align}
However, unlike the collinear case \cite{Bagchi:2023fbj}, the special conformal transformations play a role and further fix these coefficients. Consider the $L_1$ equation
\begin{align}
	&\sum_{i}z_i^2\partial_{z_i}G^3(u_i,z_i,\bar{z}_i)+2\sum_{i}z_i(h_i+\frac{u_i}{2}\partial_{u_i})G^3(u_i,z_i,\bar{z}_i)=0 \, ,  \\  \nonumber
	&\Big[(-2+2(h_1+h_3)+c+p)z_1+(2h_2+p)z_2\Big]G^3(u_i,z_i,\bar{z}_i)=0 \, ,  \\ \nonumber
	&\implies h_1+h_3=\frac{1}{2}(2-c-p), \quad h_2=-\frac{p}{2} \, .
\end{align}
Similarly from $\bar{L}_1$ we have 
\begin{equation}
	\bar{h}_1+\bar{h}_3=\frac{1}{2}(2-c-q), \quad \bar{h}_2=-\frac{q}{2} \, .
\end{equation}
From these equations, we have
\begin{equation}
	s_1 + s_3 = \dfrac{q-p}{2} \, , ~~~ s_2 = \dfrac{q-p}{2}
\end{equation}
It is clear that a scattering process with two massless scalars and one photon can never satisfy $s_1+s_2=s_3$. We will, however, see that the scattering process with three massless scalars has non-trivial contributions from the soft sector. Thus, we have the result:
\begin{equation}\label{eq:noncollinear3ptintrinsic}
	G^{(3)}(u_i,z_i,\bar{z}_i)= u^{2-\Delta_1-\Delta_3 - \frac{p+q}{2}}_{13} \, {z_{12}}^p \,{\bar{z}_{12}}^q\delta^2(z_{13}) \, .
\end{equation}

\bigskip

\subsubsection{ On $\mathbb{ R}_u\times \mathbb{C}^2$}

The three point function has a non-trivial contribution if we treat $z$ and $\bar{z}$ as independent coordinates. This amounts to working on a celestial torus \cite{Stieberger:2018edy} instead of a sphere. This is relevant for the scattering processes in the bulk when the momenta of the particles are either complexified or the bulk spacetime has a (2,2) signature.

To proceed with this set up, we set $\bar{z}_1=\bar{z}_2=\bar{z}_3$, (i.e.) our ansatz is 
\begin{equation} \label{Ansatz}
	G^{3}(u_i,z_i,\bar{z}_i)=F(u_i,z_i)\delta (\bar{z}_{12})\delta (\bar{z}_{23}) \, .
\end{equation}
The variation with respect to the supertranslations is given by
\begin{align}
	\delta_{M_{r,s}}G^{3}(u_i,z_i,\bar{z}_i)=\sum_{i=1}^{3} z_i^r\bar{z}_i^s \partial_{u_{i}} G^{3}(u_i,z_i,\bar{z}_i) \, .
\end{align}
We shall demand that the correlation function is invariant under the following generators: $M_{00}, M_{10}, M_{01}, M_{11}$. These generators, as mentioned previously, are isomorphic to the bulk translations.
Now the equations corresponding to $M_{00}$ and $M_{10}$ symmetries lead to
\begin{align} \label{super}
	\sum_{i=1}^{3} \partial_{u_i}F(u_i,z_i)=0, \quad \sum_{i=1}^{3} z_i\partial_{u_i}F(u_i,z_i)=0 \, .
\end{align}
Due to the presence of delta function in $\bar{z}$, the other two equations ($M_{01}, M_{11}$) become the same as \eqref{super}. These two above equations can be solved if
\begin{align}\label{eq:finalansatz}
	F(u_i,z_i)\cong F(U,z_i)=f(U)g(z_i) , \quad \text{where} \quad U=z_1u_{23}+z_2u_{31}+z_3u_{12} \, .
\end{align} 
There are still six more equations that would impose global superrotation invariance, which is isomorphic to the bulk Lorentz group. These equations would completely fix the form of $F(U,z_i)$. Variations with respect to the holomorphic and anti-holomorphic generators are given by 
\begin{equation}
	\begin{split}
		\delta_{L_n} G^{3}(u_i,z_i,\bar{z}_i)&=\sum_{i=1}^{3} \Big[z_{i}^{n+1}\partial_{z_{i}}+(n+1)z_{i}^{n}(h_i+\frac{1}{2}u_i\partial_{u_i})\Big]G^{3}(u_i,z_i,\bar{z}_i) \, ,  \\
		\delta_{\bar{L}_{n}} G^{3}(u_i,z_i,\bar{z}_i)&=\sum_{i=1}^{3} \Big [\bar{z_i}^{n+1}\partial_{\bar{z}_{i}}+(n+1)\bar{z_i}^{n}(\bar{h}_{i}+\frac{1}{2}u_{i}\partial_{u_i})\Big]G^{3}(u_i,z_i,\bar{z}_i) \, ,
	\end{split}
\end{equation}
and we need to solve these equations for $n=0,\pm 1$.

\medskip

Now, $n=-1$ imposes translational invariance on the spatial direction. Hence, the anti holomorphic equation $\bar{L}_{-1}$ is already solved by our ansatz \eqref{Ansatz}. The holomorphic equation $L_{-1}$ would be solved if 
\begin{equation} \label{n=-1}
	g(z_i) \cong g(z_{12},z_{23},z_{31}) \, .
\end{equation}
For $n=0$, the anti-holomorphic equation $\bar{L}_0$ would gives us
\begin{align}\label{eq:fUans}
	\sum_{i=1}^{3}(\bar{h}_i-2)+\frac{1}{2}U\partial_{U}f(U)=0 \quad \text{thus} \quad f(U)=U^{-2\Big(\sum_{i=1}^{3}\bar{h}_i-2\Big)} \,  .
\end{align}
Using \eqref{eq:fUans}, the holomorphic equation $L_0$ also simplifies to 
\begin{equation}\label{eq:L0eq}
	\sum_{i=1}^{3} z_i\partial_{z_{i}}g(z_i)+\Big(3(\bar{h}_i-2)+h_i\Big) g(z_i)=0 \, .
\end{equation}
Now from \eqref{n=-1}, we can assume $g(z_i)$ would be of the following form
\begin{equation}\label{eq:gzansatz}
	g(z_i)=z_{12}^a\,z_{23}^b\,z_{31}^{c} \, .
\end{equation} 
Plugging \eqref{eq:gzansatz} back in \eqref{eq:L0eq} would yield  
\begin{equation} \label{si}
	a+b+c= -\sum_{i=1}^{3}[3(2-\bar{h}_i)-h_i] \, .
\end{equation}
This leaves two more equations from $L_1$ and $\bar{L}_1$. Plugging in the form of solution from \eqref{eq:fUans} and \eqref{eq:gzansatz} we have obtained so far, 
\begin{equation}
	G^{3}(u_i,z_i,\bar{z}_i)=f(U)g(z_i)\delta(\bar{z}_{12})\delta(\bar{z}_{23}) \, ,
\end{equation}
in the $L_1$ equation, we would obtain the following equation
\begin{align} \label{sci}
	z_1(a+c+2h_1)+z_2(a+b+2h_2)
	+z_3(b+c+2h_3) -2\Big(\sum_{i=1}^{3}\bar{h}_i-2\Big)\sum_{i=1}z_i =0 \, .
\end{align}
Now \eqref{si} and \eqref{sci} can be solved if we choose 
\begin{subequations}\label{eq:abc}
	\begin{align}
		a=h_3-h_1-h_2-(2-\sum_{i=1}^{3}\bar{h}_i)=\Delta_3-\sigma_1-\sigma_2-2 \, , \\ 
		b= h_1-h_2-h_3-(2-\sum_{i=1}^{3}\bar{h}_i)=\Delta_1-\sigma_2-\sigma_3-2 \, , \\ 
		c=h_2-h_1-h_3-(2-\sum_{i=1}^{3}\bar{h}_i)=\Delta_2-\sigma_1-\sigma_3-2 \, .
	\end{align}
\end{subequations}
This fixes the  3-pt function exactly. Given this solution, the equation that arises from $\bar{L}_1$ is trivially solved. Thus, we have
\begin{equation}\label{eq:threepointsplit}
	G^{3}(u_i,z_i,\bar{z}_i)=(z_1u_{23}+z_2u_{31}+z_3u_{12})^{-2\Big(\sum_{i=1}^{3}\bar{h}_i-2\Big)}z_{12}^a\,z_{23}^b\,z_{31}^{c}\delta(\bar{z}_{12})\delta(\bar{z}_{23}) \, ,
\end{equation}
where $a,b,c$ are given by \eqref{eq:abc}. Though the final result depends only on the weights $h_i,\bar{h}_i$ and the helicities $\sigma_i$, it should be noted that when applied for vector and tensor primaries, there are no $u$ components because of \eqref{eq:spin1carru} and \eqref{eq:spin2carru}. Thus, the three point function receives contributions only from the spatial components of the vector and tensor primaries. 

We can also consider the other holomorphic branch by setting $z_1 = z_2 = z_3$. This is just a complex conjugation and can be implemented by switching $\sigma_i \to -\sigma_i$ in \eqref{eq:threepointsplit}. This would result in 
\begin{equation}\label{eq:threepointsplitalt}
	G^{3}(u_i,z_i,\bar{z}_i)=(\bar{z}_1u_{23}+\bar{z}_2u_{31}+\bar{z}_3u_{12})^{-2\Big(\sum_{i=1}^{3}h_i-2\Big)}\bar{z}_{12}^{\bar{a}}\,\bar{z}_{23}^{\bar{b}}\,\bar{z}_{31}^{\bar{c}}\delta(z_{12})\delta(z_{23}) \, ,
\end{equation}
where
\begin{subequations}
	\begin{align}
		\bar{a}=\bar{h}_3-\bar{h}_1-\bar{h}_2-(2-\sum_{i=1}^{3}h_i)=\Delta_3+\sigma_1+\sigma_2-2 \, , \\ 
		\bar{b}= \bar{h}_1-\bar{h}_2-\bar{h}_3-(2-\sum_{i=1}^{3}h_i)=\Delta_1+\sigma_2+\sigma_3-2 \, , \\ 
		\bar{c}=\bar{h}_2-\bar{h}_1-\bar{h}_3-(2-\sum_{i=1}^{3}h_i)=\Delta_2+\sigma_1+\sigma_3-2 \, .
	\end{align}
\end{subequations}

At this point, it is useful to mention that we should also expect this result from the point of view of scattering amplitudes. If one expresses the three point scattering amplitude in terms of spinor helicity variables, the little group scaling completely fixes the amplitude up to normalization \cite{elvang_huang_2015}. If one now applies a Modified Mellin transform on this amplitude, one should get \eqref{eq:threepointsplit} and \eqref{eq:threepointsplitalt}.

\subsection{Four point function}
\label{sec:fourpointcarroll}

For this analysis, we basically follow \cite{Banerjee:2018gce}. We use the $SL(2,\mathbb{C})$ invariance to first map the four points $z_1,z_2,z_3,z_4$ at the celestial spheres at null infinity to the following points:
\begin{equation}
	(z_1,z_2,z_3,z_4) \to (z,1,0,\infty) \, ,
\end{equation}
where $z$ is the conformal cross ratio given by
\begin{equation}
	z = \dfrac{z_{12} z_{34}}{z_{13} z_{24}} \, .
\end{equation}
The four point correlator of Carrollian primary fields with weights $(h_i,\bar{h}_i)$ is thus given by
\begin{equation}\label{eq:carrollfourpt}
	\begin{split}
		G_4(u_1,u_2,&u_3,u_4,z,\bar{z}) \\ &= \langle \Phi_{(h_1,\bar{h}_1)}(u_1,z,\bar{z}) \Phi_{(h_2,\bar{h}_2)}(u_2,1,1) \Phi_{(h_3,\bar{h}_3)}(u_3,0,0) \Phi_{(h_4,\bar{h}_4)}(u_4,\infty,\infty) \rangle \, .
	\end{split}
\end{equation}
The lowest supertranslation generator $M_{00}$ (Carroll time translations) imposes the following constraint on the four point correlator:
\begin{equation}\label{eq:fourptcarrolltime}
	\left( \dfrac{\partial}{\partial u_1} + \dfrac{\partial}{\partial u_2} + \dfrac{\partial}{\partial u_3} + \dfrac{\partial}{\partial u_4} \right) G_4(u_1,u_2,u_3,u_4,z,\bar{z}) = 0 \, .
\end{equation}
The higher supertranslation modes $M_{10}$ and $M_{01}$ (Carroll boosts) impose the following constraints
\begin{equation}\label{eq:fourptcarrollboost}
	\begin{split}
		\left( z\dfrac{\partial}{\partial u_1} + \dfrac{\partial}{\partial u_2} + \alpha \dfrac{\partial}{\partial u_4} \right) G_4(u_1,u_2,u_3,u_4,z,\bar{z}) &= 0  \, ,\\
		\left( \bar{z}\dfrac{\partial}{\partial u_1} + \dfrac{\partial}{\partial u_2} + \alpha \dfrac{\partial}{\partial u_4} \right) G_4(u_1,u_2,u_3,u_4,z,\bar{z}) &= 0 \, .
	\end{split}
\end{equation}
Here $\alpha \to \infty$. These two equations can be combined together (by suitably regulating the infinity by a limiting procedure) to obtain
\begin{equation}\label{eq:fourptcarrollboostfinal}
	(z - \bar{z}) \dfrac{\partial}{\partial u_1} G_4(u_1,u_2,u_3,u_4,z,\bar{z}) = 0 \, .
\end{equation}
If we choose the branch
\begin{equation}
	\dfrac{\partial}{\partial u_1} G_4(u_1,u_2,u_3,u_4,z,\bar{z}) = 0 \, ,
\end{equation}
then \eqref{eq:fourptcarrollboost} implies
\begin{equation}
	\left( \dfrac{\partial}{\partial u_2} + \alpha \dfrac{\partial}{\partial u_4} \right) G_4(u_1,u_2,u_3,u_4,z,\bar{z}) = 0 \, .
\end{equation}
Thus, the only possible solution is
\begin{equation}
	\dfrac{\partial \, G_4}{\partial u_2} = 0 \, , ~~~~~~ \dfrac{\partial \, G_4}{\partial u_4} = 0 \, .
\end{equation}
Thus, \eqref{eq:fourptcarrolltime} implies
\begin{equation}
	\dfrac{\partial \, G_4}{\partial u_3} = 0 \, .
\end{equation}
Hence, we essentially end up on the $u$-independent CFT branch four point correlator that depends only on the conformal cross ratio. To get the desired $u$-dependent branch of the Carrollian correlator, we must choose the other branch in \eqref{eq:fourptcarrollboostfinal} given by
\begin{equation}
	z - \bar{z} = 0 \, .
\end{equation}
Thus, the expression for the four point Carrollian correlator becomes
\begin{equation}\label{eq:fourptcarrollfinal}
	G_4(u_1,u_2,u_3,u_4,z,\bar{z}) \propto \delta(|z-\bar{z}|) \, .
\end{equation}
This is the only constraint on the four point Carrollian correlator.

\section{Flat limit of Witten diagrams}
\label{sec:flatlimitwitten}

In part I, we have argued why it is more natural to consider the Carrollian approach to flat holography as opposed to the Celestial one. We will now show how these non-trivial Carroll correlators we constructed in the previous section that encode bulk scattering arise in a carefully constructed limit of AdS/CFT correlation functions. If one keeps track of the null direction while implementing the limit, we will show that Carrollian CFT correlation functions naturally emerge in the large AdS radius limit. Before that, in this section, we will analyse the elements of the Witten diagrams. We will review AdS/CFT Witten diagrams for scalar fields in the large AdS radius limit and then proceed to work out the bulk to boundary propagator for spinning particles in the large AdS radius limit.

\subsection{AdS in the embedding space representation}

We will first quickly recap the embedding representation that is typically used in the computation of AdS Witten diagrams \cite{Dirac:1936fq,Penedones:2007ns,Penedones:2010ue}. We choose to work with a particular choice of the embedding space representation. The $d+1$-dimensional AdS  solution is given by :
\begin{equation}\label{const}
	-(X^0)^2 + \sum_{i=1}^{d}(X^i)^2 -(X^{d+1})^2= -R^2 \, ,
\end{equation}
which is  embedded in a $\mathbb{R}^{2,d}$ manifold with the metric:
\begin{equation}\label{eq:embedmetric}
	ds^2 = -(dX^0)^2 + \sum_{i=1}^{d}(dX^i)^2 -(dX^{d+1})^2 \, .
\end{equation}
This is a slightly different version of the embedding manifold when compared to the lightcone coordinates used in \cite{Bagchi:2023fbj}. It turns out that in order to discuss the flat limit of spinning propagators, it is convenient to work with this form of embedding manifold.

\medskip 

The AdS solution can be written in a parametric form as follows:
\begin{equation}\label{eq:ourembeddingcoords}
	\begin{split}
		X^0 = R \dfrac{\sin \tau}{\cos \rho} \, , ~~~ X^{d+1} = R \dfrac{\cos \tau}{\cos \rho} \, , ~~~ X^i = R \tan \rho \,  \Omega^i \, , ~~ i= 1, \dots , d
	\end{split}
\end{equation}
where $\sum_{i}\Omega^2_i= 1$ represents $S^{d-1}$. In these new global coordinates, the  AdS metric is given by:
\begin{equation}\label{eq:adsmetric}
	ds^2 =\frac{R^2}{\cos^2\rho}\left(-d\tau^2+d\rho^2+\sin^2\rho \, d\Omega_{S^{d-1}}^2\right)\, ,
\end{equation}
where $\tau\in [-\pi,\pi]$ and $\rho\in [0,\frac{\pi}{2}]$. The AdS boundary is approached by taking the limit $\rho\rightarrow\frac{\pi}{2}$ and it is given by:
\begin{equation}\label{eq:ourboundarycoords}
	\textbf{P}=\lim_{\rho\rightarrow\frac{\pi}{2}}R^{-1}\cos\rho \x \, .
\end{equation}
with $\textbf{P}^2=0$. The boundary coordinates are given by
\begin{equation}\label{eq:boundarycoords}
	P^0 = \sin \tau_p \, , ~~~ P^{d+1} = \cos \tau_p \, , ~~~ P^i = \Omega^i_p \, .
\end{equation}
Building on \cite{deGioia:2022fcn}, in \cite{Bagchi:2023fbj}, we established a general relation between AdS Witten diagrams in $d+1$ dimensions and Carrollian correlation functions in $d$ dimensions. In \cite{Bagchi:2023fbj}, we showed that AdS Witten diagrams with boundary operators inserted at specific time slices reduce to the Carrollian correlators in a carefully constructed large AdS radius (denoted by $R$ here) limit. The main assumptions were 
\begin{itemize}
	\item The boundary CFT operators are inserted at global time slices $\tau_p = \pm \frac{\pi}{2}+\frac{u}{R}$
	\item The spheres at the slices $\tau_p = \frac{\pi}{2}+\frac{u}{R}$ and $\tau_p = -\frac{\pi}{2}+\frac{u}{R}$ are antipodally identified.
\end{itemize}

We implement the flat limit in the bulk by the following rescaling \cite{Giddings:1999jq}
\begin{equation}\label{eq:bulkflatlimit}
	\tau = \dfrac{t}{R} \, ,~~~~~ \rho = \dfrac{r}{R} \, .
\end{equation}
When the flat limit of the boundary theory is performed, we crucially track the null direction of the boundary. To facilitate this, we look at what happens to the retarded time $u$ in this limit. For instance, consider
\begin{equation}
	u = t-r = R(\tau - \rho) \, .
\end{equation}
Carrollian correlators will be functions of $u$, and thus, we first relate $u$ to the boundary CFT time. In AdS, the boundary is reached as $\rho \to \frac{\pi}{2}$. Labeling the boundary time as $\tau_p$ and substituting $\rho = \frac{\pi}{2}$ above, we get
\begin{equation}
	u = R(\tau_p - \frac{\pi}{2}) \implies \tau_p = \frac{\pi}{2} + \frac{u}{R} \, .
\end{equation}
We use this input to implement the flat limit in the boundary as \cite{Bagchi:2023fbj}
\begin{equation}\label{eq:bdyflatlimit}
	\tau_p = \pm\dfrac{\pi}{2} + \dfrac{u}{R} \, .
\end{equation}
To see how the null direction emerges, we substitute this expression for $\tau_p$ back into the boundary metric:
\begin{equation}
	ds^2_{\text{bdy}} = -d\tau^2 + d\Omega^2 = -\dfrac{1}{R^2}du^2+ d\Omega^2 \to 0.du^2 + d\Omega^2 \, .
\end{equation}
Thus, under \eqref{eq:bdyflatlimit}, the metric reduces to the null/Carrollian metric at $\mathscr{I}^+$. We see that if we keep track of all the symmetries, the boundary theory does not reduce in dimension. The large AdS radius limit in the bulk is equivalent to a $c \to 0$ ultra-relativistic limit in the boundary \cite{Bagchi:2012cy}. This feature is captured by the limit implemented using \eqref{eq:bdyflatlimit}.

\medskip

Let us specialize to $d=3$. Thus implementing \eqref{eq:bulkflatlimit} and \eqref{eq:bdyflatlimit} in \eqref{eq:ourembeddingcoords} and \eqref{eq:ourboundarycoords}, we have:
\begin{equation}\label{eq:xplarger}
	\begin{split}
		X &= \left(X^0, X^i,X^{d+1}\right) \\
		&= \left(t + \mo(\dfrac{1}{R^2}),\, r \Omega_1 + \mo(\dfrac{1}{R^2}),\, r \Omega_2 + \mo(\dfrac{1}{R^2}),\, r \Omega_3 + \mo(\dfrac{1}{R^2}),\, R + \dfrac{r^2-t^2}{2R} + \mo(\dfrac{1}{R^2}) \right) \, , \\
		P &= \left(P^0, P^i, P^{d+1} \right) = \left( \pm 1 + \mo(\dfrac{1}{R^2}), \, \Omega^1_p, \, \Omega^2_p, \,\Omega^3_p, \, \mp\dfrac{u}{R} + \mo(\dfrac{1}{R^2}) \right) \, .
	\end{split}
\end{equation}
\begin{equation}
	P \cdot X = \mp t + r \Omega_i \Omega^i_p \pm u \, .
\end{equation}
Thus, we have two cases corresponding to incoming and outgoing particles:
\begin{equation}\label{eq:pdotx}
	P \cdot X = \begin{cases}-t + r \Omega_i \Omega^i_p + u = u + \tilde{q} \cdot x & \text{if $\tau_p = \frac{\pi}{2}+\frac{u}{R}$} \\ +t + r \Omega_i \Omega^i_p - u = -u -\tilde{q}^A \cdot x & \text{if $\tau_p = -\frac{\pi}{2}+\frac{u}{R}$} \end{cases}
\end{equation}
Here $\tilde{q} = (1,\mathbf{\Omega}_p)$ denotes the direction of the boundary towards which the outgoing massless particle propagates. $\tilde{q}^A = (1,\mathbf{\Omega}^A_p)$ denotes the direction of the antipodal point from which the incoming particle propagates. $x = (t,r\Omega)$. As considered in \cite{Bagchi:2023fbj}, the spheres at $\tau_p = -\frac{\pi}{2}+\frac{u}{R}$ and $\tau_p = \frac{\pi}{2}+\frac{u}{R}$ are antipodally matched, thus
\begin{equation}\label{eq:antipodal}
	\mathbf{\Omega}^A_p = - \mathbf{\Omega}_p \, .
\end{equation}
This generalized antipodal matching condition was recently proved in \cite{deGioia:2023cbd} by considering the symmetry algebra of the CFT localized in $\mathcal{O}(R^{-1})$ at different $\tau_p$'s.

\subsection{A review of scalar Witten diagrams}
\label{sec:wittendiagramscal}
A major component of the Witten diagram is the bulk to boundary propagator. The bulk to boundary propagator for scalar fields is given by \cite{Penedones:2010ue}
\begin{equation}
	\textbf{K}_\Delta(\p,\x)=\frac{C_\Delta^d}{(-P \cdot X +i\epsilon)^\Delta} \, ,
\end{equation}
\footnote{There is no factor of 2 multiplying $P.X$ in the denominator because we have chosen to rescale the boundary coordinates as in \eqref{eq:boundarycoords}.} where $C^d_{\Delta}$ with factors of $R$ is given by
\begin{equation}
	C_\Delta^d=\frac{\Gamma(\Delta)}{2 \pi^{\frac{d}{2}}\Gamma(\Delta-\frac{d}{2}+1)R^{\frac{(d-1)}{2}-\Delta}} \, .
\end{equation}
Here we have parametrized the bulk and boundary points by $x\in (\tau,\rho,\Omega)$ and $p\in (\tau_p,\Omega_p)$ respectively.  

\medskip

The explicit form of the propagators in the large $R$ limit can be derived from \eqref{eq:pdotx} \cite{Bagchi:2023fbj}:
\begin{equation}\label{eq:carrollscalwave}
	\begin{split}
		\text{incoming} \quad \textbf{K}_{\Delta_{1}}(\p_1,\x) &= N_{\Delta_1}^{d}\psi_{\Delta_1,\tilde{q}_1,u_1}^{-}(x) +\mo(R^{-1}) \, , \\
		\text{outgoing} \quad  \textbf{K}_{\Delta_{2}}(\p_2,\mathbf{Y}) &= N_{\Delta_2}^{d} \psi_{\Delta_2,\tilde{q}_2,u_2}^{+}(y)+{\mo{(R^{-1})}} \, ,\\
		\text{bulk-bulk} \quad \Pi_\Delta(\x,\y)&= G(x,y)+\mo(R^{-2}) \, ,
	\end{split}
\end{equation}
where,
\begin{subequations}
	\begin{align}
		\psi_{\Delta_1,\tilde{q}_1,u_1}^{-}(x)&=\int_{0}^{\infty}d\omega_1 \, \omega^{\Delta_1-1}_1 e^{i \omega_1 (\tilde{q}_1.x+u_1)}e^{-\epsilon\omega_1 }\, , \\
		\psi_{\Delta_2,\tilde{q}_2,u_2}^{+}(y)&=\int_{0}^{\infty}d\omega_2 \, \omega^{\Delta_2-1}_2 e^{-i \omega_2 (\tilde{q}_2.y+u_2)}e^{-\epsilon\omega_2 } \, , \\
		G(x,y)&= \int \frac{d^{d+1}k}{(2\pi)^{d+1}} \frac{e^{ik \cdot (x-y)}}{k^2+m^2+i\epsilon} \, ,
	\end{align}
\end{subequations}
and the normalisation constant $N_{\Delta_i}^{d}$ is given by
\begin{equation}\label{eq:normscalar}
	N_{\Delta_i}^{d} = \frac{(\mp i)^{\Delta_i}}{2\pi^{\frac{d}{2}}\Gamma(\Delta_i-\frac{d}{2}+1)R^{\frac{(d-1)}{2}-\Delta_i}} \, .
\end{equation}

We see that $\psi^{\pm}_{\Delta_1,\tilde{q}_1,u_1}(x)$ are just the Carrollian primary wavefunctions associated with the modified Mellin transform \eqref{eq:carrollwave} \cite{Banerjee:2018gce}. Thus, the S-matrix is covariantly encoded in terms of the basis of Carrollian primary wavefunctions. In the large $R$ limit, the bulk to boundary propagator \eqref{eq:carrollscalwave} reduces to the Carrollian primary wavefunction \eqref{eq:carrollwave} up to some overall normalization constant. 

\medskip

The works of \cite{Giddings:1999jq,Penedones:2010ue,Fitzpatrick:2011hu,Fitzpatrick:2011dm,Hijano:2019qmi,Hijano:2020szl,Li:2021snj} give an algorithm to extract the bulk S-Matrix in the centre of AdS from AdS/CFT correlators. Following the lead of \cite{deGioia:2022fcn,Bagchi:2023fbj}, we see that if one keeps track of the boundary symmetries in this limit, we end up with the correlators relevant for flat space holography. The key to this construction is the fact that the bulk to boundary propagator (restricted to specific time slices of the CFT) reduces to the Carrollian primary wavefunction that encodes flat space scattering in the center of AdS. With this information, we can now look at Witten diagrams in the large AdS radius limit.

\medskip

A general Witten diagram with appropriate vertices, internal lines, and external lines will be of the form
\begin{equation}\label{eq:genwitten}
	\begin{split}
		\langle O_{\Delta_1}(\p_1) O_{\Delta_2}(\p_2) \dots O_{\Delta_i}(\p_i) O_{\Delta_{i+1}}(\p_{i+1}) O_{\Delta_{i+2}}(\p_{i+2}) \dots O_{\Delta_j}(\p_j) \rangle \\
		= (ig)^V \int_{\text{AdS}_{d+1}}d^{d+1}\x_1 \dots d^{d+1}\x_V \Pi_{\Delta}(\x_k,\x_l) \dots \Pi_{\Delta}(\x_k',\x_l') \\
		\textbf{K}_{\Delta_1}(\p_1,\x_k) \dots \textbf{K}_{\Delta_i}(\p_i,\x_k') \,
		\textbf{K}_{\Delta_{i+1}}(\p_{i+1},\x_l) \dots \textbf{K}_{\Delta_j}(\p_j,\x_l') \, ,
	\end{split}
\end{equation}
where $V$ denotes the number of vertices, and there are an appropriate number of external lines connected to the bulk to bulk propagator. For example, a four-point function with the $\phi^3$ interaction will be of the form
\begin{equation}
	\begin{split}
		\langle O_{\Delta_1}(\p_1)O_{\Delta_2}(\p_2)O_{\Delta_3}(\p_3)O_{\Delta_4}(\p_4)\rangle= &(ig)^2\int_{\text{AdS}_{d+1}}d^{d+1}\x \, d^{d+1}\y \, \Pi_\Delta(\x,\y)\\ &\textbf{K}_{\Delta_{1}}(\p_1,\x)\textbf{K}_{\Delta_{3}}(\p_3,\y)\textbf{K}_{\Delta_{2}}(\p_2,\x)\textbf{K}_{\Delta_{4}}(\p_4,\y)	\, .
	\end{split}
\end{equation}
Now suppose, $\p_m$ are inserted at $\tau = -\frac{\pi}{2}+\frac{u_m}{R}$ for $m=1,\dots,i$ and at $\tau = \frac{\pi}{2}+\frac{u_m}{R}$ for $m=i+1,\dots,j$, then using the form of the incoming and outgoing wavefunctions given by \eqref{eq:carrollscalwave}; and also the form of the bulk to bulk propagator, we can simplify \eqref{eq:genwitten} in the large $R$ limit as
\begin{equation}\label{eq:genwittenlarger}
	\begin{split}
		\langle O_{\Delta_1}(\p_1) \dots O_{\Delta_i}(\p_i) O_{\Delta_{i+1}}(\p_{i+1}) \dots O_{\Delta_j}(\p_j) \rangle =(\prod^{j}_{a=1}N^d_{\Delta_a})(ig)^V \int_{\mathbb{R}^{1,d}} d^{d+1}x_1 \dots d^{d+1}x_V \\ \times G(x_k,,x_l) \dots G(x_k',x_l') \,
		\psi^-_{\Delta_1,\Tilde{q}_1,u_1}(x_k)  \dots \psi^-_{\Delta_i,\Tilde{q}_i,u_i}(x_k') \\
		\times \psi^+_{\Delta_{i+1},\Tilde{q}_{i+1},u_{i+1}}(x_l)  \dots \psi^+_{\Delta_j,\Tilde{q}_j,u_j}(x_l') + O(R^{-1}) \, .
	\end{split}
\end{equation}

\subsection{Spinning propagators}

We now consider spinning fields, and we are interested in studying the large AdS radius limit of spinning Witten diagrams. The embedding space representation of AdS is crucial to our flat limit, and the spinning propagators have been worked out in AdS embedding space representation in \cite{Costa:2014kfa}. We borrow those expressions and then implement the flat limit through \eqref{eq:bulkflatlimit} and \eqref{eq:bdyflatlimit}. We will consider the bulk to boundary propagator and the bulk to bulk propagator in the large AdS radius limit. The analysis of vertices is similar to \cite{deGioia:2022fcn,Bagchi:2023fbj}.

\subsubsection{Bulk to boundary propagator}
\label{sec:bbdypropspin}

The bulk to boundary propagator of a spin $J$ particle with dimension $\Delta$ is given by \cite{Costa:2014kfa,deGioia:2023cbd}
\begin{equation}\label{eq:bbdppropspin}
	\begin{split}
		K^{\Delta,J}_{\Vec{\alpha},\Vec{\beta}}\left(\p,\x\right) = \mathcal{C}_{\Delta;J} \, \partial_{\alpha_1} X^{A_1} \dots \partial_{\alpha_J} X^{A_J} \, \partial_{\beta_1} P^{B_1} \dots \partial_{\beta_J} P^{B_J} \\
		~~\times \dfrac{L_{\lfloor A_1;\lfloor B_1}\left( X;P\right)\dots L_{A_J\rfloor ;B_J\rfloor }\left(X;P\right)}{\left(-P \cdot X + i \epsilon \right)^{\Delta}} \, ,
	\end{split}
\end{equation}
where
\begin{equation}
	L_{A;B}\left(X;P\right) = \dfrac{-P \cdot X \eta_{AB} + P_A X_B}{-P \cdot X + i\epsilon} \, ,
\end{equation}
and the normalization constant is given by
\begin{equation}\label{eq:normspin}
	\mathcal{C}^d_{\Delta;J} = \dfrac{(J + \Delta -1) \, \Gamma(\Delta)}{2\pi^{\frac{d}{2}}\,(\Delta -1)\, \Gamma\left(\Delta + 1 - \frac{d}{2} \right) R^{\frac{d-1}{2}-\Delta + J}} \, .
\end{equation}
The notations used are:
\begin{equation}
	\begin{split}
		A_i,B_i &\to \text{embedding space indices} \, , \\
		\alpha_i &\to \text{bulk coordinates} \, , \\
		\beta_i &\to \text{boundary coordinates} \, , \\
		\partial_{\alpha_i} X^{A_i}, \partial_{\beta_i} P^{B_i} &\to \text{projectors onto the corresponding bulk and boundary tensors} \, , \\
		\lfloor . \rfloor &\to \text{symmetric traceless component} \, .
	\end{split}
\end{equation}
Thus, \eqref{eq:bbdppropspin} is just the scalar bulk to boundary propagator dressed with factors of $L_{A;B}(X;P)$.

\medskip

To implement the flat limit, we first rescale the $\alpha_i$ and $\beta_i$ such that $\alpha_i$ runs over $(t,r\Omega)$ and $\beta_i$ runs over $(u,\Omega)$. One can straightforwardly verify that under a large $R$ $\left( \tau_p = \pm \frac{\pi}{2} + \frac{u}{R} \right)$ limit, if we substitute \eqref{eq:xplarger} in \eqref{eq:bbdppropspin} we have two structures
\begin{equation}\label{eq:Iprelim}
	\begin{split}
		(i) ~\eta_{AB} \partial_{\alpha} X^{A} \partial_{\beta} P^{B} &= \begin{cases}
			\mo(R^{-2}) & \text{$\beta = u$ \, ,} \\
			\pm \partial_a \tilde{q}_{\alpha} + \mo(R^{-2}) & \text{$\beta = z^a = z,\bar{z}$ \, ,}
		\end{cases} \\
		(ii) ~ P_A X_B\partial_{\alpha} X^{A} \partial_{\beta} P^{B} &= \begin{cases}
			\pm \tilde{q}_{\alpha} + \mo(R^{-2}) & \text{$\beta = u$ \, ,} \\
			\pm \left( \partial_a \tilde{q} \cdot x \right) \tilde{q}_{\alpha} + \mo(R^{-2}) & \text{$\beta = z^a = z,\bar{z}$ \, .}
		\end{cases}
	\end{split}
\end{equation}
One should consider the antipodal map \eqref{eq:antipodal} for the other $\tau_p$. It is also crucial to rescale the coordinates $\mu_i$ and $\nu_i$ before implementing the flat limit, or else terms diverge as $\mathcal{O}(R)$. 

From \eqref{eq:Iprelim}, we can basically compute $\partial_{\alpha}X^{A}\partial_{\beta}P^B L_{A;B}$. For $\nu =u$, we have from \eqref{eq:pdotx}
\begin{equation}\label{eq:IABucomp}
	L_{A;B} \partial_{\alpha} X^A \partial_u P^B = \dfrac{\pm \tilde{q}_{\alpha}}{(\mp u \mp \tilde{q} \cdot x + i\epsilon)} \, .
\end{equation}
Thus, for $\beta=u$, the bulk to boundary propagator for spin 1 \eqref{eq:bbdppropspin} becomes pure gauge because
\begin{equation}\label{eq:spin1ucomp}
	K^{\Delta,1}_{\alpha,u} = \mathcal{C}^d_{\Delta;1} \dfrac{\pm \tilde{q}_{\mu}}{(\mp u \mp \tilde{q} \cdot x + i\epsilon)^{\Delta+1}} = -\dfrac{\mathcal{C}^d_{\Delta;1}}{\Delta} \partial_{\alpha} \left[ \dfrac{1}{(\mp u \mp \tilde{q} \cdot x + i\epsilon)^{\Delta}} \right] \, .
\end{equation}
This exactly matches the result of \eqref{eq:spin1carrfieldu}. 

\medskip

This is also consistent with the fact that if we take a bulk $U(1)$ gauge field $A_{\alpha}$ and push it to the boundary $\mathscr{I}^+$, the $u$ component drops out as $A_u \sim \mathcal{O}(\frac{1}{r})$ but the $z,\bar{z}$ components are non-trivial as $A_{z} \sim A_{\bar{z}} \sim \mathcal{O}(1)$ \footnote{Here $r$ is the usual radial coordinate. The limit to $\mathscr{I}^+$ is achieved by keeping $u$ constant and letting $r \to \infty$. Thus, one has to do a mode expansion of the bulk field and evaluate the boundary limit by a saddle point approximation.} \cite{Nguyen:2023vfz}. Hence, the scattering data is solely present in the spatial components. We will see that the $f_{z\z}$ correlator of \eqref{eq:carrollspin12pt} agrees with the bulk result \eqref{eq:gluon2ptfinal}. We remark that there might be sub-leading corrections to the bulk to boundary propagator in the large $R$ limit. We leave a careful analysis for future work.

\medskip

For $\beta=z^a$,
\begin{equation}\label{eq:IABzcomp}
	L_{A;B} \partial_{\alpha} X^A \partial_{a} P^B = \pm \partial_a \tilde{q}_{\alpha} + \dfrac{\pm (\partial_a \tilde{q} \cdot x)\tilde{q}_{\alpha}}{(\mp u \mp \tilde{q} \cdot x + i\epsilon)} \, . 
\end{equation}
Thus, the bulk to boundary propagator for spin 1 \eqref{eq:bbdppropspin} with the boundary index being that of the points on the Celestial sphere is given by
\begin{equation}\label{eq:mellinprimaryspin1}
	K^{\Delta,1}_{\alpha,a} =  \mathcal{C}^d_{\Delta;1} \left[ \dfrac{\pm \partial_a \tilde{q}_{\alpha}}{(\mp u \mp \tilde{q} \cdot x + i\epsilon)^{\Delta}} + \dfrac{\pm (\partial_a \tilde{q} \cdot x)\tilde{q}_{\alpha}}{(\mp u \mp \tilde{q} \cdot x + i\epsilon)^{\Delta+1}} \right] \, .
\end{equation}
This is the modified Mellin avatar of the conformal primary wavefunctions worked out in \cite{Pasterski:2017kqt}. This is the crucial difference when compared to \cite{deGioia:2023cbd}. This is due to the difference in the large $R$ limit \eqref{eq:bdyflatlimit} used, and as we will see, this will significantly change the resulting Witten diagrams.

\medskip

The form of the propagator in \eqref{eq:mellinprimaryspin1} is difficult to work with in Witten diagrams due to the explicit appearance of the bulk point $x$. Fortunately, as explained in \cite{Pasterski:2017kqt}, one can use the following gauge representative
\begin{equation}\label{eq:mellinspin1}
	K^{\Delta,1}_{\alpha,a} =  \mathcal{C}^d_{\Delta;1} \dfrac{\Delta-1}{\Delta} \dfrac{\pm \partial_a \tilde{q}_{\alpha}}{(\mp u \mp \tilde{q} \cdot x + i\epsilon)^{\Delta}} \, .
\end{equation}
This is gauge equivalent to \eqref{eq:mellinprimaryspin1} because we have \cite{Donnay:2018neh}
\begin{equation}\label{eq:gaugeqspin1}
	\dfrac{\pm \partial_a \tilde{q}_{\alpha}}{(\mp u \mp \tilde{q} \cdot x + i\epsilon)^{\Delta}} + \dfrac{\pm (\partial_a \tilde{q} \cdot x)\tilde{q}_{\alpha}}{(\mp u \mp \tilde{q} \cdot x + i\epsilon)^{\Delta+1}} = \dfrac{\Delta-1}{\Delta} \dfrac{\pm \partial_a \tilde{q}_{\alpha}}{(\mp u \mp \tilde{q} \cdot x + i\epsilon)^{\Delta}} + \partial_{\alpha} \gamma^{\Delta,\pm}_a \, ,
\end{equation}
with
\begin{equation}\label{eq:puregauge}
	\gamma^{\Delta,\pm}_a = \dfrac{\pm \partial_a \tilde{q} \cdot x}{\Delta(\mp u \mp \tilde{q} \cdot x + i\epsilon)^{\Delta}} \, .
\end{equation}
This is advantageous to us because we have a modified Mellin integral representation
\begin{equation}\label{eq:mellinintspin1}
	K^{\Delta,1}_{\alpha,a} =  \dfrac{\mathcal{C}^d_{\Delta;1}}{(\pm i)^{\Delta} \Gamma(\Delta)} \dfrac{\Delta-1}{\Delta} \int^{\infty}_0 d\omega \, \omega^{\Delta-1} \, e^{\mp i \omega \, u} e^{\mp i\omega \, \tilde{q} \cdot x} e^{-\epsilon \, \omega} (\partial_a \tilde{q}_{\alpha}) \, .
\end{equation}

\par

From \eqref{eq:Iprelim}, one can also straightforwardly obtain the spin 2 graviton primary wavefunctions from the large $R$ limit. Here, we should take into account the fact that the projections in \eqref{eq:bbdppropspin} only pick out the symmetric and traceless component of the dressing functions $L_{A;B}\left(X;P\right)$. Now, the $\beta_1=u,\beta_2=u$ component follows from \eqref{eq:IABucomp}:
\begin{equation}\label{eq:uupurediffeo}
	\begin{split}
		&K^{\Delta,2}_{\alpha_1 \alpha_2,uu} =  \mathcal{C}^d_{\Delta;2}\dfrac{\pm \tilde{q}_{\alpha_1} \tilde{q}_{\alpha_2}}{(\mp u \mp \tilde{q} \cdot x + i\epsilon)^{\Delta+2}} \\
		&= -\dfrac{\mathcal{C}^d_{\Delta;2}}{2(\Delta+1)} \left( \partial_{\alpha_1} \left[ \dfrac{\tilde{q}_{\alpha_2}}{(\mp u \mp \tilde{q} \cdot x + i\epsilon)^{\Delta+1}} \right] + \partial_{\alpha_2} \left[ \dfrac{\tilde{q}_{\alpha_1}}{(\mp u \mp \tilde{q} \cdot x + i\epsilon)^{\Delta+1}} \right] \right) \\
		&= \partial_{\alpha_1} \xi_{\alpha_2} + \partial_{\alpha_2} \xi_{\alpha_1} \, .
	\end{split}
\end{equation}
Thus, analogous to \eqref{eq:spin1ucomp}, the $\beta_1=u,\beta_2=u$ component becomes pure diffeomorphism. Also, one can use \eqref{eq:IABucomp}, \eqref{eq:IABzcomp} and \eqref{eq:gaugeqspin1} to argue that the $\beta_1 =u,\beta_2=z^a$ component becomes pure diffeomorphsim analogous to \eqref{eq:uupurediffeo}. This can be generalized to higher spin by noting that because of \eqref{eq:IABucomp}, the bulk to boundary propagator \eqref{eq:bbdppropspin} becomes pure gauge when one of the boundary indices is taken to be $u$.

\medskip

For the $\beta_1 = z^a,\beta_2=z^a$ term, we can follow a similar analysis as \eqref{eq:gaugeqspin1} \cite{Pasterski:2017kqt} to show that we can use the Mellin gauge representative (up to terms that are pure diffeomorphism) given by 
\begin{equation}\label{eq:gravitonmellin}
	K^{\Delta,2}_{\alpha_1 \alpha_2,a_1 a_2} =  \mathcal{C}^d_{\Delta;2} \dfrac{\Delta-1}{\Delta} \dfrac{\pm P^{b_1 b_2}_{a_1 a_2} \, \partial_{b_1} \tilde{q}_{\alpha_1} \, \partial_{b_2} \tilde{q}_{\alpha_2}}{(\mp u \mp \tilde{q} \cdot x + i\epsilon)^{\Delta}} \, .
\end{equation}
Here $P^{b_1 b_2}_{a_1 a_2}$ is the projector given by
\begin{equation}
	P^{b_1 b_2}_{a_1 a_2} = \delta^{b_1}_{(a_1}\delta^{b_2}_{a_2)} - \dfrac{1}{d-1}g_{a_1 a_2} g^{b_1 b_2} \, .
\end{equation}
This projector ensures that the bulk to boundary propagator is transverse and traceless.

\subsubsection{Bulk to bulk propagator}

For the bulk to bulk propagator, we simply follow the analysis of \cite{deGioia:2023cbd}. The spin $J$ bulk-bulk propagator satisfies the following equations \cite{Costa:2014kfa}:
\begin{equation}
	\begin{split}
		\left( \Box_{\text{AdS}} - \dfrac{\Delta(\Delta-d)}{R^2} + \dfrac{J}{R^2} \right) \Pi_{\alpha_1 \dots \alpha_J,\beta_1 \dots \beta_J}(X,\bar{X}) &= -g_{\alpha_1 \lfloor \beta_1} \dots g_{|\alpha_J|\beta_J\rfloor } \delta_{\text{AdS}}\left(X,\Bar{X}\right) \, , \\
		\nabla^{\alpha_1} \Pi_{\alpha_1 \dots \alpha_J,\beta_1 \dots \beta_J}(X,\bar{X}) &= 0 \, .
	\end{split}
\end{equation}
One can follow a similar analysis of \cite{deGioia:2022fcn,Bagchi:2023fbj} to arrive at
\begin{equation}
	\Pi_{\alpha_1 \dots \alpha_J,\beta_1 \dots \beta_J}(X,\bar{X}) = G_{\alpha_1 \dots \alpha_J,\beta_1 \dots \beta_J}(x,\bar{x}) + \mo(R^{-2}) \, ,
\end{equation}
where $G_{\alpha_1 \dots \alpha_J,\beta_1 \dots \beta_J}(x,\bar{x})$ is the Feynman propagator for a symmetric traceless tensor of spin $J$ in $\mathbb{R}^{1,d}$.

\bigskip \bigskip

\section{Explicit calculations of Witten diagrams}
\label{sec:wittendiagex}

In this section, we will use the analysis of Sec. \ref{sec:flatlimitwitten} to construct the AdS Witten diagrams focussing on spinning particles and obtain a large $R$ series expansion. We show that the spinning Carrollian CFT correlators of Sec. \ref{sec:carrollcorrelators} arise naturally as the leading order term in the spinning equivalent of \eqref{eq:genwittenlarger}. To compare with the intrinsic Carrollian CFT correlators, we work with $3+1$ dimensions because $\text{BMS}_{4}$ symmetry group was used to constrain the correlators living on the null boundary of $\mathbb{R}^{1,3}$ in Sec. \ref{sec:carrollcorrelators}. However, we emphasize that the formalism is valid for arbitrary spacetime dimensions.

\medskip

In $3+1$ dimensions, we use the following parametrization for the null vector in the direction of $\mathbf{\Omega}_p$:
\begin{equation}\label{eq:4dparamq}
	\tilde{q}^{\mu} = \left[1,\dfrac{z+\bar{z}}{1+z\bar{z}},\dfrac{-i(z-\bar{z})}{1+z\bar{z}},\dfrac{1-z\bar{z}}{1+ z\bar{z}} \right] \, .
\end{equation}
The topology of the null boundary is $\mathbb{R}\times S^2$ and $(z,\bar{z})$, which characterize the coordinates on the celestial sphere. As considered in \cite{Bagchi:2023fbj}, we must carefully ensure that the split of the momentum conserving delta functions is Lorentz invariant to evaluate Witten diagrams. In the below subsection, we will summarize the diagrams that we have computed. We also highlight the salient features of each diagram.

\bigskip \bigskip

\subsection{Summary of the diagrams computed}
\label{sec:summarywitten}

We list the various bulk Witten diagrams (twelve in total) that we have computed and matched with the field theory analysis of Section \ref{sec:carrollcorrelators}:

\begin{enumerate}
	\item{\em{Scalar three point function: non collinear case}} \\
	A non-trivial branch of the three point scattering where one of the particles is soft and the remaining two scatter collinearly. This branch was not considered in \cite{Bagchi:2023fbj}. Here, we see how the $\Delta \to 1$ subsector captures the soft sector.
	\item{\em{Scalar three point function: split signature}} \\
	Three point scattering amplitude is non-zero for bulk $(2,2)$ split signature spacetimes. We get a non-trivial result by working in the celestial torus by treating $z$ and $\bar{z}$ independently.
	\item{\em{Scalar four point function: contact diagram}} \\
	A contact Witten diagram arising from a $\phi^4$ interaction in the bulk effective theory.
	\item{\em{Scalar four point function: exchange diagram}} \\
	A t-channel scalar exchange Witten diagram that arises through a $\phi^3$ interaction.
	\item{\em{Gluon two point function}}: The diagram that captures free gluon propagation.
	\item{\em{Scalar-gluon-scalar three point function: split signature}} \\
	A contact Witten diagram that arises through a cubic interaction vertex involving a spin 1 field and two scalar fields.
	\item{\em{Gluon MHV three point amplitude: split signature}} \\
	A contact Witten diagram that arises through a cubic gluon interaction vertex. We get a finite result, which should be contrasted with the Celestial case \cite{Pasterski:2017ylz,Stieberger:2018edy}, where one has distributional delta functions in the conformal weights of primaries.
	\item{\em{Gluon three point amplitude from $F^3$: split signature}} \\
	A non-MHV $(+++)/(---)$ amplitude arising from bulk loop corrections. These amplitudes cannot be captured by the 2D Celestial CFT because they are quadratically divergent (as they are loop amplitudes) in the Mellin basis. This is even more severe than the linearly divergent MHV graviton scattering amplitude \cite{Stieberger:2018edy}. We capture this scattering in a finite Carroll CFT correlator.
	\item{\em{Graviton two point function:}} The diagram that captures free graviton propagation.
	\item{\em{Scalar-graviton-scalar three point function: split signature}} \\
	A contact Witten diagram that arises through a cubic interaction vertex involving a spin 2 field and two scalar fields.
	\item{\em{Graviton MHV three point amplitude: split signature}} \\
	A contact Witten diagram that arises through a cubic graviton interaction vertex. Graviton amplitudes are well known to be divergent in the Mellin basis \cite{Stieberger:2018edy}. We show that such scattering processes could be captured by a finite Carroll CFT correlator \cite{Banerjee:2019prz}.
	\item{\em{Graviton-gluon-gluon three point function: split signature}} \\
		A contact Witten diagram that arises in Einstein-Yang-Mills theory with a vertex involving a spin 2 field and two spin 1 fields.
\end{enumerate}

We now proceed to evaluate all the diagrams.

\subsection{Scalar three point function: non collinear case}

We consider the contact Witten diagram in $\phi^3$ theory:
\begin{equation}
	\langle O_{\Delta_1}(\p_1) O_{\Delta_2}(\p_2) O_{\Delta_3}(\p_3) \rangle = \int_{\text{AdS}_4} d\x  \,\textbf{K}_{\Delta_{1}}(\p_1,\x) \, \textbf{K}_{\Delta_{2}}(\p_2,\x) \, \textbf{K}_{\Delta_{3}}(\p_3,\x) \, .
\end{equation}
In the large $R$ limit, with $\p_1$ inserted at $\tau_p = -\frac{\pi}{2}+\frac{u_1}{R}$ and $\p_2$ and $\p_3$ inserted at $\tau_p = \frac{\pi}{2}+\frac{u_i}{R}$, we have
\begin{equation}\label{eq:threepointscalsoft}
	\begin{split}
		\langle O_{\Delta_1}(\p_1) O_{\Delta_2}(\p_2) O_{\Delta_3}(\p_3) \rangle &= \mathcal{A}_{(3)s}\int_{\mathbb{R}^{1,3}} d^4x \, \psi^{-}_{\Delta_1,\tilde{q}_1,u_1}(x)\,\psi^{+}_{\Delta_2,\tilde{q}_2,u_2}(x)\,\psi^{+}_{\Delta_3,\tilde{q}_3,u_3}(x) \\
		&= \mathcal{A}_{(3)s}\int^{\infty}_0 d\omega_1 d\omega_2 d\omega_3 \, \omega^{\Delta_1-1}_1 \omega^{\Delta_2-1}_2 \omega^{\Delta_3-1}_3 e^{i\omega_1 \, u_1 - i\omega_2 \, u_2 - i \omega_3 \, u_3} \\
		& \hspace{5cm} \times \delta^{(4)}(\omega_1 \tilde{q}_1 - \omega_2 \tilde{q}_2 - \omega_3 \tilde{q}_3) \, .
	\end{split}
\end{equation}
The normalization constant is given by
\begin{equation}
	\mathcal{A}_{(3)s} = (i\mu) N^3_{\Delta_1} N^3_{\Delta_2} N^3_{\Delta_3} = \dfrac{(i\mu)\, (i)^{\Delta_1} (-i)^{\Delta_2+\Delta_3}}{8\pi^{\frac{9}{2}} \prod^3_{i=1}\Gamma\left(\Delta_i - \frac{1}{2}\right) R^{3-\Delta_1-\Delta_2-\Delta_3}} \, .
\end{equation}

We will use a soft factorization of the momentum conserving delta function to get a non zero answer. This branch was not considered in \cite{Bagchi:2023fbj}. The split of the delta function is given by
\begin{equation}
	\delta^{(4)}(\omega_1 \tilde{q}_1 -\omega_2 \tilde{q}_2 - \omega_3 \tilde{q}_3) = \dfrac{1}{\omega^2_3}\dfrac{1}{z_{12}\Bar{z}_{12}}\delta(\omega_2)\delta(\omega_1 - \omega_3)\delta^2(z_{13}) \, .
\end{equation}
Because of the crucial soft factor $\delta(\omega_2)$, we have
\begin{equation}
	\int^{\infty}_0 d\omega_2 \, \omega^{\Delta_2-1}_2 e^{i\omega_2 \, u_2} \delta(\omega_2) = \delta_{\Delta_2,1} \, .
\end{equation}
Due to $\delta(\omega_2)$, we need to regularize the integral by choosing a small negative value for the lower bound as $-\epsilon$ and then let $\epsilon\to 0$. The remaining integrals of \eqref{eq:threepointscalsoft} can be straightforwardly simplified to
\begin{equation}
	\langle O_{\Delta_1}(\p_1) O_{\Delta_2}(\p_2) O_{\Delta_3}(\p_3) \rangle = \mathcal{A}_{(3)s}\dfrac{\Gamma(\Delta_1 + \Delta_3 - 3)\delta_{\Delta_2,1} \, \delta^2(z_{13})}{\left(i(u_1 - u_3)^{\Delta_1 + \Delta_3 -3}\right) \, z_{12} \, \Bar{z}_{12}} \, .
\end{equation}
If we compare comparing with the intrinsic result given by \eqref{eq:noncollinear3ptintrinsic}, we have
\begin{equation}
	c = 3 - \Delta_1 -\Delta_3 \, , ~~~ p=-1 \, , ~~~ q=-1 \, .
\end{equation}
This exactly matches with the intrinsic result \eqref{eq:noncollinear3ptintrinsic} (as $\Delta_2 = 1$) from which we have
\begin{equation}
	\Delta_2 = -\dfrac{q+p}{2} \, , ~~~ \Delta_1 + \Delta_3 = 2 - c - p \, .
\end{equation}

\subsection{Scalar three point function: split signature}

The three point function is also non-zero when we consider the split signature for the space-time \cite{Pasterski:2017ylz}. This branch of the three point function was not considered in \cite{Bagchi:2023fbj}. We consider the following split of the momentum conserving delta function that is suited for the celestial torus \cite{Stieberger:2018edy}
\begin{equation}\label{eq:celestialtorusdelta}
	\delta^{(4)}(\omega_1 \tilde{q}_1 + \omega_2 \tilde{q}_2 - \omega_3 \tilde{q}_3) = \dfrac{4}{\omega^2_3 z_{23}z_{31}}\delta\left(\omega_1 - \omega_3 \dfrac{z_{32}}{z_{12}} \right) \delta\left(\omega_2 - \omega_3 \dfrac{z_{31}}{z_{21}} \right)\delta(\Bar{z}_{13})\delta(\Bar{z}_{23}) \, .
\end{equation}
As explained in \cite{Pasterski:2017ylz,Stieberger:2018edy}, we must choose one of the orderings $z_1 < z_3 < z_2$ or $z_2 < z_3 < z_1$ to ensure that the energies are positive. The proof of the Lorentz invariance of the split is given in Appendix \ref{ap:lorentzdelta}. Consider the three point contact Witten diagram in $\phi^3$ theory:
\begin{equation}\label{eq:threepointsplitfirst}
	\langle O_{\Delta_1}(\p_1) O_{\Delta_2}(\p_2) O_{\Delta_3}(\p_3) \rangle = \int_{\text{AdS}_4} d\x  \,\textbf{K}_{\Delta_{1}}(\p_1,\x) \, \textbf{K}_{\Delta_{2}}(\p_2,\x) \, \textbf{K}_{\Delta_{3}}(\p_3,\x) \, .
\end{equation}
In the large $R$ limit, with $\p_1$ and $\p_2$ inserted at $\tau_p = - \frac{\pi}{2}+\frac{u_i}{R}$ and $\p_3$ inserted at $\tau_p = \frac{\pi}{2}+\frac{u_3}{R}$, we have
\begin{equation}
	\begin{split}
		\langle O_{\Delta_1}(\p_1) O_{\Delta_2}(\p_2) O_{\Delta_3}(\p_3) \rangle &= \mathcal{A}_{(3)sp}\int_{\mathbb{R}^{1,3}} d^4x \, \psi^{-}_{\Delta_1,\tilde{q}_1,u_1}(x)\,\psi^{-}_{\Delta_2,\tilde{q}_2,u_2}(x)\,\psi^{+}_{\Delta_3,\tilde{q}_3,u_3}(x) \\
		&= \mathcal{A}_{(3)sp}\int^{\infty}_0 d\omega_1 d\omega_2 d\omega_3 \, \omega^{\Delta_1-1}_1 \omega^{\Delta_2-1}_2 \omega^{\Delta_3-1}_3 e^{i\omega_1 u_1 + i\omega_2 u_2 - i \omega_3 u_3} \\
		& \hspace{5cm} \times \delta^{(4)}(\omega_1 \tilde{q}_1 + \omega_2 \tilde{q}_2 - \omega_3 \tilde{q}_3) \, .
	\end{split}
\end{equation}
Using the celestial torus split of the delta function and evaluating the resulting integrals, we get
\begin{equation}\label{eq:threepointsplitresult}
	\begin{split}
		\langle O_{\Delta_1}(\p_1) O_{\Delta_2}(\p_2) O_{\Delta_3}(\p_3) \rangle &= \mathcal{A}_{(3)sp} (-1)^{\Delta_1 + \Delta_2 - 2}\Gamma(\Delta_1 + \Delta_2 + \Delta_3 -4) \, \delta(\Bar{z}_{12}) \,\delta(\Bar{z}_{23}) \\&\times
		\left[-i(z_1 u_{23} + z_2 u_{31} + z_3 u_{12}) \right]^{4-\Delta_1 -\Delta_2 -\Delta_3} z^{\Delta_3 -2}_{12} z^{\Delta_1 - 2}_{32} z^{\Delta_2 - 2}_{31} \, .
	\end{split}
\end{equation}
\begin{equation}
	\mathcal{A}_{(3)sp} = (i\mu) N^3_{\Delta_1} N^3_{\Delta_2} N^3_{\Delta_3} = \dfrac{(i\mu)\, (i)^{\Delta_1 +\Delta_2} (-i)^{\Delta_3}}{8\pi^{\frac{9}{2}} \prod^3_{i=1}\Gamma\left(\Delta_i - \frac{1}{2}\right) R^{3-\Delta_1-\Delta_2-\Delta_3}} \, .
\end{equation}
\eqref{eq:threepointsplitresult} matches the intrinsic result \eqref{eq:threepointsplit} once we use the fact that we are working with scalars with zero helicities ($\sigma_i = 0$). The calculations leading to \eqref{eq:threepointsplitresult} are given in Appendix \ref{ap:threepointsplit}. We will see that one can extend this by considering three point gluon and graviton amplitudes. In that case, non-trivial helicities will play a role. 

\medskip

Note that the signature of AdS$_4$ for the bulk spacetime is still $(-,+,+,+)$. The three point function is non-zero because the momenta of the particles become complex valued \cite{Pasterski:2017ylz}.

\subsection{Scalar four point function: contact diagram}
\label{sec:scalfourpointcontact}

We now consider a couple of four point scalar diagrams. Let us first evaluate the contact Witten diagram for the four point function given by
\begin{equation}
	\begin{split}
		\langle O_{\Delta_1}(\p_1)O_{\Delta_2}(\p_2)&O_{\Delta_3}(\p_3)O_{\Delta_4}(\p_4)\rangle \\
		& =(ig_2)\int_{AdS_{4}}d^{4}\x \, \textbf{K}_{\Delta_{1}}(\p_1,\x)\textbf{K}_{\Delta_{2}}(\p_2,\x)\textbf{K}_{\Delta_{3}}(\p_3,\x)\textbf{K}_{\Delta_{4}}(\p_4,\x) \, .
	\end{split}
\end{equation}
We need a $\phi^4$ interaction in the bulk for this contact diagram. We will now insert $\p_1$ and $\p_2$ at $\tau=-\frac{\pi}{2}+ \frac{u_i}{R}$ and $\p_3$ and $\p_4$ at $\tau=\frac{\pi}{2}+\frac{u_i}{R}$. Thus, we have
\begin{equation}\label{eq:fourptcontactfirst}
	\begin{split}
		\langle O_{\Delta_1}(\p_1)&O_{\Delta_2}(\p_2)O_{\Delta_3}(\p_3)O_{\Delta_4}(\p_4)\rangle \\
		&= \mathcal{A}_{(4)c} \int d\omega_1 \, d\omega_2 \, d\omega_3 \, d\omega_4 \, e^{i(\omega_1 u_1 + \omega_2 u_2 - \omega_3 u_3 - \omega_4 u_4)} (ig_2) \\
		&\times \delta^{4}(\omega_3 \Tilde{q}_3 + \omega_4 \Tilde{q}_4 - \omega_1 \Tilde{q}_1 - \omega_2 \Tilde{q}_2) \, \omega^{\Delta_1-1}_1 \omega^{\Delta_2-1}_2 \omega^{\Delta_3-1}_3\omega^{\Delta_4-1}_4e^{-\epsilon(\omega_1+\omega_2+\omega_3+\omega_4)} \, .
	\end{split}
\end{equation}
This just becomes the modified Mellin transform of the flat space scattering amplitude $ig_2$. The integrals in $\omega_i$ can be simplified by changing variables to simplex variables as described in \cite{Pasterski:2017ylz,Banerjee:2019prz}. This is reviewed in Appendix \ref{ap:fourpointcontact}. The expression of \eqref{eq:fourptcontactfirst} can be simplified to 
\begin{equation}\label{eq:fourptcontactresult}
	\begin{split}
		\dfrac{\mathcal{A}_{(4)c} \, ig_2 \, \Gamma(\Delta_1 + \Delta_2 + \Delta_3 + \Delta_4 - 4)}{ \left[-i(\sigma^*_1 u_1 + \sigma^*_2 u_2 - \sigma^*_3 u_3 - \sigma^*_4 u_4 )\right]^{\Delta_1 + \Delta_2 + \Delta_3 + \Delta_4 - 4}} \prod_{i=1}^4 (\sigma^*_i)^{\Delta_i-1} \dfrac{\delta(|z-\Bar{z}|)}{4 z_{12} \Bar{z}_{13} z_{24} \Bar{z}_{24}} \prod_{i=1}^4 \mathbbm{1}_{[0,1]}(\sigma^*_i) \, .
	\end{split}
\end{equation}
If we had set $u=0$ (the Celestial case \cite{deGioia:2022fcn}) before the calculation, we would not have gotten the $u$ dependent factors. The $u$ independent factors agree with the equation above eq(3.11) of \cite{Banerjee:2017jeg}. The constant of proportionality is given by
\begin{equation}
	\mathcal{A}_{(4)c} = \dfrac{(-i)^{\Delta_1 + \Delta_2} (i)^{\Delta_3 + \Delta_4}}{16 \pi^6 \prod_{i=1}^4 \Gamma\left(\Delta_i - \frac{1}{2} \right) R^{4- (\Delta_1+\Delta_2+\Delta_3+\Delta_4)}} \, .
\end{equation}
The indicator function is defined as
\begin{equation}
	\mathbbm{1}_{[0,1]}(x) = \begin{cases}
		1 \, , & \text{if $x\in [0,1]$} \\
		0 \, , & \text{otherwise}
	\end{cases}
\end{equation}
$z$ and $\Bar{z}$ are the cross ratios given by
\begin{equation}
	z = \dfrac{z_{12} z_{34}}{z_{13} z_{24}} \, , ~~~~~ \Bar{z} = \dfrac{\Bar{z}_{12}\Bar{z}_{34}}{\Bar{z}_{13}\Bar{z}_{24}} \, .
\end{equation}
The remaining terms in \eqref{eq:fourptcontactresult} are given by
\begin{gather}
	\sigma^*_1 = \dfrac{1}{D} \dfrac{z_{24} \Bar{z}_{34}}{z_{12} \Bar{z}_{13}} \, , ~~ \sigma^*_2 = -\dfrac{1}{D} \dfrac{z_{34} \Bar{z}_{14}}{z_{23} \Bar{z}_{12}} \, , ~~ \sigma^*_3 = -\dfrac{1}{D} \dfrac{z_{24} \Bar{z}_{14}}{z_{23} \Bar{z}_{13}} \, , ~~ \sigma^*_4 = \dfrac{1}{D} \nonumber \\
	D = 2 \dfrac{z_{24} \Bar{z}_{34}}{z_{12} \Bar{z}_{13}} - 2 \dfrac{z_{34}\Bar{z}_{14}}{z_{23}\Bar{z}_{12}}
\end{gather}
One can show that the combination
\begin{equation}
	\sigma^*_1 u_1 + \sigma^*_2 u_2 - \sigma^*_3 u_3 - \sigma^*_4 u_4 \, ,
\end{equation}
is proportional to $(z-\Bar{z})$ under a global space-time translation $u_i \to u_i + A + B z_i + \Bar{B} \Bar{z}_i + C z_i \Bar{z}_i$. Thus, the amplitude is invariant under global space-time translations due to the $\delta(|z-\Bar{z}|)$ constraint \cite{Banerjee:2018gce}. This shows that the four point function satisfies the Carrollian Ward identities.

\subsection{Scalar four point function: exchange diagram}
The analysis is identical to section \ref{sec:scalfourpointcontact}, but the only difference is that we have a bulk to bulk propagator in the Witten diagram. We consider the t-channel massless exchange. This diagram has been used to illustrate the flat space limit in \cite{Penedones:2010ue}. For this diagram, we need a $\phi^3$ term in the bulk.
\begin{equation}
	\begin{split}
		\langle O_{\Delta_1}(\p_1)O_{\Delta_2}(\p_2)O_{\Delta_3}(\p_3)O_{\Delta_4}(\p_4)\rangle&= (ig)^2\int_{AdS_{d+1}}d^{d+1}\x d^{d+1}\y\Pi_\Delta(\x,\y)\\ &~~\textbf{K}_{\Delta_{1}}(\p_1,\x)\textbf{K}_{\Delta_{3}}(\p_3,\x)\textbf{K}_{\Delta_{2}}(\p_2,\y)\textbf{K}_{\Delta_{4}}(\p_4,\y)	\, .
	\end{split}
\end{equation}
The general expression of \eqref{eq:genwittenlarger} evaluates to 
\begin{equation}\label{eq:scalfourpointexfirst}
	\begin{split}
		\langle O_{\Delta_1}(\p_1)O_{\Delta_2}(\p_2)O_{\Delta_3}(\p_3)O_{\Delta_4}(\p_4)\rangle & \simeq \left(\prod_{i=1}^{4}N_{\Delta_i}^{d}\right)\left((ig)^2\int_{\mathbb{R}^{1,d}}d^{d+1}\x d^{d+1}\y G(x,y)\right. \\ & \left. \times\psi_{\Delta_1,\tilde{q}_1,u_1}^{-}(x)\psi_{\Delta_3,\tilde{q}_3,u_3}^{-}(x)\psi_{\Delta_2,\tilde{q}_2,u_2}^{+}(y)\psi_{\Delta_4,\tilde{q}_4,u_4}^{+}(y)\right) \, .
	\end{split}
\end{equation}
Due to momentum conservation, the internal momenta $k$ takes the value given by
\begin{equation}
	\begin{split}
		k &= - \omega_1 \tilde{q}_1 - \omega_3 \tilde{q}_3 \\
		\implies k^2 &= 2 \omega_1 \omega_3 \tilde{q}_1 \cdot \tilde{q}_3 = - \dfrac{4 \omega_1 \omega_3 z_{13} \bar{z}_{13}}{(1+z_1 \bar{z}_1)(1+ z_3 \bar{z}_3)} = s_{13} \, ,
	\end{split}
\end{equation}
where $s_{13}$ is the Mandelstam invariant. Thus, if we substitute this in \eqref{eq:scalfourpointexfirst}, we get
\begin{equation}
	\begin{split}
		\mathcal{A}_{(4)ex} (ig)^2 \int d\omega_1 d\omega_2 d\omega_3 d\omega_4 \, \omega^{\Delta_1 - 1}_1 \omega^{\Delta_2 -1}_2 \omega^{\Delta_3-1}_3 \omega^{\Delta_4-1}_4 \, e^{i(\omega_1 u_1 + \omega_3 u_3 - \omega_2 u_2 - \omega_4 u_4)} \\
		\times \delta^{(4)}(\omega_1 \tilde{q}_1 + \omega_3 \tilde{q}_3 - \omega_2 \tilde{q}_2 - \omega_4 \tilde{q}_4) \dfrac{1}{s_{13} + i \epsilon} \, .
	\end{split}
\end{equation}
$\frac{1}{s_{13}}$ is the on-shell scattering amplitude of the t-channel exchange process in flat space. Thus, in the large $R$ limit, we essentially get a modified Mellin transformation of the scattering amplitude. This is what one expects from the general proposal of \cite{Bagchi:2022emh}.

\medskip

Introducing simplex variables and doing a similar analysis as Appendix \ref{ap:fourpointcontact}, we get the final result for the four point correlator as
\begin{equation}\label{eq:fourpointexresult}
	\mathcal{A}_{(4)ex} \dfrac{(1+z_1 \bar{z}_1)(1+z_3 \bar{z}_3)}{4 z^2_{13} \bar{z}^2_{13} z_{24} \bar{z}_{24}}\delta(|z - \bar{z}|) \dfrac{(\sigma^*_1)^{\Delta_1-2}(\sigma^*_2)^{\Delta_2-1}(\sigma^*_3)^{\Delta_3-2}(\sigma^*_4)^{\Delta_4-1}}{[-i(\sigma^*_1 u_1 + \sigma^*_3 u_3 - \sigma^*_2 u_2 - \sigma^*_4 u_4)]^{\Delta_1 + \Delta_2 + \Delta_3 + \Delta_4 -6}} \, .
\end{equation}
Here
\begin{equation}
	\mathcal{A}_{(4)ex} = \dfrac{(-i)^{\Delta_1 + \Delta_3} (i)^{\Delta_2 + \Delta_4}}{16 \pi^6 \prod_{i=1}^4 \Gamma\left(\Delta_i - \frac{1}{2} \right) R^{4- (\Delta_1+\Delta_2+\Delta_3+\Delta_4)}} \, .
\end{equation}
The other quantities are
\begin{gather}
	\sigma^*_1 = \dfrac{1}{D} \dfrac{z_{24} \Bar{z}_{34}}{z_{12} \Bar{z}_{13}} \, , ~~ \sigma^*_2 = \dfrac{1}{D} \dfrac{z_{34} \Bar{z}_{14}}{z_{23} \Bar{z}_{12}} \, , ~~ \sigma^*_3 = \dfrac{1}{D} \dfrac{z_{24} \Bar{z}_{14}}{z_{23} \Bar{z}_{13}} \, , ~~ \sigma^*_4 = \dfrac{1}{D} \nonumber \\
	D = 2 \dfrac{z_{24} \Bar{z}_{34}}{z_{12} \Bar{z}_{13}} + 2 \dfrac{z_{24}\Bar{z}_{14}}{z_{23}\Bar{z}_{13}}
\end{gather}
The presence of $\delta(|z-\bar{z}|)$ where $z$ is the cross ratio ensures that the four point correlator remains invariant under global spacetime translations. This factor in the results of \eqref{eq:fourptcontactresult} and \eqref{eq:fourpointexresult} agrees with the general intrinsic analysis of the four point Carroll correlators given in section \ref{sec:fourpointcarroll} (in particular \eqref{eq:fourptcarrollfinal}).

\subsection{Gluon two point function}

We now proceed to consider non-trivial spinning Witten diagrams. We first consider the gluon two point function. The gluon two point function is given by the following diagram \footnote{Note that here we do a bulk integral to obtain the two point function. This is unlike the case in \cite{Witten:1998qj}, where one does a boundary integral. The difference between the two computations is just an overall normalization constant.}
\begin{equation}
	\langle O_{\Delta_1;\beta_1}(\p_1) O_{\Delta_2;\beta_2}(\p_2) \rangle = \int d\x \, K^{\Delta_1,1}_{\alpha_1;\beta_1}(\p_1,\x) \, \eta^{\alpha_1\alpha_2} \, K^{\Delta_2,1}_{\alpha_2;\beta_2}(\p_2,\x) \, .
\end{equation}
$O_{\Delta_2;\beta}$ corresponds to the operator with spin 1. Here $\beta = z,\bar{z}$. Most importantly, $\beta \neq u$ because the bulk to boundary propagator becomes pure gauge \eqref{eq:spin1ucomp}. We insert $\p_1$ at $\tau_p = -\frac{\pi}{2}+\frac{u_1}{R}$ and $\p_2$ at $\tau_p = +\frac{\pi}{2}+\frac{u_2}{R}$. In the large $R$ limit, using \eqref{eq:mellinspin1} and \eqref{eq:mellinintspin1} for $\beta_i = z^a$ we have
\begin{equation}
	\begin{split}
		\langle O_{\Delta_1;\beta_1}(\p_1) O_{\Delta_2;\beta_2}(\p_2) \rangle &= \mathcal{D}_{(1)} \partial_{\beta_1} \tilde{q}_{1\alpha} \partial_{\beta_2} \tilde{q}^{\alpha}_2 \int d^4x \, \psi^-_{\Delta_1,\tilde{q}_1,u_1}(x) \,  \psi^+_{\Delta_2,\tilde{q}_2,u_2}(x) \, , \\
		&= \mathcal{D}_{(1)} \partial_{\beta_1} \tilde{q}_{1\alpha} \partial_{\beta_2} \tilde{q}^{\alpha}_2 \int d^4x \int d\omega_1 d\omega_2 \omega^{\Delta_1 - 1}_1 \omega^{\Delta_2-1}_2 \\ & ~~~~~~~\times e^{i\omega_1u_1 -i\omega_2u_2} e^{i(\omega_1\tilde{q}_1-\omega_2 \tilde{q}_2)\cdot x} \, , \\
		&= \mathcal{D}_{(1)} \partial_{\beta_1} \tilde{q}_{1\alpha} \partial_{\beta_2} \tilde{q}^{\alpha}_2 \dfrac{\Gamma(\Delta_1+\Delta_2-2)\,\delta^2(z_{12})}{[i(u_2-u_1)]^{\Delta_1+\Delta_2-2}} \, .
	\end{split}
\end{equation}
Thus,
\begin{equation}\label{eq:gluon2pt}
	\langle O_{\Delta_1;\beta_1}(\p_1) O_{\Delta_2;\beta_2}(\p_2) \rangle = \mathcal{D}_{(1)} \partial_{\beta_1} \tilde{q}_{1\alpha} \partial_{\beta_2} \tilde{q}^{\alpha}_2 \dfrac{\Gamma(\Delta_1+\Delta_2-2)\,\delta^2(z_{12})}{[i(u_2-u_1)]^{\Delta_1+\Delta_2-2}} \, .
\end{equation}
Here
\begin{equation}
	\mathcal{D}_{(1)} = \dfrac{\mathcal{C}^3_{\Delta_1;1}}{(-i)^{\Delta_1}\Gamma(\Delta_1)}\dfrac{\Delta_1-1}{\Delta_1} \dfrac{\mathcal{C}^3_{\Delta_2;1}}{(i)^{\Delta_2}\Gamma(\Delta_2)}\dfrac{\Delta_2-1}{\Delta_2} = \dfrac{(i)^{\Delta_1} (-i)^{\Delta_2}}{4\pi^3 \Gamma\left(\Delta_1 -\frac{1}{2}\right) \Gamma\left(\Delta_2 -\frac{1}{2}\right)R^{4-\Delta_1-\Delta_2}} \, .
\end{equation}
This result is identical to the scalar two point function of \cite{Bagchi:2022emh,Bagchi:2023fbj} except for normalization and a crucial factor of $\partial_{\beta_1} \tilde{q}_{1\alpha} \partial_{\beta_2} \tilde{q}^{\alpha}_2$. This crucial factor for various cases of $\nu_i$ is given by
\begin{equation}
	\begin{split}
		\partial_{\bar{z}_1} \tilde{q}_{1\alpha} \partial_{\bar{z}_2} \tilde{q}^{\alpha}_2 &=  \dfrac{-2\,z^2_{12}}{(1+z_1\bar{z}_1)^2(1+z_2\bar{z}_2)^2} \, , \\
		\partial_{z_1} \tilde{q}_{1\alpha} \partial_{z_2} \tilde{q}^{\alpha}_2 &=  \dfrac{-2\,\bar{z}^2_{12}}{(1+z_1\bar{z}_1)^2(1+z_2\bar{z}_2)^2} \, , \\
		\partial_{z_1} \tilde{q}_{1\alpha} \partial_{\bar{z}_2} \tilde{q}^{\alpha}_2 &= \dfrac{2(1+z_2 \bar{z}_1)^2}{(1+z_1\bar{z}_1)^2(1+z_2\bar{z}_2)^2} \, .
	\end{split}
\end{equation}
Now, because of the delta function $\delta^2(z_{12})$, the two point function for spin 1 particles will be non zero only if $\nu_1=z_1$ and $\nu_2=\bar{z}_2$ and the result is given by
\begin{equation}\label{eq:gluon2ptfinal}
	\langle O_{\Delta_1;z_1}(\p_1) O_{\Delta_2;\z_2}(\p_2) \rangle = \mathcal{D}_{(1)} \dfrac{2}{(1+z_1\bar{z}_1)(1+z_2\bar{z}_2)} \dfrac{\Gamma(\Delta_1+\Delta_2-2)\,\delta^2(z_{12})}{[i(u_2-u_1)]^{\Delta_1+\Delta_2-2}} \, .
\end{equation}
We are learning that the $z$ component of the spinning primary behaves like a helicity $1$ particle, and the $\bar{z}$ component of the spinning primary behaves like a helicity $-1$ particle. This implies $\sigma_1+\sigma_2=0$ for a non-zero two point function. Also, another non-trivial result is that the correlators involving the $u$-component of the field become zero because the bulk to boundary propagator becomes pure gauge \eqref{eq:spin1ucomp}. \eqref{eq:gluon2ptfinal} agrees with the intrinsic result of \eqref{eq:carrollspin12pt} if we construct representations of vector primaries with upper indices. Our result also agrees with the recently obtained result from the embedding space approach \cite{Salzer:2023jqv} and the bulk result of \cite{Nguyen:2023miw}. 

\subsection{Scalar-gluon-scalar three point function: split signature}

We now consider different cases of the three point function. The simplest possible local cubic interaction vertex in AdS involving a spin $J$ field and two scalars $\phi_1$ and $\phi_2$ is given by \cite{Costa:2014kfa}
\begin{equation}\label{eq:scalspinscalvertex}
	g_{\phi_1 \phi_2 h} \int_{\text{AdS}_{d+1}} d\x \left( \phi_1 \nabla_{\mu_1} \dots \nabla_{\mu_J} \phi_2 \right) h^{\mu_1 \dots \mu_J} \, .
\end{equation}
The derivatives can act on either of the scalar fields because we consider the spin $J$ field to have vanishing divergence.

\medskip

We now consider spin 1, and the contact Witten diagram is given by
\begin{equation}
	\begin{split}
		\langle O_{\Delta_1}(\p_1) O_{\Delta_2;\beta}(\p_2) &O_{\Delta_3}(\p_3) \rangle \\
		&= g_{\phi_1 \phi_2 h}\int_{\text{AdS}_{d+1}} d\x \, K_{\Delta_1}(\p_1,\x) \, K^{\Delta_2,1}_{\alpha,\beta}(\p_2,\x) \, K_{\Delta_3}(\p_3,\x) (k^{\alpha}_1 + k^{\alpha}_3) \, .
	\end{split}
\end{equation}
Here $k^{\alpha}_1$ and $k^{\alpha}_3$ correspond to the ingoing and outgoing momenta of the massless scalars. Inserting $\p_1$ at $\tau_p = -\frac{\pi}{2}+\frac{u_1}{R}$, $\p_2$ at $\tau_p = +\frac{\pi}{2}+\frac{u_2}{R}$, $\p_3$ at $\tau_p = +\frac{\pi}{2}+\frac{u_3}{R}$ and then taking the large $R$ limit, we have using \eqref{eq:carrollscalwave}, \eqref{eq:mellinspin1} and \eqref{eq:mellinintspin1}
\begin{equation}\label{eq:scalphotscalfirst}
	\begin{split}
		\langle O_{\Delta_1}(\p_1) O_{\Delta_2;\beta}(\p_2) O_{\Delta_3}(&\p_3) \rangle= \mathcal{B}_{(1)sp} \int_{\mathbb{R}^{1,3}} d^4x \, \psi^-_{\Delta_1,\tilde{q}_1,u_1}(x) \, \psi^-_{\Delta_2,\tilde{q}_2,u_2}(x) \, \psi^+_{\Delta_3,\tilde{q}_3,u_3}(x)  \\
		&\hspace{3cm} \times \partial_{\beta} \tilde{q}_{2\alpha}(k^{\alpha}_1 + k^{\alpha}_3) \\
		&= \mathcal{B}_{(1)sp} \int d\omega_1 \, d\omega_2 \, d\omega_3 \, \omega^{\Delta_1 -1}_1\omega^{\Delta_2-1}_2\omega^{\Delta_3-1}_3 e^{i(\omega_1 u_1 + \omega_2 u_2 -\omega_3 u_3)} \\
		&~~~(\omega_1 \tilde{q}^{\alpha}_1 + \omega_3 \tilde{q}^{\alpha}_3)(\partial_{\beta}\tilde{q}_{2\alpha}) \delta^{(4)}(\omega_1 \tilde{q}_1 + \omega_2 \tilde{q}_2 - \omega_3 \tilde{q}_3) \, .
	\end{split}
\end{equation}
Here, the normalization constant is given by
\begin{equation}
	\mathcal{B}_{(1)sp} = N^3_{\Delta_1} \dfrac{\mathcal{C}^d_{\Delta_2;1}}{(i)^{\Delta_2}\Gamma(\Delta_2)} \dfrac{\Delta_2 -1}{\Delta_2} N^3_{\Delta_3} = \dfrac{i^{\Delta_1} \,(-i)^{\Delta_2+\Delta_3}}{8 \pi^{\frac{9}{2}} \, \prod^{3}_{i=1} \Gamma\left(\Delta_i - \frac{1}{2} \right) R^{4-(\Delta_1 + \Delta_2 + \Delta_3)}} \, .
\end{equation}
We have used \eqref{eq:normscalar} and \eqref{eq:normspin}.

\medskip

We now use the split of \eqref{eq:celestialtorusdelta} for the momentum conserving delta function. If we use that, we have \footnote{ \label{paramfootnote} For these vertex factors, we take the momenta $\tilde{q}_i$ of the polarization vectors in the bulk to boundary propagator of \eqref{eq:mellinspin1} in the standard parametrization given by \eqref{eq:4dparamq}. For the $\tilde{q}_i$ that appears in the momentum conserving delta function in the combination $\omega_i \tilde{q}_i$, we choose $$\tilde{q}_i = (1+z_i \bar{z}_i, z_i + \bar{z}_i, -i(z_i - \bar{z}_i), 1- z_i \bar{z}_i ) \, .$$ We can always do this by suitably redefining $\omega_i \to \omega_i(1+z_i \bar{z}_i)$. This ensures that the vertex factors are simple. This will be extremely helpful when we look at graviton three point amplitudes.} 
\begin{equation}\label{eq:v3gloun-}
	V_{3 \, \text{gluon}(-)} =(\omega_1 \tilde{q}^{\alpha}_1 + \omega_3 \tilde{q}^{\alpha}_3)(\partial_{\bar{z_2}} \, \tilde{q}_{2\alpha}) = \dfrac{4 \, z_{32} \, \omega_3}{(1+z_2 \bar{z}_2)} \, .
\end{equation}
We finally get
\begin{equation}\label{eq:scalphotscalsplitresult}
	\begin{split}
		\langle O_{\Delta_1}(\p_1) O_{\Delta_2;\bar{z}_2}(\p_2) O_{\Delta_3}(\p_3) \rangle = 32 \, \mathcal{B}_{(1)sp} \, C_z \, (-1)^{\Delta_2} z^{\Delta_3-1}_{12} z^{\Delta_1-1}_{32} z^{\Delta_2-2}_{31} \\ \times \dfrac{\Gamma(\Delta_1+\Delta_2+\Delta_3-3)\,\delta(\bar{z}_{12})\,\delta(\bar{z}_{13})}{[-i(z_1 u_{23} + z_2 u_{31} + z_3 u_{12})]^{\Delta_1+\Delta_2+\Delta_3-3}} \, .
	\end{split}
\end{equation}
$C_z$ is a normalization constant that is given by
\begin{equation}
	C_z = \dfrac{1}{1+z_2 \bar{z}_2} \, .
\end{equation}
\eqref{eq:scalphotscalsplitresult} matches with the intrinsic result \eqref{eq:threepointsplit} once we use the fact that the helicities are given by $\sigma_2=-1,\sigma_1=0,\sigma_3=0$. \eqref{eq:scalphotscalsplitresult} also agrees with \cite{Salzer:2023jqv}. We can also consider the equivalent $z$ component of \eqref{eq:scalphotscalsplitresult}. This would just be given by the complex conjugate of \eqref{eq:scalphotscalsplitresult}. The detailed calculation resulting in \eqref{eq:scalphotscalsplitresult} is given in Appendix \ref{ap:scalphotonscalsplit}.

\subsection{Gluon MHV three point amplitude: split signature}
\label{sec:gluonmhv3pt}

The gluon three point vertex with three incoming momenta $k,p,q$ is given by \cite{schwartz_2013}
\begin{equation}
	g \, f^{abc} [ g^{\mu\nu}(k-p)^{\rho} + g^{\nu\rho}(p-q)^{\mu} + g^{\rho\mu}(q-k)^{\nu} ] \, .
\end{equation}
Here $g$ is the coupling constant, and $a,b,c$ denote the colour indices. Thus $f^{abc}$ denotes the structure constants of the $SU(N)$ gauge group. We have two incoming momenta $q_1,q_2$ and one outgoing momenta $q_3$. Thus, the relevant vertex function for us is given by
\begin{equation}
	V^{\alpha_1\alpha_2\alpha_3}_{3g} = g f^{abc} [ g^{\alpha_1\alpha_2}(q_1-q_2)^{\alpha_3} + g^{\alpha_2\alpha_3}(q_2+q_3)^{\alpha_1} + g^{\alpha_3\alpha_1}(-q_3-q_1)^{\alpha_2} ] \, .
\end{equation}
As explained in \cite{Witten:1998qj}, we can take the vertex factor from the usual QFT analysis to evaluate contact Witten diagrams. For us, in the flat limit, the metric factors $g^{\mu\nu} \to \eta^{\mu\nu}$.

\medskip

Thus, the contact three point Witten diagram is given by
\begin{equation}\label{eq:gluon3ptwitten}
	\begin{split}
		\langle O_{\Delta_1;\beta_1}(\p_1) O_{\Delta_2;\beta_2}(\p_2) &O_{\Delta_3;\beta_3}(\p_3) \rangle \\&= \int d\x \, K^{\Delta_1,1}_{\alpha_1;\beta_1}(\p_1,\x) \, K^{\Delta_2,1}_{\alpha_2;\beta_2}(\p_2,\x) \, K^{\Delta_3,1}_{\alpha_3;\beta_3}(\p_3,\x) V^{\alpha_1\alpha_2\alpha_3}_{3g} \, .
	\end{split}
\end{equation}
We consider the $(--+)$ MHV amplitude $\beta_1=\bar{z}_1$, $\beta_2=\bar{z}_2$, $\beta_3=z_3$. Thus, in the large $R$ limit, using \eqref{eq:mellinspin1} and \eqref{eq:mellinintspin1} we have
\begin{equation}\label{eq:gluonmhv3pt}
	\begin{split}
		\langle O_{\Delta_1;\bar{z}_1}(\p_1) O_{\Delta_2;\bar{z}_2}(\p_2) &O_{\Delta_3;z_3}(\p_3) \rangle \\ &= \mathcal{C}_{(3)g} \int d\omega_1 d\omega_2 d\omega_3 \omega^{\Delta_1-1}_1 \omega^{\Delta_2-1}_2 \omega^{\Delta_3-1}_3 e^{i(\omega_1 u_1 + \omega_2 u_2 -\omega_3 u_3)} \\
		& ~~~~\times \partial_{\bar{z}_1} \tilde{q}_{1\alpha_1} \partial_{\bar{z}_2} \tilde{q}_{2\alpha_2} \partial_{z_3} \tilde{q}_{3\alpha_3} V^{\alpha_1 \alpha_2 \alpha_3}_{3g} \delta^{(4)}(\omega_1 \tilde{q}_1 + \omega_2 \tilde{q}_2 - \omega_3 \tilde{q}_3) \, .
	\end{split}
\end{equation}
Using the momentum conserving delta function \eqref{eq:celestialtorusdelta}, we evaluate the vertex factor according to the parametrization explained in footnote \ref{paramfootnote}:
\begin{equation}\label{eq:vertexcontract3pt}
	\begin{split}
		\partial_{\bar{z}_1} \tilde{q}_{1\alpha_1} \partial_{\bar{z}_2} \tilde{q}_{2\alpha_2} \partial_{z_3} \tilde{q}_{3\alpha_3} V^{\alpha_1 \alpha_2 \alpha_3}_{3g} &= g f^{abc} \left[ \partial_{\bar{z}_1} \tilde{q}_{1\mu} \partial_{\bar{z}_2}  \tilde{q}^{\mu}_2 \partial_{z_3} \tilde{q}_{3\nu}(\omega_1 \tilde{q}_1 +  \omega_2 \tilde{q}_2)^{\nu} + \partial_{\bar{z}_2}\tilde{q}_{2\mu} \partial_{z_3} \tilde{q}^{\mu}_3 \partial_{\bar{z}_1} \tilde{q}_{1\nu}  \right. \\ &~~~\left. \times (\omega_2 \tilde{q}_2 + \omega_3 \tilde{q}_3)^{\nu}  + \partial_{\bar{z}_1}\tilde{q}_{1\mu} \partial_{z_3} \tilde{q}^{\mu}_3 \partial_{\bar{z}_2} \tilde{q}_{2\nu}(-\omega_3 \tilde{q}_3 - \omega_1 \tilde{q}_1)^{\nu} \right] \\
		&= \dfrac{-8 g\, f^{abc} \, z_{12} \, \omega_3 }{(1+z_1\bar{z}_1) (1+z_2\bar{z}_2)(1+z_3\bar{z}_3) } \\
		&= g f^{abc} D_z z_{12} \, \omega_3 = V_{3 \,\text{gloun}\,(--+)} \, .
	\end{split}
\end{equation}
Here $D_z$ is given by
\begin{equation}\label{eq:Dz}
	D_z = \dfrac{1}{(1+z_1\bar{z}_1) (1+z_2\bar{z}_2)(1+z_3\bar{z}_3)} \, .
\end{equation}
At this point it is useful to recall the MHV $(--+)$ gluon amplitude in terms of spinor helicity variables \cite{schwartz_2013}:
\begin{equation}\label{eq:shgluon--+}
	\mathcal{A}^{\text{gloun}}_{--+} = \dfrac{\langle12 \rangle^3}{ \langle 23 \rangle \langle 31 \rangle } \delta^{(4)}(p^{\mu}_1 + p^{\mu}_2 + p^{\mu}_3 ) \, .
\end{equation}
Following the conventions for spinor-helicity variables of \cite{Pasterski:2017ylz} and using the momentum conserving delta function \eqref{eq:celestialtorusdelta}, we have
\begin{equation}
	\dfrac{\langle12 \rangle^3}{ \langle 23 \rangle \langle 31 \rangle } \propto \omega_3 z_{12} \propto V_{3 \,\text{gloun}\,(--+)} \, .
\end{equation}
Thus, in the large $R$ limit, the Witten diagram \eqref{eq:gluonmhv3pt} essentially becomes the modified Mellin transform of the flat space scattering amplitude given by \eqref{eq:vertexcontract3pt}. 

\medskip

Using \eqref{eq:celestialtorusdelta}, we can simplify \eqref{eq:gluonmhv3pt} using similar manipulations as Appendix \ref{ap:threepointsplit} and \ref{ap:scalphotonscalsplit}. The final answer is given by
\begin{equation}
	\begin{split}
		\langle O_{\Delta_1;\bar{z}_1}(\p_1) O_{\Delta_2;\bar{z}_2}(\p_2) O_{\Delta_3;z_3}(\p_3) \rangle &= 32g f^{abc}\mathcal{C}_{(3)g} \, D_z \, (-1)^{\Delta_1 +\Delta_2} z^{\Delta_3}_{12} z^{\Delta_1-2}_{23} z^{\Delta_2-2}_{31} \\
		& ~~~ \times \dfrac{\Gamma(\Delta_1+\Delta_2+\Delta_3-3) \, \delta(\bar{z}_{13}) \delta(\bar{z}_{23})}{[-i(z_1 u_{23} + z_2 u_{31} + z_{3} u_{12})]^{\Delta_1 + \Delta_2 + \Delta_3 - 3}} \, .
	\end{split}
\end{equation}
Here $\mathcal{C}_{(3)g}$ is given by
\begin{equation}\label{eq:c3gloun3pt}
	\begin{split}
		\mathcal{C}_{(3)g} &= \dfrac{\mathcal{C}^3_{\Delta_1;1}}{(-i)^{\Delta_1}\Gamma(\Delta_1)}\dfrac{\Delta_1-1}{\Delta_1} \dfrac{\mathcal{C}^3_{\Delta_2;1}}{(-i)^{\Delta_2}\Gamma(\Delta_2)}\dfrac{\Delta_2-1}{\Delta_2} \dfrac{\mathcal{C}^3_{\Delta_3;1}}{(i)^{\Delta_3}\Gamma(\Delta_3)}\dfrac{\Delta_3-1}{\Delta_3} \\
		&= \dfrac{i^{\Delta_1+\Delta_2} (-i)^{\Delta_3}}{8\pi^{\frac{9}{2}} \prod_{i=1}^{3}\Gamma\left(\Delta_i-\frac{1}{2}\right) R^{6-\Delta_1-\Delta_2-\Delta_3}} \, .
	\end{split}
\end{equation}
This answer matches with the intrinsic result \eqref{eq:threepointsplit} as we have $\sigma_1 = -1$, $\sigma_2 = -1$ and $\sigma_3=+1$.

\medskip

We can also consider the other $(++-)$ MHV amplitude. In \eqref{eq:gluon3ptwitten}, we basically consider $\beta_1 = z_1, \beta_2 = z_2, \beta_3 = \bar{z}_3$. Thus, in the large $R$ limit, we have
\begin{equation}
	\begin{split}
		\langle O_{\Delta_1;z_1}(\p_1) O_{\Delta_2;z_2}(\p_2) &O_{\Delta_3;\bar{z}_3}(\p_3) \rangle \\ &= \mathcal{C}_{(3)} \int d\omega_1 d\omega_2 d\omega_3 \omega^{\Delta_1-1}_1 \omega^{\Delta_2-1}_2 \omega^{\Delta_3-1}_3 e^{i(\omega_1 u_1 + \omega_2 u_2 -\omega_3 u_3)} \\
		& ~~~~\times \partial_{z_1} \tilde{q}_{1\alpha_1} \partial_{z_2} \tilde{q}_{2\alpha_2} \partial_{\bar{z}_3} \tilde{q}_{3\alpha_3} V^{\alpha_1 \alpha_2 \alpha_3}_{3g} \delta^{(4)}(\omega_1 \tilde{q}_1 + \omega_2 \tilde{q}_2 - \omega_3 \tilde{q}_3)
	\end{split}
\end{equation}
The vertex factor analogous to \eqref{eq:vertexcontract3pt} essentially becomes the complex conjugate
\begin{equation}\label{eq:gluon++-3pt}
	\begin{split}
		\partial_{\bar{z}_1} \tilde{q}_{1\alpha_1} \partial_{\bar{z}_2} \tilde{q}_{2\alpha_2} \partial_{z_3} \tilde{q}_{3\alpha_3} V^{\alpha_1 \alpha_2 \alpha_3}_{3g} &= \dfrac{-8 g\, f^{abc} \, \bar{z}_{12} \, \omega_3 }{(1+z_1\bar{z}_1) (1+z_2\bar{z}_2)(1+z_3\bar{z}_3) } \\
		&= g f^{abc} D_z \bar{z}_{12} \, \omega_3 = V_{3 \,\text{gloun}\,(++-)} \, .
	\end{split}
\end{equation}
Now, because of this complex conjugate, we should crucially consider the complex conjugate of the momentum conserving delta function split \eqref{eq:celestialtorusdelta} to ensure that we get a non-trivial result:
\begin{equation}\label{eq:celestialtorusdelta2}
	\delta^{(4)}(\omega_1 \tilde{q}_1 + \omega_2 \tilde{q}_2 - \omega_3 \tilde{q}_3) = \dfrac{4}{\omega^2_3 \bar{z}_{23} \bar{z}_{31}}\delta\left(\omega_1 - \omega_3 \dfrac{\bar{z}_{32}}{\bar{z}_{12}} \right) \delta\left(\omega_2 - \omega_3 \dfrac{\bar{z}_{31}}{\bar{z}_{21}} \right)\delta(z_{13})\delta(z_{23}) \, .
\end{equation}
We should choose the orderings $\bar{z}_1 < \bar{z}_3 < \bar{z}_2$ or $\bar{z}_2 < \bar{z}_3 < \bar{z}_1$ to ensure that the energies are positive. Thus, the vertex factor of the bulk Witten diagram determines what branch of the delta function we should take to ensure a non-trivial answer. The final result can be obtained as
\begin{equation}
	\begin{split}
		\langle O_{\Delta_1;z_1}(\p_1) O_{\Delta_2;z_2}(\p_2) O_{\Delta_3;\bar{z}_3}(\p_3) \rangle &= 8g f^{abc}\mathcal{C}_{(3)} \, D_z \, (-1)^{\Delta_1 +\Delta_2} \bar{z}^{\Delta_3}_{12} \bar{z}^{\Delta_1-2}_{23} \bar{z}^{\Delta_2-2}_{31}  \\
		& ~~~ \times \dfrac{\Gamma(\Delta_1+\Delta_2+\Delta_3-3) \, \delta(z_{13}) \, \delta(z_{23})}{[-i(\bar{z}_1 u_{23} + \bar{z}_2 u_{31} + \bar{z}_{3} u_{12})]^{\Delta_1 + \Delta_2 + \Delta_3 - 3}} \, .
	\end{split}
\end{equation}
This answer matches the alternate branch intrinsic result \eqref{eq:threepointsplitalt} if we use $\sigma_1=1,\sigma_2=1,\sigma_3=-1$. 

\medskip

We do not have MHV amplitudes for $(+++)$ and $(---)$ cases. That is reflected from the vertex factors for both cases vanishing when we use the momentum conserving delta function of \eqref{eq:celestialtorusdelta2} and \eqref{eq:celestialtorusdelta} respectively i.e.,
\begin{equation}
	\begin{split}
		\partial_{z_1} \tilde{q}_{1\alpha_1} \partial_{z_2} \tilde{q}_{2\alpha_2} \partial_{z_3} \tilde{q}_{3\alpha_3} V^{\alpha_1 \alpha_2 \alpha_3}_{3g} &= 0 \, , \\
		\partial_{\bar{z}_1} \tilde{q}_{1\alpha_1} \partial_{\bar{z}_2} \tilde{q}_{2\alpha_2} \partial_{\bar{z}_3} \tilde{q}_{3\alpha_3} V^{\alpha_1 \alpha_2 \alpha_3}_{3g} &= 0 \, .
	\end{split}
\end{equation}

\subsection{Gluon three point amplitude from $F^3$: split signature}

We see that from the Carrollian field theory analysis; we do have counterparts for correlators that have the helicity combinations $(+++)$ or $(---)$ in \eqref{eq:threepointsplit}. The little group scaling would imply that the tree level S-Matrix vanishes for these helicity configurations if we only have renormalizable interactions \cite{elvang_huang_2015}. From the bulk perspective, these arise from higher derivative terms or loop corrections. For instance, in our case, they arise from \cite{Dixon:1993xd,Dixon:2004za}
\begin{equation}
	f^{abc} F^{a\mu}_{~~\nu} F^{b\nu}_{~~\rho} F^{c\rho}_{~~\mu} \, ,
\end{equation}
where $F^a_{\mu\nu}$ is the $SU(N)$ Yang-Mills field strength tensor. The vertex factor for three ingoing momenta $p_1,p_2,p_3$ can be worked out as
\begin{equation}
	V_{3\,g_2}^{\alpha_1 \alpha_2 \alpha_3} = f^{abc} \left( p^{\alpha_1}_1 p^{\alpha_2}_2 p^{\alpha_3}_3 - p^{\alpha_3}_1 p^{\alpha_1}_2 p^{\alpha_2}_3 \right) \, .
\end{equation}
Using the split signature momentum conserving delta function \eqref{eq:celestialtorusdelta} for the $(---)$ case, we have
\begin{equation}\label{eq:gluon---3pt}
	\begin{split}
		\partial_{\bar{z}_1} \tilde{q}_{1\alpha_1} \partial_{\bar{z}_2} \tilde{q}_{2\alpha_2} \partial_{\bar{z}_3} \tilde{q}_{3\alpha_3} V^{\alpha_1 \alpha_2 \alpha_3}_{3g} &= \dfrac{8 z^2_{13} z^2_{23} \omega^3_3}{z_{12}} \dfrac{1}{(1+z_1\bar{z}_1)(1+z_2\bar{z}_2)(1+z_3\bar{z}_3)} \\
		&=\dfrac{z^2_{13} z^2_{23} \omega^3_3}{z_{12}} D_{z} \, .
	\end{split}
\end{equation}
Here $D_z$ is given by \eqref{eq:Dz}. 

\medskip

It is now straightforward to work out the three point function/ amplitude by doing similar manipulations as Appendix \ref{ap:threepointsplit} and \ref{ap:scalphotonscalsplit}. The final result is given by
\begin{equation}
	\begin{split}
		\langle O_{\Delta_1;\bar{z}_1}(\p_1) O_{\Delta_2;\bar{z}_2}(\p_2) O_{\Delta_3;\bar{z}_3}(\p_3) \rangle &= 8g f^{abc}\mathcal{C}_{(3)} \, D_{z} \, (-1)^{\Delta_2} z^{\Delta_3}_{12} z^{\Delta_1}_{32} z^{\Delta_2}_{31} \delta(\bar{z}_{13}) \delta(\bar{z}_{23}) \\
		& ~~~ \times \dfrac{\Gamma(\Delta_1+\Delta_2+\Delta_3-1)}{[-i(z_1 u_{23} + z_2 u_{31} + z_{3} u_{12})]^{\Delta_1 + \Delta_2 + \Delta_3 - 1}} \, .
	\end{split}
\end{equation}
$\mathcal{C}_{(3)}$ is given by \eqref{eq:c3gloun3pt}. This matches the result of \eqref{eq:threepointsplit} for $\sigma_1 = -1,\sigma_2 = -1, \sigma_3 = -1$. This was also recently worked out in \cite{Salzer:2023jqv}. The $(+++)$ case is just the complex conjugate of the above answer with $\delta(z_{12}) \delta(z_{23})$. It is instructive to note that this amplitude has not been focussed in the celestial amplitudes literature \cite{Pasterski:2021rjz}. As we remarked before, these diagrams are quadratically divergent in the Mellin basis. Hence, \cite{PhysRevD.96.085006,Stieberger:2018edy} have mostly considered MHV amplitudes. In the Carrollian approach, these different helicity configuration amplitudes are naturally encoded in the holographic correlation functions, which are crucially finite.

\subsection{Graviton two point function}
\label{sec:gravprop}

We will see that features of the gluon two point function \eqref{eq:gluon2pt} are also present for the graviton two point function. The graviton two point function is given by
\begin{equation}
	\langle O_{\Delta_1;\beta_1 \beta_2}(\p_1) O_{\Delta_2;\beta_3\beta_4}(\p_2) \rangle = \int_{\text{AdS}_4} d\x \, K^{\Delta_1,2}_{\alpha_1\alpha_2;\beta_1\beta_2}(\p_1,\x) V^{\alpha_1\alpha_2\alpha_3\alpha_4}_{2(grav)}  K^{\Delta_1,2}_{\alpha_3\alpha_4;\beta_3\beta_4}(\p_2,\x) \, .
\end{equation}
The vertex factor arises from the graviton propagator \cite{elvang_huang_2015}:
\begin{equation}
	V^{\alpha_1\alpha_2\alpha_3\alpha_4}_{2(grav)} = g^{\alpha_1\alpha_3}g^{\alpha_2\alpha_4}+g^{\alpha_1\alpha_4}g^{\alpha_2\alpha_3} - g^{\alpha_1\alpha_2}g^{\alpha_3\alpha_4} \, .
\end{equation}

\medskip

Due to \eqref{eq:uupurediffeo}, the indices $\beta_i \neq u$ as they will correspond to pure gauge modes. Thus, we consider $\beta_i = z^a$, and the bulk to boundary propagators in the flat limit have been worked in \eqref{eq:gravitonmellin}. The modified Mellin gauge representative is given by
\begin{equation}\label{eq:gravitonmellinint}
	K^{\Delta,2}_{\alpha_1\alpha_2;\beta_1\beta_2}(\p,\x)=\dfrac{\mathcal{C}^d_{\Delta;2}}{(\pm i)^{\Delta} \Gamma(\Delta)} \dfrac{\Delta-1}{\Delta} (\pm P^{\rho_1 \rho_2}_{\beta_1 \beta_2} \, \partial_{\rho_1} \tilde{q}_{\alpha_1} \, \partial_{\rho_2} \tilde{q}_{\alpha_2})\int^{\infty}_0 d\omega \, \omega^{\Delta-1} \, e^{\mp i \omega \, u} e^{\mp i\omega \, \tilde{q} \cdot x} \, .
\end{equation}
Here $P^{\rho_1 \rho_2}_{\beta_1 \beta_2}$ is the projector that projects to the transverse and traceless part of the propagator
\begin{equation}
	P^{\rho_1 \rho_2}_{\beta_1 \beta_2} = \delta^{\rho_1}_{(\beta_1}\delta^{\rho_2}_{\beta_2)} - \dfrac{1}{2}g_{\beta_1 \beta_2} g^{\rho_1 \rho_2} \, .
\end{equation}
The graviton two point function thus becomes
\begin{equation}
	\begin{split}
		\langle O_{\Delta_1;\beta_1 \beta_2}(\p_1) O_{\Delta_2;\beta_3\beta_4}&(\p_2) \rangle \\ &= \mathcal{A}_{(2)grav}(-P^{\rho_1 \rho_2}_{\beta_1 \beta_2} \, \partial_{\rho_1} \tilde{q}_{\alpha_1} \, \partial_{\rho_2} \tilde{q}_{\alpha_2}) (P^{\rho_3 \rho_4}_{\beta_3 \beta_4} \, \partial_{\rho_3} \tilde{q}_{\alpha_3} \, \partial_{\rho_4} \tilde{q}_{\alpha_4}) V^{\alpha_1\alpha_2\alpha_3\alpha_4}_{2(grav)} \\
		&~~~\times \int d^4x \, \psi^{-}_{\Delta_1,\tilde{q}_1,u_1}(x) \psi^{+}_{\Delta_2,\tilde{q}_2,u_2}(x) \\
		& =\mathcal{A}_{(2)grav}(-P^{\rho_1 \rho_2}_{\beta_1 \beta_2} \, \partial_{\rho_1} \tilde{q}_{\alpha_1} \, \partial_{\rho_2} \tilde{q}_{\alpha_2}) (P^{\rho_3 \rho_4}_{\beta_3 \beta_4} \, \partial_{\rho_3} \tilde{q}_{\alpha_3} \, \partial_{\rho_4} \tilde{q}_{\alpha_4}) V^{\alpha_1\alpha_2\alpha_3\alpha_4}_{2(grav)} \\
		& ~~~ \times \dfrac{\Gamma(\Delta_1+\Delta_2-2) \, \delta^2(z_{12})}{[-i(u_1-u_2)]^{\Delta_1+\Delta_2-2}} \, .
	\end{split}
\end{equation}
Here, the normalization constant is given by
\begin{equation}
	\begin{split}
		\mathcal{A}_{(2)grav} &= \dfrac{\mathcal{C}^3_{\Delta_1;2}}{(-i)^{\Delta_1}\Gamma(\Delta_1)}\dfrac{\Delta_1-1}{\Delta_1} \dfrac{\mathcal{C}^3_{\Delta_2;2}}{(i)^{\Delta_2}\Gamma(\Delta_2)}\dfrac{\Delta_2-1}{\Delta_2} \\ &= \dfrac{(i)^{\Delta_1} (-i)^{\Delta_2}(\Delta_1+1)(\Delta_2+1)}{4\pi^3 \Delta_1 \Delta_2 \Gamma\left(\Delta_1 -\frac{1}{2}\right) \Gamma\left(\Delta_2 -\frac{1}{2}\right)R^{6-\Delta_1-\Delta_2}} \, .
	\end{split}
\end{equation}

\medskip

Thus, the behaviour of the two point function crucially depends on
\begin{equation}
	G_{(2)} = (-P^{\rho_1 \rho_2}_{\beta_1 \beta_2} \, \partial_{\rho_1} \tilde{q}_{\alpha_1} \, \partial_{\rho_2} \tilde{q}_{\alpha_2}) (P^{\rho_3 \rho_4}_{\beta_3 \beta_4} \, \partial_{\rho_3} \tilde{q}_{\alpha_3} \, \partial_{\beta_4} \tilde{q}_{\alpha_4}) V^{\alpha_1\alpha_2\alpha_3\alpha_4}_{2(grav)} \, .
\end{equation}
We will evaluate $G_{(2)}$ for various cases. Let us consider the case when $\beta_1=\beta_2=z_1$ and $\beta_3=\beta_4=\bar{z}_2$. We have
\begin{equation}
	G_{(2)} = \dfrac{8(1+z_2 \bar{z}_1)^4}{(1+z_1\bar{z}_1)^4(1+z_2\bar{z}_2)^4} \, .
\end{equation}
Thus, the two point function is non-zero. 

Let us consider the case when $\beta_1=\beta_2=z_1$ and $\beta_3=\beta_4=z_2$. We have
\begin{equation}
	G_{(2)} = \dfrac{8(\bar{z}_{12})^4}{(1+z_1\bar{z}_1)^4(1+z_2\bar{z}_2)^4} \, .
\end{equation}
The two point function vanishes for this case due to the presence of $\delta^2(z_{12})$. A similar analysis also holds for the antiholomorphic components, which will have a factor of $(z_{12})^4$, leading to a vanishing two point function.

Finally, let us consider the case when $\beta_1=z_1$ and $\beta_2=\bar{z}_1$. For this, however, we trivially have
\begin{equation}
	\begin{split}
		P^{\rho_1\rho_2}_{z_1\bar{z}_1} \partial_{\rho_1}\tilde{q}_{1\alpha_1} \partial_{\rho_2}\tilde{q}_{1\alpha_2} &= \dfrac{1}{2}(\partial_{z_1} \tilde{q}_{\alpha_1} \partial_{\bar{z}_1} \tilde{q}_{1\alpha_2} + \partial_{\bar{z}_1} \tilde{q}_{1\alpha_1} \partial_{z_1} \tilde{q}_{1\alpha_2} ) \\
		&~~~ -\dfrac{1}{2}(\partial_{z_1} \tilde{q}_{\alpha_1} \partial_{\bar{z}_1} \tilde{q}_{1\alpha_2} + \partial_{\bar{z}_1} \tilde{q}_{1\alpha_1} \partial_{z_1} \tilde{q}_{1\alpha_2} ) =0 \, .
	\end{split}
\end{equation}
This is because we project onto the transverse and traceless components only. Thus, the two point function vanishes in this case. Thus, the only non-trivial two point function is
\begin{equation}\label{eq:grav2ptfinal}
	\langle O_{\Delta_1;z_1 z_2}(\p_1) O_{\Delta_2;\z_3\z_4}(\p_2) \rangle = \dfrac{8}{(1+z_1\bar{z}_1)^2(1+z_2\bar{z}_2)^2} \dfrac{\Gamma(\Delta_1+\Delta_2-2) \, \delta^2(z_{12})}{[-i(u_1-u_2)]^{\Delta_1+\Delta_2-2}} \, .
\end{equation}
Our result is consistent with the standard one \cite{Bagchi:2022emh} since we respect $\delta_{\sigma_1+\sigma_2,0}$ i.e. it agrees with \eqref{eq:carrollspin22pt}. The graviton two point function is similar to the gluon two point function \eqref{eq:gluon2pt}. This is related to the fact that both photon and graviton have only two polarizations corresponding to positive and negative helicity.

\subsection{Scalar-graviton-scalar three point amplitude: split signature}
\label{sec:scalgravscal}

We consider the following AdS Witten contact diagram arising from \eqref{eq:scalspinscalvertex}
\begin{equation}
	\begin{split}
		\langle O_{\Delta_1}(\p_1) O_{\Delta_2;\bar{z}_2 \bar{z}_2}(\p_2) O_{\Delta_3}(\p_3) \rangle = \int d\x K_{\Delta_1}(\p_1,\x) K^{\Delta_2,2;\alpha_1\alpha_2}_{\hspace{1.4cm};\bar{z}_2 \bar{z}_2}(\p_2,\x) \\ \times K_{\Delta_3}(\p_3,\x) V_{3 \, \alpha_1 \alpha_2} \, .
	\end{split}  
\end{equation}
The vertex factor is given by \eqref{eq:scalspinscalvertex}
\begin{equation}
	V_{3\,\alpha_1\alpha_2} = k_{1\alpha_1} k_{3\alpha_2} + k_{3\alpha_1} k_{1\alpha_2} - \eta_{\alpha_1\alpha_2} k_1 \cdot k_3 \, .
\end{equation}
Using the bulk to boundary propagator of the graviton in the large $R$ limit \eqref{eq:gravitonmellin} and using \eqref{eq:celestialtorusdelta},
\begin{equation}
	V_{3\,\text{grav}(-)} = \partial_{\bar{z}_2} \tilde{q}^{\alpha_1}_2 \partial_{\bar{z}_2} \tilde{q}^{\alpha_2}_2 V_{3\,\alpha_1\alpha_2} = \dfrac{8 z^2_{23}\omega^2_3}{(1+z_2\bar{z}_2)^2} = \dfrac{\left(V_{3 \, \text{gluon}(-)}\right)^2}{2} \, ,
\end{equation} 
where $V_{3 \, \text{gluon}(-)}$ is given in \eqref{eq:v3gloun-}. This is the expected result, and one can evaluate the Witten diagram in the large $R$ limit in a straightforward way to be
\begin{equation}
	\begin{split}
		\langle O_{\Delta_1}(\p_1)& O_{\Delta_2;\bar{z}_2 \bar{z}_2}(\p_2)  O_{\Delta_3}(\p_3) \rangle \\
		&= \dfrac{32 \mathcal{C}_{(1)grav}}{(1+z_2\bar{z}_2)^2} z^{\Delta_3}_{12} z^{\Delta_1}_{32} z^{\Delta_2-2}_{13} \dfrac{\Gamma(\Delta_1+\Delta_2+\Delta_3-2) \, \delta(\bar{z}_{12}) \, \delta(\bar{z}_{23}) }{[-i(z_1 u_{23} + z_2 u_{31} + z_3 u_{12})]^{\Delta_1+\Delta_2+\Delta_3-2}} \, .
	\end{split}
\end{equation}
This agrees with the intrinsic result \eqref{eq:threepointsplit} if we set $\sigma_1=0,\sigma_2=-2,\sigma_3=0$. The normalization constant $\mathcal{C}_{(1)grav}$ is given by
\begin{equation}
	\begin{split}
		\mathcal{C}_{(1)grav} &= N^3_{\Delta_1} \dfrac{\mathcal{C}^3_{\Delta_2;2}}{(-i)^{\Delta_2}\Gamma(\Delta_2)} \dfrac{\Delta_2-1}{\Delta_2} N^3_{\Delta_3} 
		= \dfrac{i^{\Delta_1+\Delta_2}(-i)^{\Delta_3}(\Delta_2+1)}{8\pi^{\frac{9}{2}} \prod_{i=1}^{3} \Gamma \left( \Delta_i - \dfrac{1}{2}\right) \, \Delta_2 \, R^{5-\Delta_1-\Delta_2-\Delta_3} } \, .
	\end{split}
\end{equation}

\subsection{Graviton MHV three point amplitude: split signature}

Much of the literature has focussed on how the MHV amplitudes for graviton scattering undergo an enormous simplification, resulting in simple formulas for their expression \cite{Nguyen:2009jk,Hodges:2011wm,Hodges:2012ym}. These works were inspired by the famous Parke-Taylor formula for $n$-gloun scattering amplitude \cite{Parke:1986gb}. We now obtain these simple answers for MHV amplitudes from the limit of AdS Witten diagrams. It is well known that the standard Mellin transform of tree level graviton amplitudes is UV divergent \cite{Stieberger:2018edy,Puhm:2019zbl}. However, it was pointed out that the modified Mellin transform of MHV graviton scattering amplitudes is UV finite \cite{Banerjee:2019prz} since $u$ acts as a UV regulator. Here, we show how to obtain the modified Mellin transformation of the three point MHV graviton amplitude worked out in \cite{Banerjee:2019prz} from an AdS Witten contact diagram.

\medskip

The three point vertex has been derived in \cite{DeWitt:1967uc}, and it is given by
\begin{equation}\label{eq:threeptgravvertex}
	\begin{split}
		V^{3(\text{grav})}_{\mu\alpha,\nu\beta,\sigma\gamma} = \lambda_3 \text{sym} \Bigg[ -\dfrac{1}{2} P_3(k_1 \cdot k_2 \, \eta_{\mu\alpha} \eta_{\nu\beta} \eta_{\sigma\gamma}) - \dfrac{1}{2} P_6( k_{1\nu} k_{1\beta} \eta_{\mu\alpha} \eta_{\sigma\gamma} ) \\ + \dfrac{1}{2} P_3(k_1 \cdot k_2 \, \eta_{\mu\nu} \eta_{\alpha\beta} \eta_{\sigma\gamma})
		+ P_6 (k_1 \cdot k_2 \, \eta_{\mu\alpha} \eta_{\nu\sigma} \eta_{\beta\gamma}) +2 P_3( k_{1\nu} k_{1\gamma} \eta_{\mu\alpha} \eta_{\beta\sigma} ) \\ - P_3( k_{1\beta} k_{2\mu} \eta_{\alpha\nu} \eta_{\sigma\gamma} ) + P_3( k_{1\sigma} k_{2\gamma} \eta_{\mu\nu} \eta_{\alpha\beta} )  		- P_6( k_{1\sigma} k_{1\gamma} \, \eta_{\mu\nu} \eta_{\alpha\beta} ) \\ + 2 P_6( k_{1\nu} k_{2\gamma} \eta_{\beta\mu} \eta_{\alpha\sigma} ) - 2 P_3( k_{1\nu} k_{2\mu} \eta_{\beta\sigma} \eta_{\gamma\alpha} ) - 2 P_3(k_1 \cdot k_2 \eta_{\alpha\nu} \eta_{\beta\sigma} \eta_{\gamma\mu}) \Bigg] \, .
	\end{split}
\end{equation}
We have adopted the normalization of \cite{PhysRevD.34.1749} and $\lambda_3 \propto \sqrt{G_N}$ where $G_N$ is the Newton's constant. We follow the conventions of \cite{DeWitt:1967uc}: $\text{sym}$ implies that we must symmetrize the three point vertex in each pair of indices $\mu\alpha,\nu\beta,\sigma\beta$. $P_k$ indicates that we should sum over the distinct permutations of the momentum index combinations $k_1-\mu\alpha,k_2-\nu\beta,k_3-\sigma\gamma$ and the subscript $k$ denotes the number of such terms in each class. Altogether, we have 171 terms that need to be summed over. One crucial difference between the expressions of \cite{DeWitt:1967uc,PhysRevD.34.1749} is that the 8th term and 10th term of \eqref{eq:threeptgravvertex} have an overall minus sign. We need that sign to obtain the correct answer, which was explicitly checked in Mathematica.

\medskip

The graviton MHV $(--+)$ three point function is given by following the contact AdS Witten diagram
\begin{equation}\label{eq:gravwitt3pt}
	\begin{split}
		&\langle O_{\Delta_1;\bar{z}_1 \bar{z}_1}(\p_1) O_{\Delta_2;\bar{z}_2 \bar{z}_2}(\p_2) O_{\Delta_3;z_3 z_3}(\p_3) \rangle = \\ &\int d \x K^{\Delta_1,2;\alpha_1\alpha_2}_{\hspace{1.4cm};\bar{z}_1 \bar{z}_1}(\p_1,\x) K^{\Delta_2,2;\alpha_3\alpha_4}_{\hspace{1.4cm};\bar{z}_2 \bar{z}_2}(\p_2,\x)
		K^{\Delta_3,2;\alpha_5\alpha_6}_{\hspace{1.4cm};z_3 z_3}(\p_3,\x)V^{3(\text{grav})}_{\alpha_1\alpha_2,\alpha_3\alpha_4,\alpha_5\alpha_6} \, .
	\end{split}
\end{equation}
Here $K^{\Delta_2,2}_{\alpha_1\alpha_2;\beta_1\beta_2}(\p_i,\x)$ denotes the bulk to boundary propagator of the graviton field and in the large $R$ limit, it is given by the Mellin representative \eqref{eq:gravitonmellin}. The action of the projector is analyzed in the graviton propagator calculation of Section \ref{sec:gravprop}. Thus, in the large $R$ limit, the three point function of \eqref{eq:gravwitt3pt} becomes
\begin{equation}\label{eq:gravwitt3ptlarger}
	\begin{split}
		\mathcal{C}^{grav}_{(3)} \int d\omega_1 d\omega_2 d\omega_3 \omega^{\Delta_1-1}_1 \omega^{\Delta_2-1}_2 \omega^{\Delta_3-1}_3 e^{i(\omega_1 u_1 + \omega_2 u_2 - \omega_3 u_3)} \\ ~~~\times \delta^{(4)}(\omega_1 \tilde{q}_1 + \omega_2 \tilde{q}_2 - \omega_3 \tilde{q}_3) V_{3 \, \text{grav}(--+)} \, .
	\end{split}
\end{equation}
The normalization constant $\mathcal{C}^{grav}_{(3)}$ is given by ($\mathcal{C}^d_{\Delta;J}$ is given in \eqref{eq:normspin})
\begin{equation}
	\begin{split}
		\mathcal{C}^{grav}_{(3)} &= \dfrac{\mathcal{C}^3_{\Delta_1;2}}{(-i)^{\Delta_1}\Gamma(\Delta_1)}\dfrac{\Delta_1-1}{\Delta_1}\dfrac{\mathcal{C}^3_{\Delta_2;2}}{(-i)^{\Delta_2}\Gamma(\Delta_2)}\dfrac{\Delta_2-1}{\Delta_2}\dfrac{\mathcal{C}^3_{\Delta_3;2}}{(i)^{\Delta_3}\Gamma(\Delta_3)}\dfrac{\Delta_3-1}{\Delta_3} \\
		&= \dfrac{i^{\Delta_1+\Delta_2} (-i)^{\Delta_3}(\Delta_1+1)(\Delta_2+1)(\Delta_3+1)}{8 \pi^{\frac{9}{2}} \prod_{i=1}^n \Gamma \left(\Delta_i - \frac{1}{2}\right) \, \Delta_1 \, \Delta_2 \, \Delta_3 \, R^{9-(\Delta_1+\Delta_2+\Delta_3)}} \, .
	\end{split}
\end{equation}
The vertex factor $V_{3 \, \text{grav}(--+)}$ is given by
\begin{equation}\label{eq:vertexfactorgrav3pt}
	\begin{split}
		V_{3 \, \text{grav}(--+)} = \partial_{\bar{z}_1} \tilde{q}^{\alpha_1}_1 \partial_{\bar{z}_1} \tilde{q}^{\alpha_2}_1 \partial_{\bar{z}_2} \tilde{q}^{\alpha_3}_2 \partial_{\bar{z}_2} \tilde{q}^{\alpha_4}_2 \partial_{z_3} \tilde{q}^{\alpha_5}_3 \partial_{z_3} \tilde{q}^{\alpha_6}_3 V^{3(\text{grav})}_{\alpha_1\alpha_2,\alpha_3\alpha_4,\alpha_5\alpha_6} 
	\end{split}
\end{equation}
Using the parametrization explained in footnote \ref{paramfootnote} and employing the momentum conserving delta function \eqref{eq:celestialtorusdelta}, we can evaluate this crucial factor using Mathematica to be \footnote{The terms 1-6 of \eqref{eq:threeptgravvertex} vanish trivially because the ingoing momenta are null. Terms 7-11 are non-trivial and sum together intricately to give this result. This is where the signs of Term 8 and Term 10 become all important to give a clean result.}
\begin{equation}\label{eq:vertexcontractgrav3pt}
	V_{3 \, \text{grav}(--+)} = \dfrac{-16 \lambda_3 \, z^2_{12} \, \omega^2_3 }{(1+z_1\bar{z}_1)^2 (1+z_2\bar{z}_2)^2 (1+z_3\bar{z}_3)^2 } = -\dfrac{1}{4} \lambda_3 D^2_z z^2_{12} \omega^2_3 \, ,
\end{equation}
where $D_z$ was defined in \eqref{eq:vertexcontract3pt}. Thus, we get
\begin{equation}
	V_{3 \, \text{grav}(--+)} = -\dfrac{\lambda_3}{4 g^2 (f^{abc})^2 } \left(V_{3\, \text{gluon}(--+)} \right)^2 \, .
\end{equation}
This is the standard result one obtains using spinor helicity variables \cite{Hodges:2011wm,elvang_huang_2015} (i.e.) the amplitude is given by
\begin{equation}\label{eq:shgrav--+}
	\mathcal{A}^{\text{grav}}_{--+} = \dfrac{\langle12 \rangle^6}{ \langle 23 \rangle^2 \langle 31 \rangle^2 } \delta^{(4)}(p^{\mu}_1 + p^{\mu}_2 + p^{\mu}_3 ) \propto \left(\mathcal{A}^{\text{gluon}}_{--+}\right)^2 \, ,
\end{equation}
where $\mathcal{A}^{\text{gluon}}_{--+}$ is given in \eqref{eq:shgluon--+}. The existence of this squared structure through the Feynman diagrams has not been explicitly worked out before \cite{Brandhuber:2022qbk} \footnote{In order to get to this standard result, one needs to incorporate the crucial overall negative signs in Term 8 and Term 10 of \eqref{eq:threeptgravvertex}. The parameterisation of footnote \ref{paramfootnote} is also crucial. We cannot comment on whether these signs will change the analysis of \cite{PhysRevD.34.1749}.}. From \eqref{eq:shgrav--+}, we learn that \eqref{eq:gravwitt3ptlarger} essentially becomes the Modified Mellin transformation of the scattering amplitude as expected.

\medskip

Thus, using \eqref{eq:vertexcontractgrav3pt} in \eqref{eq:gravwitt3ptlarger}, we can simplify it using \eqref{eq:celestialtorusdelta} to obtain the final result to be
\begin{equation}\label{eq:grav3ptfinal}
	\begin{split}
		&\langle O_{\Delta_1;\bar{z}_1 \bar{z}_1}(\p_1) O_{\Delta_2;\bar{z}_2 \bar{z}_2}(\p_2) O_{\Delta_3;z_3 z_3}(\p_3) \rangle \\ &= \mathcal{C}^{grav}_{(3)} (-1)^{\Delta_1+\Delta_2+2} z^{\Delta_3+2}_{12} z^{\Delta_1-2}_{23} z^{\Delta_2-2}_{31} \dfrac{\Gamma(\Delta_1+\Delta_2+\Delta_3-2) \, \delta(\bar{z}_{12}) \, \delta(\bar{z}_{23})}{[-i(z_1 u_{23} + z_2 u_{31} + z_3 u_{12})]^{\Delta_1+\Delta_2+\Delta_3-2}} \, .	 
	\end{split}
\end{equation}
This matches with (4.20) of \cite{Banerjee:2019prz}. This also matches the expected intrinsic result \eqref{eq:threepointsplit} once we use $\sigma_1=-2,\sigma_2=-2,\sigma_3=2$. 

\medskip

For the $(++-)$ case, the vertex factor involved will just be the complex conjugate, and it is given by
\begin{equation}
	V_{3 \, \text{grav}(++-)} = -\dfrac{\lambda_3}{4 g^2 (f^{abc})^2 } \left(V_{3\, \text{gluon}(++-)} \right)^2 \, ,
\end{equation}
where $V_{3\, \text{gluon}(++-)}$ is given by \eqref{eq:gluon++-3pt}. As for the final expression of the amplitude, it is just given by the complex conjugate of \eqref{eq:grav3ptfinal} and that matches with the alternate branch intrinsic result \eqref{eq:threepointsplitalt} if we use $\sigma_1=2,\sigma_2=2,\sigma_3=-2$.

\medskip

Similar to the gluon case, there is no $(+++)$ or $(---)$ amplitudes because the vertex factor \eqref{eq:threeptgravvertex} vanishes for the combinations. This is consistent with the expectation that there are no MHV amplitudes of the form $(+++)$ or $(---)$. We must consider the vertex from a higher derivative term like $R^3$, where $R$ is the Ricci scalar to get a non-zero result. However, from the Carrollian perspective, there will be a non-zero result for the $(+++)$ and $(---)$ correlators from \eqref{eq:threepointsplitalt} and \eqref{eq:threepointsplit} respectively. They will correspond to holographic higher derivative terms in the bulk. For the Witten diagram, we can only predict that the vertex factor will be the square of the gluon $(---)$ vertex factor given in \eqref{eq:gluon---3pt}. The computation of the vertex factor from the higher derivative term like $R^3$ will be tedious.

\subsection{Graviton-gluon-gluon three point function: split signature}

There is an interesting three point diagram involving two gluons and one graviton that arises when Yang-Mills theory is minimally coupled to Einstein gravity \cite{PhysRevLett.84.3531,Fu:2017uzt,Plefka:2018zwm}. Einstein-Yang-Mills theory serves as a theoretical laboratory to explore the deep connections between graviton and gauge amplitudes following upon the establishment of the KLT relations \cite{Kawai:1985xq}. The three point vertex factor for two gluons and one graviton is given by \cite{Plefka:2018zwm}:
\begin{equation}
	\begin{split}
		V^{\alpha\beta \mu_a \mu_b}_{\text{gl-gl-grav}} = \lambda_3 \left[ \dfrac{1}{2}\eta^{\alpha \mu_a} \eta^{\beta \mu_b} p_a \cdot p_b + \eta^{\mu_a \mu_b} p^{(\alpha}_b p^{\beta)}_a \right.
		\left. - \eta^{\mu_a(\alpha}p^{\beta)}_b p^{\mu_b}_a - \eta^{\mu_b(\alpha}p^{\beta)}_a p^{\mu_a}_b + \dfrac{1}{2}\eta^{\alpha\beta} p^{\mu_a}_a p^{\mu_b}_b \right]
	\end{split}
\end{equation}
Here $\alpha$ and $\beta$ are the graviton indices and $\mu_i$ are the gauge indices (the subscript $a,b$ denote the colour indices). 

\medskip

The contact Witten diagram is given by
\begin{equation}\label{eq:glglgrav3ptwit}
	\begin{split}
		&\langle O_{\Delta_1;\beta_1}(\p_1) O_{\Delta_2;\beta_2}(\p_2) O_{\Delta_3;\beta_3 \beta_4}(\p_3) \rangle \\
		&= \int d \x \, K^{\Delta_1,1}_{\alpha_1;\beta_1}(\p_1,\x) \, K^{\Delta_2,1}_{\alpha_2;\beta_2}(\p_2,\x) \, K^{\Delta_3,2}_{\alpha_3 \alpha_4;\beta_3 \beta_4}(\p_3,\x) \, V^{\alpha_3 \alpha_4 \alpha_1 \alpha_2}_{\text{gl-gl-grav}} \, ,
	\end{split} 
\end{equation}
where the gluon and graviton bulk to boundary propagators in the large $R$ limit are given by \eqref{eq:mellinspin1} and \eqref{eq:gravitonmellin}. For the diagram \eqref{eq:glglgrav3ptwit} to result in a non-trivial answer, the gluons should be of opposite helicities. Let us consider the $(+--)$ case where the graviton is of negative helicity. In the large $R$ limit, \eqref{eq:glglgrav3ptwit} results in
\begin{equation}\label{eq:glglgrav3ptwitlarger}
	\begin{split}
		&\langle O_{\Delta_1;z_1}(\p_1) O_{\Delta_2;\bar{z}_2}(\p_2) O_{\Delta_3;\bar{z}_3 \bar{z}_4}(\p_3) \rangle \\
		&= \mathcal{C}^{gl-gl-grav}_{(3)} \int d\omega_1 d\omega_2 d\omega_3 \omega^{\Delta_1-1}_1 \omega^{\Delta_2-1}_2 \omega^{\Delta_3-1}_3 e^{i(\omega_1 u_1 + \omega_2 u_2 - \omega_3 u_3)} \\
		& ~~~ \times \partial_{z_1} q_{1\alpha_1} \partial_{\bar{z}_2} q_{2\alpha_2} \partial_{\bar{z}_3} q_{3\alpha_3} \partial_{\bar{z}_4} q_{4\alpha_4} V^{\alpha_3 \alpha_4 \alpha_1 \alpha_2}_{\text{gl-gl-grav}} \delta^{(4)}(\omega_1 \tilde{q}_1 + \omega_2 \tilde{q}_2 - \omega_3 \tilde{q}_3) \, ,
	\end{split}
\end{equation}
where $\mathcal{C}^{gl-gl-grav}_{(3)}$ is given by
\begin{equation}
	\begin{split}
		\mathcal{C}^{gl-gl-grav}_{(3)} &= \dfrac{\mathcal{C}^3_{\Delta_1;1}}{(-i)^{\Delta_1}\Gamma(\Delta_1)}\dfrac{\Delta_1-1}{\Delta_1} \dfrac{\mathcal{C}^3_{\Delta_2;1}}{(-i)^{\Delta_2}\Gamma(\Delta_2)}\dfrac{\Delta_2-1}{\Delta_2} \dfrac{\mathcal{C}^3_{\Delta_3;1}}{(i)^{\Delta_3}\Gamma(\Delta_3)}\dfrac{\Delta_3-1}{\Delta_3} \\
		&= \dfrac{(i)^{\Delta_1+\Delta_2} (-i)^{\Delta_3} (\Delta_3+1)}{8 \pi^{\frac{9}{2}} \Delta_3 R^{7-(\Delta_1+\Delta_2+\Delta_3)} \prod_{i=1}^{3} \Gamma \left( \Delta_i- \frac{1}{2} \right)} \, .
	\end{split}
\end{equation}
Now
\begin{equation}\label{eq:v3glglgrav}
      \begin{split}
      		V^{\text{gl-gl-grav}}_{(3)} &= \partial_{z_1} q_{1\alpha_1} \partial_{\bar{z}_2} q_{2\alpha_2} \partial_{\bar{z}_3} q_{3\alpha_3} \partial_{\bar{z}_4} q_{4\alpha_4} V^{\alpha_3 \alpha_4 \alpha_1 \alpha_2}_{\text{gl-gl-grav}} \\
      		&= \dfrac{8 \omega_1 \omega_2 z^2_{23} (1+ 2 \bar{z}_1 z_2 + z_1 \bar{z}^2_1 z_2)}{(1+z_1\bar{z}_1)^2 (1+z_2 \bar{z}_2)^2} \\
      		&= \omega_1 \omega_2 z^2_{23} \, G_z
      \end{split}
\end{equation}
Using \eqref{eq:v3glglgrav} in \eqref{eq:glglgrav3ptwitlarger} and simplifying the integrals following analogous steps of Appendix \ref{ap:scalphotonscalsplit}, we get
\begin{equation}\label{eq:glglgrav3ptwitfinal}
	\begin{split}
		\langle O_{\Delta_1;z_1}(\p_1) &O_{\Delta_2;\bar{z}_2}(\p_2) O_{\Delta_3;\bar{z}_3 \bar{z}_4}(\p_3) \rangle \\
		&= \mathcal{C}^{gl-gl-grav}_{(3)} G_z z^{\Delta_3-2}_{12} z^{\Delta_1+1}_{23} z^{\Delta_2-1}_{31} \dfrac{\Gamma(\Delta_1+\Delta_2+\Delta_3-2) \, \delta(\bar{z}_{12}) \delta(\bar{z}_{13})}{[-i(z_1 u_{23} + z_2 u_{31} + z_3 u_{12})]^{\Delta_1+\Delta_2+\Delta_3-2}} 
	\end{split}
\end{equation}
This agrees with the intrinsic result eq.\eqref{eq:threepointsplit} if we set $\sigma_1 =1, \sigma_2= -1, \sigma_3 = -2$. We again see the Carrollian correlators that encode gravitational scattering in the bulk are finite.

\bigskip 

This completes our main section of computing various non-trivial Witten diagrams, which form the bulk analysis. The expressions non-trivially match the correlators obtained through intrinsic field theory analysis. This completes the check of holography as seen through the lens of the limit of AdS/CFT.

\section{Conclusions}
\label{sec:conclusions}

\subsection{Summary of our results}

In this paper, we have tried to highlight the various advantages of working with the Carrollian approach to holography in asymptotically flat spacetimes. After an extensive review of our approach in the first part of the paper and comparing and contrasting it first to the Celestial holography programme and later to a different approach in the Carrollian programme, in the second part, our primary focus has been to make a case for Carroll stronger by working out various non-trivial examples. In particular, we showed how the co-dimension \textit{one} Carrollian correlators that encode bulk scattering of spinning particles naturally arise from the large AdS radius limit of Witten diagrams. One crucial consequence of working with the co-dimension one Carrollian correlators is that one does not need to artificially regulate the UV divergences of the Mellin transform \eqref{mellin} by resorting to distributional delta functions. The $u$ of the modified Mellin transform \eqref{modmellin} naturally acts as a UV regulator, resulting in correlation functions ultra localized across the \textit{entire} future null infinity. This serves as an indication that it is perhaps the more appropriate way to go about flat holography. 

\medskip

Our work is also an extension of the results of \cite{Bagchi:2023fbj} to spinning particles. In this regard, we looked at some new scalar Witten diagrams not considered in \cite{Bagchi:2023fbj}. These are the four point function, the soft limit (non-collinear case), and the split signature cases of the three point function. The soft limit of the scalar three point function was crucially captured by the $\Delta \to 1$ subsector of Carrollian CFT. The split signature result was a non-trivial branch of the three point function where one works with the Celestial torus in which $z$ and $\bar{z}$ were independent.

\medskip

In the context of spinning particles, we looked at the canonical free particle propagation and also non-trivial three point amplitudes. We obtained the gluon and graviton MHV amplitudes \cite{Pasterski:2017ylz,Stieberger:2018edy} from contact Witten diagrams and showed how they become finite in the modified Mellin basis. We worked out the amplitude from a cubic interaction vertex involving a spinning field and two scalar fields. We then proceeded to look at Einstein-Yang-Mills theory and worked out the three point amplitude involving one graviton and two gluons. We also looked at a non-trivial example of an amplitude that arises from loop corrections of the form $F^3$ where $F$ represents the gauge field strength tensor. This was not previously considered in the Celestial approach. In all the above examples, the large AdS limit of bulk Witten diagrams matched the Carrollian correlators obtained from the intrinsic Ward identities of the co-dimension one field theory.

\medskip

The field theory analysis involved working out the highest weight representations of the spinning Carrollian primaries. These primaries were fields with upper indices. One has to distinguish fields with upper and lower indices because the Carrollian metric is degenerate. This distinction ensured that the correlators with time indices turned out to be zero in the leading order. This matched with the bulk result in that the spinning bulk to boundary propagator with one of the boundary indices being the time index becomes a pure gauge at the leading order of the large AdS radius limit. The highest weight representations then manifest as Ward identities that completely constrain the two and three point correlation function for spinning primaries. It is well known that the little group scaling completely fixes the three point scattering amplitude for particles of any helicity in Minkowski space \cite{elvang_huang_2015}. For the first time in the literature, we derive this result directly from the Ward identities of the co-dimension one Carrollian field theory. All these results validate our claims for seriously considering the Carrollian approach to flat space holography.

\medskip

\subsection{Discussions}

We list out some things that we still need to understand fully and hope to return to in the near future and clarify. 

\medskip

{\em{Normalization constant:}} The normalization constants in the conformal primary wavefunctions \eqref{eq:normscalar} and \eqref{eq:normspin} diverge as $R \to \infty$ depending on the value of $\Delta_i$.  These constants arise in the computation of the correlators through Witten diagrams. We expect that these divergences will be resolved through holographic renormalization.

\medskip

{\em{Sub-leading corrections to the bulk to boundary propagator:}} We don't fully understand the sub-leading large $R$ corrections to the bulk to boundary propagator \eqref{eq:bbdppropspin}. These sub-leading corrections are crucial if we are looking at operators with raised indices. Initial efforts at calculating these corrections have run into difficulties in comparison with boundary answers. 

\medskip

{\em{Precise meaning of $\Delta$:}} $\Delta$ labels the boundary conformal field. It arises because we have chosen a unitary representation that diagonalizes the action of $L_0$ and $\bar{L}_0$. However, purely from the point of view of the boundary theory, it arises from the Casimir of the boundary algebra. It would be interesting to make this notion more precise from the point of view of the representation theory of the Conformal Carroll theory.

\subsection{Future Directions}

There are, of course, numerous avenues of immediate research. The above-listed points serve as the first things to clear up in our current programme. But we hope to do much more. 

\medskip

As we pointed out before, there seems to be some tension between the two different approaches of Carrollian road to flat space holography, the one motivated by an asymptotic expansion \cite{Donnay:2022aba,Donnay:2022wvx} and our approach where the modified Mellin transformation plays a central role. We have pointed out some reservations we have with the former in Sec. \ref{sec:celestialvscarroll}. But clearly, something deeper is at play as the approach advocated in \cite{Donnay:2022aba,Donnay:2022wvx} apriori seems perfectly fine. It would be good to clarify this apparent tension between Carrollian methods going forward. 

\medskip 

Understanding the soft sector in the light of modified Mellin transformation from the perspective of a purely Carrollian theory is a priority. This has already been done from the point of view of the modified Mellin in \cite{Banerjee:2019prz,Banerjee:2020zlg}. We need to just have Carrollian interpretations for the same. 

\medskip

In AdS/CFT, the boundary EM tensor sources the bulk graviton; hence, EM tensor correlations are related to graviton amplitudes. Graviton amplitudes should carry over from AdS to flat space in the infinite radius limit, and there should be a similar story relating Carrollian EM tensors and flat space graviton amplitudes. However, in flat spacetime, there is the additional complication of radiation leaking out of null infinity. This is what instigated the use of sourced Carroll Ward identities to match with soft theorems in \cite{Donnay:2022wvx} {\footnote{See however \cite{Saha:2023hsl} for a discussion of the same without using sources.}}. It would be interesting to understand how to take the limit on AdS/CFT analysis in this context and get to graviton amplitudes in asymptotically flat spacetimes. We hope to report on this in the near future. 

\medskip

It is of interest to understand massive scattering and relate them to Carrollian structures. For this, the Carrollian nature of $i^\pm$ is important to understand. The limit from AdS/CFT in this case would also be very interesting to understand. 

\medskip

To conclude, we re-emphasise that it seems the Carrollian co-dimension one approach to holography in asymptotically flat spacetimes has several advantages over the co-dimension two Celestial picture. We have highlighted that in the first part of our paper and, in this second half, given several examples from the flat limit of AdS/CFT to strengthen our point. 

\bigskip \bigskip

\subsection*{Acknowledgements}
We would like to thank Daniel Grumiller, Shahin Sheikh-Jabbari, Romain Ruzziconi and especially Shamik Banerjee for helpful discussions. 

\medskip

PD also thanks Tim Adamo, Dionysios Anninos, Alejandra Castro, Chandramouli Chowdhury, Jelle Hartong, Nabil Iqbal, Arthur Lipstein, Prahar Mitra, Silvia Nagy, Gerben Oling, Simon Ross, Joan Simon, Marika Taylor and Mritunjay Verma for comments on the work during and after it was presented in places mentioned below. PD acknowledges the warm hospitality of the University of Cambridge, University of Southampton, Durham University, King's College London, and the University of Edinburgh during the course of this work.

\medskip

AB is partially supported by a Swarnajayanti Fellowship from the Science and Engineering Research Board
(SERB) under grant SB/SJF/2019-20/08 and SERB grant CRG/2020/002035 and further by a Royal Society of London international exchange grant with the University of Edinburgh. SD thanks partial support from grant SB/SJF/2019-20/08 of AB. PD would like to duly acknowledge the Council of Scientific and Industrial Research (CSIR), New Delhi, for financial assistance through the Senior Research Fellowship (SRF) scheme and the partial support from the Royal Society of London international exchange grant with the University of Edinburgh.

\newpage

\appendix

\section*{APPENDICES}

\section{Details of computations}
In the various subsections of this appendix, we present details of computations that we have omitted in the main text for ease of reading. 

\subsection{Scalar four point function: contact diagram analysis}
\label{ap:fourpointcontact}
We will need the following integral for most of the computations
\begin{equation}\label{eq:omegaint}
     \int d\omega \, \omega^{\Delta - k} e^{i\omega u} = (-iu)^{-1+k-\Delta} \Gamma(1-k+\Delta) \hspace{1cm} \text{if}~~\text{Re}(k-\Delta)<1\, ,\text{Im}(u)>0 \, .
 \end{equation}
The $i\epsilon$ prescription always ensures $\text{Im}(u)>0$. We will now simplify \eqref{eq:fourptcontactfirst} to \eqref{eq:fourptcontactresult}. To evaluate the Mellin integrals, we will use a convenient change of variables. We will change the variables to $s = \sum_i \omega_i$ and a set of ``simplex" variables $\sigma_i = s^{-1} \omega_i \in [0,1]$ with $\sum_{i=1}^{n} \sigma_i = 1$ \cite{Pasterski:2017ylz,Banerjee:2019prz}. We thus have
\begin{equation}\label{eq:simplexint}
    \prod_{i=1}^n \int^{\infty}_0 d\omega_i \omega^{\Delta_i -1}_i [\dots] = \int ds s^{n-1+ \sum_{i=1}^n(\Delta_i -1)} \prod_{i=1}^n \int d\sigma_i \sigma^{\Delta_i-1}_i \delta\left(\sum_{i=1}^n \sigma_i -1 \right) [\dots] \, .
\end{equation}
 We can now combine this integral with the delta function in momentum conservation ($\varepsilon_i = \pm 1$ is used to denote incoming and outgoing particles respectively) given by
 \begin{equation}
     \delta^{(4)}\left(\sum_i \varepsilon_i \omega_i \tilde{q}_i \right) = \delta^{(4)}\left(\sum_i \varepsilon_i \sigma_i s \tilde{q}_i \right) = \dfrac{1}{s^4} \delta^{(4)}\left(\sum_i \varepsilon_i \sigma_i \tilde{q}_i \right) \, .
 \end{equation}
We have used the fact that $d=3$. Now 
\begin{equation}
    \delta^{(4)}\left(\sum_i \varepsilon_i \sigma_i \tilde{q}_i \right)\delta\left(\sum_i \sigma_i - 1\right) = C(z_i,\Bar{z}_i) \prod_{i=1}^{n\leq 5} \delta(\sigma_i - \sigma^*_i) \,.
\end{equation}
The form of $C(z_i,\Bar{z}_i)$ and $\sigma^*_i$ for three point (for split signature) and four point amplitudes is given in \cite{Pasterski:2017ylz}. For the four point amplitude in Minkowski signature, we have
\begin{equation}\label{eq:fourptsimplexdelta}
    \delta^{(4)}\left(\sum_{i=1}^4 \varepsilon_i \sigma_i \tilde{q}_i \right)\delta\left(\sum_{i=1}^4 \sigma_i - 1\right) = C(z_i,\Bar{z}_i) \prod_{i=1}^{n=4} \delta(\sigma_i - \sigma^*_i) \,,
\end{equation}
where the various quantities are given by
\begin{equation}
    C(z_i,\Bar{z}_i) = \dfrac{1}{4} \delta(|z_{12}z_{34}\Bar{z}_{13}\Bar{z}_{24} - \Bar{z}_{12}\Bar{z}_{34}z_{13}z_{24}|) = \dfrac{\delta(|z-\Bar{z}|)}{4 z_{12} \Bar{z}_{13} z_{24} \Bar{z}_{24}} \, ,
\end{equation}
where $z$ and $\Bar{z}$ are the conformal cross ratios
\begin{equation}
    z = \dfrac{z_{12} z_{34}}{z_{13} z_{24}} \, , ~~~~~ \Bar{z} = \dfrac{\Bar{z}_{12}\Bar{z}_{34}}{\Bar{z}_{13}\Bar{z}_{24}} \, .
\end{equation}
\begin{gather}
    \sigma^*_1 = -\dfrac{\varepsilon_1 \varepsilon_4}{D} \dfrac{z_{24} \Bar{z}_{34}}{z_{12} \Bar{z}_{13}} \, , ~~ \sigma^*_2 = \dfrac{\varepsilon_2 \varepsilon_4}{D} \dfrac{z_{34} \Bar{z}_{14}}{z_{23} \Bar{z}_{12}} \, , ~~ \sigma^*_3 = -\dfrac{\varepsilon_3 \varepsilon_4}{D} \dfrac{z_{24} \Bar{z}_{14}}{z_{23} \Bar{z}_{13}} \, , ~~ \sigma^*_4 = \dfrac{1}{D} \nonumber \\
    D = (1-\varepsilon_1\varepsilon_4)\dfrac{z_{24} \Bar{z}_{34}}{z_{12} \Bar{z}_{13}} + (\varepsilon_2 \varepsilon_4 -1) \dfrac{z_{34}\Bar{z}_{14}}{z_{23}\Bar{z}_{12}} + (1- \varepsilon_3 \varepsilon_4) \dfrac{z_{24}\Bar{z}_{14}}{z_{23}\Bar{z}_{13}}
\end{gather}
Note that the parametriation for the null momenta used in \cite{Pasterski:2017ylz} differs from our parametrization \eqref{eq:4dparamq} by an overall factor:
\begin{equation}
    q^{\mu} = (1+z\Bar{z}, z+ \Bar{z}, -i(z-\Bar{z}), 1- z\Bar{z}) = (1+z\Bar{z}) \tilde{q}^{\mu} \, .
\end{equation}
But this won't affect the above results of the product of delta functions since we can always suitably redefine the energy $\omega_i$.

We will now use \eqref{eq:simplexint} in \eqref{eq:fourptcontactfirst}. We get
\begin{equation}
    \begin{split}
        \langle O_{\Delta_1}(\p_1)&O_{\Delta_2}(\p_2) O_{\Delta_3}(\p_3)O_{\Delta_4}(\p_4)\rangle \\
        &= \mathcal{A}_{(4)c} \int^{\infty}_0 ds s^{3+\sum_{i=1}^4(\Delta_i -1)} \prod_{i=1}^4 \int d\sigma_i \sigma^{\Delta_i-1}_i e^{i \sigma_1 s u_1 + i \sigma_2 s u_2 - i \sigma_3 s u_3 - i \sigma_3 s u_4} \\
         &~~~\times \dfrac{1}{s^4} \delta^{(4)}(\sigma_1 \tilde{q}_1 + \sigma_2 \tilde{q}_2 - \sigma_3 \tilde{q}_3 - \sigma_4 \tilde{q}_4) \delta\left(\sum_{i=1}^4 \sigma_i - 1 \right) \, .
    \end{split}
\end{equation}
Thus, we can now use \eqref{eq:fourptsimplexdelta} to solve the integrals over $\sigma_i$ from the product of delta functions. Thus, we finally get
\begin{equation}
    \begin{split}
        \langle O_{\Delta_1}(\p_1)O_{\Delta_2}(\p_2) O_{\Delta_3}(\p_3)O_{\Delta_4}(\p_4)\rangle &= \int^{\infty}_0 ds s^{\Delta_1 + \Delta_2 + \Delta_3 + \Delta_4 - 5} e^{i(\sigma^*_1 u_1 + \sigma^*_2 u_2 - \sigma^*_3 u_3 - \sigma^*_4 u_4)s} \\
    & ~~~\times \prod_{i=1}^4 (\sigma^*_i)^{\Delta_i-1} \dfrac{\delta(|z-\Bar{z}|)}{4 z_{12} \Bar{z}_{13} z_{24} \Bar{z}_{24}} \prod_{i=1}^4 \mathbbm{1}_{[0,1]}(\sigma^*_i)
    \end{split}
\end{equation}
Doing the $s$ integral using \eqref{eq:omegaint}, we get \eqref{eq:fourptcontactresult} in a straight-forward way.

\subsection{Scalar three point function: split signature}
\label{ap:threepointsplit}

In this subsection, we will derive \eqref{eq:threepointsplitresult} from \eqref{eq:threepointsplitfirst}. We start by substituting \eqref{eq:celestialtorusdelta} in \eqref{eq:threepointsplitfirst}:
\begin{equation}
    \begin{split}
        \dfrac{\mathcal{A}_{3} \, \delta(\Bar{z}_{12}) \,\delta(\Bar{z}_{23})}{z_{23} z_{31}} &\int^{\infty}_0 d\omega_3 \, \omega^{\Delta_3-3}_3 e^{-i\omega_3 \, u_3} \int^{\infty}_0 d\omega_1 \, \omega^{\Delta_1 -1}_1 e^{i\omega_1 \, u_1} \delta\left(\omega_1 - \omega_3 \dfrac{z_{32}}{z_{12}} \right) \\ &\times\int^{\infty}_0 d\omega_2 \, \omega^{\Delta_2-1}_2 e^{i\omega_2 \, u_2} \delta\left(\omega_2 - \omega_3 \dfrac{z_{31}}{z_{21}} \right)  \\
        &= \dfrac{\mathcal{A}_{3} \delta(\Bar{z}_{12})\delta(\Bar{z}_{23})}{z_{23} z_{31}} \int^{\infty}_0 d\omega_3 \, \omega^{\Delta_3-3}_3 e^{-i\omega_3 \, u_3} \left(\omega_3 \dfrac{z_{32}}{z_{12}} \right)^{\Delta_1 - 1} \text{exp}\left[i\omega_3 \dfrac{z_{32}}{z_{12}}u_1 \right] \\
        & \hspace{6cm}\times\left(\omega_3 \dfrac{z_{31}}{z_{21}} \right)^{\Delta_2-1} \text{exp}\left[i\omega_3 \dfrac{z_{31}}{z_{21}} u_2 \right] \\
        &= \int^{\infty}_0 d\omega_3 \, \omega^{\Delta_1 + \Delta_2 + \Delta_3 - 5}_3 \text{exp}\left[ i\omega_3 \left(\dfrac{z_{32}u_1 + z_{13}u_2 + z_{21}u_3 }{z_{12}} \right) \right] \\
        &~~~\times \dfrac{(-1)^{\Delta_1+\Delta_2 -2}\mathcal{A}_{3} \, \delta(\Bar{z}_{12}) \,\delta(\Bar{z}_{23}) \,  z^{\Delta_1-2}_{32}z^{\Delta_2-2}_{31}}{z_{12}^{\Delta_1+\Delta_2 - 2}} \, .
    \end{split}
\end{equation}
Now we have
\begin{equation}
    z_{32}u_1 + z_{13}u_2 + z_{21}u_3 = z_1 u_{23} + z_2 u_{31} + z_3 u_{12} \, .
\end{equation}
Thus, using \eqref{eq:omegaint}, we get \eqref{eq:threepointsplitresult} in a straight-forward way.

\subsection{Scalar-gluon-scalar three point function: Split signature}
\label{ap:scalphotonscalsplit}

In this subsection, we will derive \eqref{eq:scalphotscalsplitresult} from \eqref{eq:scalphotscalfirst}. If we substitute \eqref{eq:celestialtorusdelta} in \eqref{eq:scalphotscalfirst}, we get
\begin{equation}
	\begin{split}
		&= \mathcal{B}_{(1)sp} \int d\omega_1 d\omega_2 d\omega_3 \omega^{\Delta_1-1}_1 \omega^{\Delta_2-1}_2 \omega^{\Delta_3-1}_3 e^{i\omega_1 \, u_1} e^{i \omega_2 \, u_2} e^{-i\omega_3 \, u_3} \\
		& ~~~\times (\omega_1 \tilde{q}^{\mu}_1 + \omega_3 \tilde{q}^{\mu}_3)(\partial_{\bar{z}_2}\tilde{q}_{2\mu}) \dfrac{4}{\omega^2_3 z_{23} z_{31}} \delta\left(\omega_1 -\omega_3 \dfrac{z_{32}}{z_{12}} \right)\delta\left(\omega_2 -\omega_3 \dfrac{z_{31}}{z_{21}} \right) \delta(\bar{z}_{12})\delta(\bar{z}_{13}) \\
		&= \dfrac{4 \mathcal{B}_{(1)sp}}{z_{23} z_{31}} \left(\dfrac{z_{31}}{z_{21}} \right)^{\Delta_2-1} \left(\dfrac{z_{32}}{z_{12}} \right)^{\Delta_1 -1} \int d\omega_3 \omega^{\Delta_1+\Delta_2+\Delta_3-5}_3 \, \delta(\bar{z}_{12}) \delta(\bar{z}_{13}) \\
		&~~~ \times \left[ \omega_3 \dfrac{{z}_{32}}{z_{12}} \tilde{q}^{\mu}_1 + \omega_3 \tilde{q}^{\mu}_3\right] \partial_{\bar{z}_2}\tilde{q}_{2\mu} \, \text{exp}\left[ i\omega_3 u_1 \dfrac{z_{32}}{z_{12}} + i \omega_3 u_2 \dfrac{z_{31}}{z_{21}} - i \omega_3 u_3 \right] \, .
	\end{split}
\end{equation}
Using the parametrization of footnote \ref{paramfootnote}, we have 
\begin{equation}
	 \left[\dfrac{z_{32}}{z_{12}}\tilde{q}^{\mu}_1 + \tilde{q}^{\mu}_3\right] \partial_{\bar{z}_2} \tilde{q}_{2\mu} = \dfrac{4 \, z_{32} \, \omega_3}{(1+z_2 \bar{z}_2)} = 4 z_{32} \, C_z \, .
 \end{equation}
Doing the $\omega_3$ integral, we straightforwardly get \eqref{eq:scalphotscalsplitresult}. This is similar to the calculations of Appendix \ref{ap:threepointsplit}.

\subsection{Lorentz invariance of delta functions}
\label{ap:lorentzdelta}
We check the Lorentz invariance of the split given in \eqref{eq:celestialtorusdelta}
\begin{equation}
	\delta^{(4)}(\omega_1 \tilde{q}_1 + \omega_2 \tilde{q}_2 - \omega_3 \tilde{q}_3) = \dfrac{4}{\omega^2_3 z_{23}z_{31}}\delta\left(\omega_1 - \omega_3 \dfrac{z_{32}}{z_{12}} \right) \delta\left(\omega_2 - \omega_3 \dfrac{z_{31}}{z_{21}} \right)\delta(\Bar{z}_{13})\delta(\Bar{z}_{23}) \, .
\end{equation}
Let us consider a non-trivial case of the SL$(2,\mathbb{C})$ transformation given by
\begin{equation}
	z_i \to -\dfrac{1}{z_i} \, .
\end{equation}
Thus after using $\bar{z}_1=\bar{z}_2=\bar{z}_3$, we have
\begin{equation}
	\begin{split}
		\delta\left(\omega_1 - \omega_3 \dfrac{z_{32}}{z_{12}} \right) &\to \dfrac{1}{z_1 \bar{z}_1} \delta\left(\omega_1 - \omega_3 \dfrac{z_{32}}{z_{12}} \right) \\
		\delta\left(\omega_2 - \omega_3 \dfrac{z_{31}}{z_{21}} \right) &\to \dfrac{1}{z_2 \bar{z}_2} \delta\left(\omega_2 - \omega_3 \dfrac{z_{31}}{z_{21}} \right) \\
		\delta(\bar{z}_{13}) &\to \bar{z}_1 \bar{z}_3\delta(\bar{z}_{13}) \\
		\delta(\bar{z}_{23}) &\to \bar{z}_2 \bar{z}_3 \delta(\bar{z}_{23}) \\
		\dfrac{1}{\omega^2_3 z_{23} z_{31}} &\to \dfrac{z_1 z_2}{\bar{z}^2_3} \dfrac{1}{\omega^2_3 z_{23} z_{31}} \, .
	\end{split}
\end{equation}
Combining them we see that the split of the delta function given in \eqref{eq:celestialtorusdelta} is indeed Lorentz invariant.

 \newpage

\bibliographystyle{JHEP}
\bibliography{References}

\providecommand{\href}[2]{#2}\begingroup\raggedright\begin{thebibliography}{10}

\bibitem{Strominger:2013jfa}
A.~Strominger, \emph{{On BMS Invariance of Gravitational Scattering}},
  \href{http://dx.doi.org/10.1007/JHEP07(2014)152}{\emph{JHEP} {\bfseries 07}
  (2014) 152}, [\href{https://arxiv.org/abs/1312.2229}{{\ttfamily 1312.2229}}].

\bibitem{He:2014laa}
T.~He, V.~Lysov, P.~Mitra and A.~Strominger, \emph{{BMS supertranslations and
  Weinberg\textquoteright{}s soft graviton theorem}},
  \href{http://dx.doi.org/10.1007/JHEP05(2015)151}{\emph{JHEP} {\bfseries 05}
  (2015) 151}, [\href{https://arxiv.org/abs/1401.7026}{{\ttfamily 1401.7026}}].

\bibitem{Cachazo:2014fwa}
F.~Cachazo and A.~Strominger, \emph{{Evidence for a New Soft Graviton
  Theorem}},  \href{https://arxiv.org/abs/1404.4091}{{\ttfamily 1404.4091}}.

\bibitem{Kapec:2014opa}
D.~Kapec, V.~Lysov, S.~Pasterski and A.~Strominger, \emph{{Semiclassical
  Virasoro symmetry of the quantum gravity $ \mathcal{S}$-matrix}},
  \href{http://dx.doi.org/10.1007/JHEP08(2014)058}{\emph{JHEP} {\bfseries 08}
  (2014) 058}, [\href{https://arxiv.org/abs/1406.3312}{{\ttfamily 1406.3312}}].

\bibitem{Strominger:2014pwa}
A.~Strominger and A.~Zhiboedov, \emph{{Gravitational Memory, BMS
  Supertranslations and Soft Theorems}},
  \href{http://dx.doi.org/10.1007/JHEP01(2016)086}{\emph{JHEP} {\bfseries 01}
  (2016) 086}, [\href{https://arxiv.org/abs/1411.5745}{{\ttfamily 1411.5745}}].

\bibitem{He:2014cra}
T.~He, P.~Mitra, A.~P. Porfyriadis and A.~Strominger, \emph{{New Symmetries of
  Massless QED}}, \href{http://dx.doi.org/10.1007/JHEP10(2014)112}{\emph{JHEP}
  {\bfseries 10} (2014) 112},
  [\href{https://arxiv.org/abs/1407.3789}{{\ttfamily 1407.3789}}].

\bibitem{Strominger:2017zoo}
A.~Strominger, \emph{{Lectures on the Infrared Structure of Gravity and Gauge
  Theory}}.
\newblock 3, 2017.

\bibitem{Pasterski:2021rjz}
S.~Pasterski, \emph{{Lectures on celestial amplitudes}},
  \href{http://dx.doi.org/10.1140/epjc/s10052-021-09846-7}{\emph{Eur. Phys. J.
  C} {\bfseries 81} (2021) 1062},
  [\href{https://arxiv.org/abs/2108.04801}{{\ttfamily 2108.04801}}].

\bibitem{Raclariu:2021zjz}
A.-M. Raclariu, \emph{{Lectures on Celestial Holography}},
  \href{https://arxiv.org/abs/2107.02075}{{\ttfamily 2107.02075}}.

\bibitem{Pasterski:2021raf}
S.~Pasterski, M.~Pate and A.-M. Raclariu, \emph{{Celestial Holography}},  in
  \emph{{Snowmass 2021}}, 11, 2021.
\newblock \href{https://arxiv.org/abs/2111.11392}{{\ttfamily 2111.11392}}.

\bibitem{Pasterski:2016qvg}
S.~Pasterski, S.-H. Shao and A.~Strominger, \emph{{Flat Space Amplitudes and
  Conformal Symmetry of the Celestial Sphere}},
  \href{http://dx.doi.org/10.1103/PhysRevD.96.065026}{\emph{Phys. Rev. D}
  {\bfseries 96} (2017) 065026},
  [\href{https://arxiv.org/abs/1701.00049}{{\ttfamily 1701.00049}}].

\bibitem{Pasterski:2017kqt}
S.~Pasterski and S.-H. Shao, \emph{{Conformal basis for flat space
  amplitudes}}, \href{http://dx.doi.org/10.1103/PhysRevD.96.065022}{\emph{Phys.
  Rev. D} {\bfseries 96} (2017) 065022},
  [\href{https://arxiv.org/abs/1705.01027}{{\ttfamily 1705.01027}}].

\bibitem{Pasterski:2017ylz}
S.~Pasterski, S.-H. Shao and A.~Strominger, \emph{{Gluon Amplitudes as 2d
  Conformal Correlators}},
  \href{http://dx.doi.org/10.1103/PhysRevD.96.085006}{\emph{Phys. Rev. D}
  {\bfseries 96} (2017) 085006},
  [\href{https://arxiv.org/abs/1706.03917}{{\ttfamily 1706.03917}}].

\bibitem{deBoer:2003vf}
J.~de~Boer and S.~N. Solodukhin, \emph{{A Holographic reduction of Minkowski
  space-time}},
  \href{http://dx.doi.org/10.1016/S0550-3213(03)00494-2}{\emph{Nucl. Phys. B}
  {\bfseries 665} (2003) 545--593},
  [\href{https://arxiv.org/abs/hep-th/0303006}{{\ttfamily hep-th/0303006}}].

\bibitem{Barnich:2010eb}
G.~Barnich and C.~Troessaert, \emph{{Aspects of the BMS/CFT correspondence}},
  \href{http://dx.doi.org/10.1007/JHEP05(2010)062}{\emph{JHEP} {\bfseries 05}
  (2010) 062}, [\href{https://arxiv.org/abs/1001.1541}{{\ttfamily 1001.1541}}].

\bibitem{Bagchi:2010zz}
A.~Bagchi, \emph{{Correspondence between Asymptotically Flat Spacetimes and
  Nonrelativistic Conformal Field Theories}},
  \href{http://dx.doi.org/10.1103/PhysRevLett.105.171601}{\emph{Phys. Rev.
  Lett.} {\bfseries 105} (2010) 171601},
  [\href{https://arxiv.org/abs/1006.3354}{{\ttfamily 1006.3354}}].

\bibitem{Bagchi:2012cy}
A.~Bagchi and R.~Fareghbal, \emph{{BMS/GCA Redux: Towards Flatspace Holography
  from Non-Relativistic Symmetries}},
  \href{http://dx.doi.org/10.1007/JHEP10(2012)092}{\emph{JHEP} {\bfseries 10}
  (2012) 092}, [\href{https://arxiv.org/abs/1203.5795}{{\ttfamily 1203.5795}}].

\bibitem{Duval:2014uva}
C.~Duval, G.~W. Gibbons and P.~A. Horvathy, \emph{{Conformal Carroll groups and
  BMS symmetry}},
  \href{http://dx.doi.org/10.1088/0264-9381/31/9/092001}{\emph{Class. Quant.
  Grav.} {\bfseries 31} (2014) 092001},
  [\href{https://arxiv.org/abs/1402.5894}{{\ttfamily 1402.5894}}].

\bibitem{Bagchi:2016bcd}
A.~Bagchi, R.~Basu, A.~Kakkar and A.~Mehra, \emph{{Flat Holography: Aspects of
  the dual field theory}},
  \href{http://dx.doi.org/10.1007/JHEP12(2016)147}{\emph{JHEP} {\bfseries 12}
  (2016) 147}, [\href{https://arxiv.org/abs/1609.06203}{{\ttfamily
  1609.06203}}].

\bibitem{Bagchi:2012yk}
A.~Bagchi, S.~Detournay and D.~Grumiller, \emph{{Flat-Space Chiral Gravity}},
  \href{http://dx.doi.org/10.1103/PhysRevLett.109.151301}{\emph{Phys. Rev.
  Lett.} {\bfseries 109} (2012) 151301},
  [\href{https://arxiv.org/abs/1208.1658}{{\ttfamily 1208.1658}}].

\bibitem{Bagchi:2012xr}
A.~Bagchi, S.~Detournay, R.~Fareghbal and J.~Sim\'on, \emph{{Holography of 3D
  Flat Cosmological Horizons}},
  \href{http://dx.doi.org/10.1103/PhysRevLett.110.141302}{\emph{Phys. Rev.
  Lett.} {\bfseries 110} (2013) 141302},
  [\href{https://arxiv.org/abs/1208.4372}{{\ttfamily 1208.4372}}].

\bibitem{Afshar:2013vka}
H.~Afshar, A.~Bagchi, R.~Fareghbal, D.~Grumiller and J.~Rosseel, \emph{{Spin-3
  Gravity in Three-Dimensional Flat Space}},
  \href{http://dx.doi.org/10.1103/PhysRevLett.111.121603}{\emph{Phys. Rev.
  Lett.} {\bfseries 111} (2013) 121603},
  [\href{https://arxiv.org/abs/1307.4768}{{\ttfamily 1307.4768}}].

\bibitem{Gonzalez:2013oaa}
H.~A. Gonzalez, J.~Matulich, M.~Pino and R.~Troncoso, \emph{{Asymptotically
  flat spacetimes in three-dimensional higher spin gravity}},
  \href{http://dx.doi.org/10.1007/JHEP09(2013)016}{\emph{JHEP} {\bfseries 09}
  (2013) 016}, [\href{https://arxiv.org/abs/1307.5651}{{\ttfamily 1307.5651}}].

\bibitem{Bagchi:2014iea}
A.~Bagchi, R.~Basu, D.~Grumiller and M.~Riegler, \emph{{Entanglement entropy in
  Galilean conformal field theories and flat holography}},
  \href{http://dx.doi.org/10.1103/PhysRevLett.114.111602}{\emph{Phys. Rev.
  Lett.} {\bfseries 114} (2015) 111602},
  [\href{https://arxiv.org/abs/1410.4089}{{\ttfamily 1410.4089}}].

\bibitem{Hartong:2015usd}
J.~Hartong, \emph{{Holographic Reconstruction of 3D Flat Space-Time}},
  \href{http://dx.doi.org/10.1007/JHEP10(2016)104}{\emph{JHEP} {\bfseries 10}
  (2016) 104}, [\href{https://arxiv.org/abs/1511.01387}{{\ttfamily
  1511.01387}}].

\bibitem{Bagchi:2015wna}
A.~Bagchi, D.~Grumiller and W.~Merbis, \emph{{Stress tensor correlators in
  three-dimensional gravity}},
  \href{http://dx.doi.org/10.1103/PhysRevD.93.061502}{\emph{Phys. Rev. D}
  {\bfseries 93} (2016) 061502},
  [\href{https://arxiv.org/abs/1507.05620}{{\ttfamily 1507.05620}}].

\bibitem{Jiang:2017ecm}
H.~Jiang, W.~Song and Q.~Wen, \emph{{Entanglement Entropy in Flat Holography}},
  \href{http://dx.doi.org/10.1007/JHEP07(2017)142}{\emph{JHEP} {\bfseries 07}
  (2017) 142}, [\href{https://arxiv.org/abs/1706.07552}{{\ttfamily
  1706.07552}}].

\bibitem{Hijano:2017eii}
E.~Hijano and C.~Rabideau, \emph{{Holographic entanglement and Poincar\'e
  blocks in three-dimensional flat space}},
  \href{http://dx.doi.org/10.1007/JHEP05(2018)068}{\emph{JHEP} {\bfseries 05}
  (2018) 068}, [\href{https://arxiv.org/abs/1712.07131}{{\ttfamily
  1712.07131}}].

\bibitem{Apolo:2020bld}
L.~Apolo, H.~Jiang, W.~Song and Y.~Zhong, \emph{{Swing surfaces and holographic
  entanglement beyond AdS/CFT}},
  \href{http://dx.doi.org/10.1007/JHEP12(2020)064}{\emph{JHEP} {\bfseries 12}
  (2020) 064}, [\href{https://arxiv.org/abs/2006.10740}{{\ttfamily
  2006.10740}}].

\bibitem{Banerjee:2018gce}
S.~Banerjee, \emph{{Null Infinity and Unitary Representation of The Poincare
  Group}}, \href{http://dx.doi.org/10.1007/JHEP01(2019)205}{\emph{JHEP}
  {\bfseries 01} (2019) 205},
  [\href{https://arxiv.org/abs/1801.10171}{{\ttfamily 1801.10171}}].

\bibitem{Bagchi:2022emh}
A.~Bagchi, S.~Banerjee, R.~Basu and S.~Dutta, \emph{{Scattering Amplitudes:
  Celestial and Carrollian}},
  \href{http://dx.doi.org/10.1103/PhysRevLett.128.241601}{\emph{Phys. Rev.
  Lett.} {\bfseries 128} (2022) 241601},
  [\href{https://arxiv.org/abs/2202.08438}{{\ttfamily 2202.08438}}].

\bibitem{Bagchi:2023fbj}
A.~Bagchi, P.~Dhivakar and S.~Dutta, \emph{{AdS Witten diagrams to Carrollian
  correlators}}, \href{http://dx.doi.org/10.1007/JHEP04(2023)135}{\emph{JHEP}
  {\bfseries 04} (2023) 135},
  [\href{https://arxiv.org/abs/2303.07388}{{\ttfamily 2303.07388}}].

\bibitem{Donnay:2022aba}
L.~Donnay, A.~Fiorucci, Y.~Herfray and R.~Ruzziconi, \emph{{Carrollian
  Perspective on Celestial Holography}},
  \href{http://dx.doi.org/10.1103/PhysRevLett.129.071602}{\emph{Phys. Rev.
  Lett.} {\bfseries 129} (2022) 071602},
  [\href{https://arxiv.org/abs/2202.04702}{{\ttfamily 2202.04702}}].

\bibitem{Donnay:2022wvx}
L.~Donnay, A.~Fiorucci, Y.~Herfray and R.~Ruzziconi, \emph{{Bridging Carrollian
  and Celestial Holography}},
  \href{https://arxiv.org/abs/2212.12553}{{\ttfamily 2212.12553}}.

\bibitem{Nguyen:2023vfz}
K.~Nguyen and P.~West, \emph{{Carrollian conformal fields and flat
  holography}},  \href{https://arxiv.org/abs/2305.02884}{{\ttfamily
  2305.02884}}.

\bibitem{Dixon:1993xd}
L.~J. Dixon and Y.~Shadmi, \emph{{Testing gluon selfinteractions in three jet
  events at hadron colliders}},
  \href{http://dx.doi.org/10.1016/0550-3213(94)90563-0}{\emph{Nucl. Phys. B}
  {\bfseries 423} (1994) 3--32},
  [\href{https://arxiv.org/abs/hep-ph/9312363}{{\ttfamily hep-ph/9312363}}].

\bibitem{Dixon:2004za}
L.~J. Dixon, E.~W.~N. Glover and V.~V. Khoze, \emph{{MHV rules for Higgs plus
  multi-gluon amplitudes}},
  \href{http://dx.doi.org/10.1088/1126-6708/2004/12/015}{\emph{JHEP} {\bfseries
  12} (2004) 015}, [\href{https://arxiv.org/abs/hep-th/0411092}{{\ttfamily
  hep-th/0411092}}].

\bibitem{PhysRevD.96.085006}
S.~Pasterski, S.-H. Shao and A.~Strominger, \emph{Gluon amplitudes as $2d$
  conformal correlators},
  \href{http://dx.doi.org/10.1103/PhysRevD.96.085006}{\emph{Phys. Rev. D}
  {\bfseries 96} (Oct, 2017) 085006}.

\bibitem{Schreiber:2017jsr}
A.~Schreiber, A.~Volovich and M.~Zlotnikov, \emph{{Tree-level gluon amplitudes
  on the celestial sphere}},
  \href{http://dx.doi.org/10.1016/j.physletb.2018.04.010}{\emph{Phys. Lett. B}
  {\bfseries 781} (2018) 349--357},
  [\href{https://arxiv.org/abs/1711.08435}{{\ttfamily 1711.08435}}].

\bibitem{Banerjee:2019prz}
S.~Banerjee, S.~Ghosh, P.~Pandey and A.~P. Saha, \emph{{Modified celestial
  amplitude in Einstein gravity}},
  \href{http://dx.doi.org/10.1007/JHEP03(2020)125}{\emph{JHEP} {\bfseries 03}
  (2020) 125}, [\href{https://arxiv.org/abs/1909.03075}{{\ttfamily
  1909.03075}}].

\bibitem{Bhattacharjee:2023sfd}
A.~Bhattacharjee and M.~Saha, \emph{{Entropy of Flat Space Cosmologies from
  Celestial dual}},  \href{https://arxiv.org/abs/2310.02682}{{\ttfamily
  2310.02682}}.

\bibitem{Salzer:2023jqv}
J.~Salzer, \emph{{An embedding space approach to Carrollian CFT correlators for
  flat space holography}},
  \href{http://dx.doi.org/10.1007/JHEP10(2023)084}{\emph{JHEP} {\bfseries 10}
  (2023) 084}, [\href{https://arxiv.org/abs/2304.08292}{{\ttfamily
  2304.08292}}].

\bibitem{Saha:2023hsl}
A.~Saha, \emph{{Carrollian approach to 1 + 3D flat holography}},
  \href{http://dx.doi.org/10.1007/JHEP06(2023)051}{\emph{JHEP} {\bfseries 06}
  (2023) 051}, [\href{https://arxiv.org/abs/2304.02696}{{\ttfamily
  2304.02696}}].

\bibitem{Saha:2023abr}
A.~Saha, \emph{{$w_{1+\infty}$ and Carrollian Holography}},
  \href{https://arxiv.org/abs/2308.03673}{{\ttfamily 2308.03673}}.

\bibitem{Bondi:1962px}
H.~Bondi, M.~G.~J. van~der Burg and A.~W.~K. Metzner, \emph{{Gravitational
  waves in general relativity. 7. Waves from axisymmetric isolated systems}},
  \href{http://dx.doi.org/10.1098/rspa.1962.0161}{\emph{Proc. Roy. Soc. Lond.
  A} {\bfseries 269} (1962) 21--52}.

\bibitem{Sachs:1962wk}
R.~K. Sachs, \emph{{Gravitational waves in general relativity. 8. Waves in
  asymptotically flat space-times}},
  \href{http://dx.doi.org/10.1098/rspa.1962.0206}{\emph{Proc. Roy. Soc. Lond.}
  {\bfseries A270} (1962) 103--126}.

\bibitem{Brown:1986nw}
J.~D. Brown and M.~Henneaux, \emph{{Central Charges in the Canonical
  Realization of Asymptotic Symmetries: An Example from Three-Dimensional
  Gravity}}, \href{http://dx.doi.org/10.1007/BF01211590}{\emph{Commun. Math.
  Phys.} {\bfseries 104} (1986) 207--226}.

\bibitem{Penedones:2010ue}
J.~Penedones, \emph{{Writing CFT correlation functions as AdS scattering
  amplitudes}}, \href{http://dx.doi.org/10.1007/JHEP03(2011)025}{\emph{JHEP}
  {\bfseries 03} (2011) 025},
  [\href{https://arxiv.org/abs/1011.1485}{{\ttfamily 1011.1485}}].

\bibitem{Fitzpatrick:2011hu}
A.~L. Fitzpatrick and J.~Kaplan, \emph{{Analyticity and the Holographic
  S-Matrix}}, \href{http://dx.doi.org/10.1007/JHEP10(2012)127}{\emph{JHEP}
  {\bfseries 10} (2012) 127},
  [\href{https://arxiv.org/abs/1111.6972}{{\ttfamily 1111.6972}}].

\bibitem{Hijano:2019qmi}
E.~Hijano, \emph{{Flat space physics from AdS/CFT}},
  \href{http://dx.doi.org/10.1007/JHEP07(2019)132}{\emph{JHEP} {\bfseries 07}
  (2019) 132}, [\href{https://arxiv.org/abs/1905.02729}{{\ttfamily
  1905.02729}}].

\bibitem{Nguyen:2023miw}
K.~Nguyen, \emph{{Carrollian conformal correlators and massless scattering
  amplitudes}},  \href{https://arxiv.org/abs/2311.09869}{{\ttfamily
  2311.09869}}.

\bibitem{Banerjee:2020kaa}
S.~Banerjee, S.~Ghosh and R.~Gonzo, \emph{{BMS symmetry of celestial OPE}},
  \href{http://dx.doi.org/10.1007/JHEP04(2020)130}{\emph{JHEP} {\bfseries 04}
  (2020) 130}, [\href{https://arxiv.org/abs/2002.00975}{{\ttfamily
  2002.00975}}].

\bibitem{Dutta:2022vkg}
S.~Dutta, \emph{{Stress tensors of 3d Carroll CFTs}},
  \href{https://arxiv.org/abs/2212.11002}{{\ttfamily 2212.11002}}.

\bibitem{Fiorucci:2023lpb}
A.~Fiorucci, D.~Grumiller and R.~Ruzziconi, \emph{{Logarithmic Celestial
  Conformal Field Theory}},  \href{https://arxiv.org/abs/2305.08913}{{\ttfamily
  2305.08913}}.

\bibitem{LevyLeblond}
L.~Leblond, \emph{{Une nouvelle limite non-relativiste du group de
  Poincar{\'e}}}, {\emph{Annales Poincare Phys.Theor.} {\bfseries 3} (1965) }.

\bibitem{NDS}
N.~Sen~Gupta, \emph{{On an Analogue of the Galileo Group}}, {\emph{Nuovo Cim.
  54 (1966) 512 * DOI: 10.1007/BF02740871} }.

\bibitem{Bagchi:2019xfx}
A.~Bagchi, A.~Mehra and P.~Nandi, \emph{{Field Theories with Conformal
  Carrollian Symmetry}},  \href{https://arxiv.org/abs/1901.10147}{{\ttfamily
  1901.10147}}.

\bibitem{Bagchi:2019clu}
A.~Bagchi, R.~Basu, A.~Mehra and P.~Nandi, \emph{{Field Theories on Null
  Manifolds}}, \href{http://dx.doi.org/10.1007/JHEP02(2020)141}{\emph{JHEP}
  {\bfseries 02} (2020) 141},
  [\href{https://arxiv.org/abs/1912.09388}{{\ttfamily 1912.09388}}].

\bibitem{Henneaux:2021yzg}
M.~Henneaux and P.~Salgado-Rebolledo, \emph{{Carroll contractions of
  Lorentz-invariant theories}},
  \href{http://dx.doi.org/10.1007/JHEP11(2021)180}{\emph{JHEP} {\bfseries 11}
  (2021) 180}, [\href{https://arxiv.org/abs/2109.06708}{{\ttfamily
  2109.06708}}].

\bibitem{deBoer:2021jej}
J.~de~Boer, J.~Hartong, N.~A. Obers, W.~Sybesma and S.~Vandoren, \emph{{Carroll
  Symmetry, Dark Energy and Inflation}},
  \href{http://dx.doi.org/10.3389/fphy.2022.810405}{\emph{Front. in Phys.}
  {\bfseries 10} (2022) 810405},
  [\href{https://arxiv.org/abs/2110.02319}{{\ttfamily 2110.02319}}].

\bibitem{Barnich:2022bni}
G.~Barnich, K.~Nguyen and R.~Ruzziconi, \emph{{Geometric action for extended
  Bondi-Metzner-Sachs group in four dimensions}},
  \href{http://dx.doi.org/10.1007/JHEP12(2022)154}{\emph{JHEP} {\bfseries 12}
  (2022) 154}, [\href{https://arxiv.org/abs/2211.07592}{{\ttfamily
  2211.07592}}].

\bibitem{deBoer:2023fnj}
J.~de~Boer, J.~Hartong, N.~A. Obers, W.~Sybesma and S.~Vandoren, \emph{{Carroll
  stories}}, \href{http://dx.doi.org/10.1007/JHEP09(2023)148}{\emph{JHEP}
  {\bfseries 09} (2023) 148},
  [\href{https://arxiv.org/abs/2307.06827}{{\ttfamily 2307.06827}}].

\bibitem{Chen:2021xkw}
B.~Chen, R.~Liu and Y.-f. Zheng, \emph{{On higher-dimensional Carrollian and
  Galilean conformal field theories}},
  \href{http://dx.doi.org/10.21468/SciPostPhys.14.5.088}{\emph{SciPost Phys.}
  {\bfseries 14} (2023) 088},
  [\href{https://arxiv.org/abs/2112.10514}{{\ttfamily 2112.10514}}].

\bibitem{Stieberger:2018edy}
S.~Stieberger and T.~R. Taylor, \emph{{Strings on Celestial Sphere}},
  \href{http://dx.doi.org/10.1016/j.nuclphysb.2018.08.019}{\emph{Nucl. Phys. B}
  {\bfseries 935} (2018) 388--411},
  [\href{https://arxiv.org/abs/1806.05688}{{\ttfamily 1806.05688}}].

\bibitem{elvang_huang_2015}
H.~Elvang and Y.-t. Huang, \emph{Scattering Amplitudes in Gauge Theory and
  Gravity}.
\newblock Cambridge University Press, 2015,
  \href{http://dx.doi.org/10.1017/CBO9781107706620}{10.1017/CBO9781107706620}.

\bibitem{Dirac:1936fq}
P.~A.~M. Dirac, \emph{{Wave equations in conformal space}},
  \href{http://dx.doi.org/10.2307/1968455}{\emph{Annals Math.} {\bfseries 37}
  (1936) 429--442}.

\bibitem{Penedones:2007ns}
J.~Penedones, \emph{{High Energy Scattering in the AdS/CFT Correspondence}},
  other thesis, 12, 2007.

\bibitem{deGioia:2022fcn}
L.~P. de~Gioia and A.-M. Raclariu, \emph{{Eikonal approximation in celestial
  CFT}}, \href{http://dx.doi.org/10.1007/JHEP03(2023)030}{\emph{JHEP}
  {\bfseries 03} (2023) 030},
  [\href{https://arxiv.org/abs/2206.10547}{{\ttfamily 2206.10547}}].

\bibitem{Giddings:1999jq}
S.~B. Giddings, \emph{{Flat space scattering and bulk locality in the AdS / CFT
  correspondence}},
  \href{http://dx.doi.org/10.1103/PhysRevD.61.106008}{\emph{Phys. Rev. D}
  {\bfseries 61} (2000) 106008},
  [\href{https://arxiv.org/abs/hep-th/9907129}{{\ttfamily hep-th/9907129}}].

\bibitem{deGioia:2023cbd}
L.~P. de~Gioia and A.-M. Raclariu, \emph{{Celestial Sector in CFT: Conformally
  Soft Symmetries}},  \href{https://arxiv.org/abs/2303.10037}{{\ttfamily
  2303.10037}}.

\bibitem{Fitzpatrick:2011dm}
A.~L. Fitzpatrick and J.~Kaplan, \emph{{Unitarity and the Holographic
  S-Matrix}}, \href{http://dx.doi.org/10.1007/JHEP10(2012)032}{\emph{JHEP}
  {\bfseries 10} (2012) 032},
  [\href{https://arxiv.org/abs/1112.4845}{{\ttfamily 1112.4845}}].

\bibitem{Hijano:2020szl}
E.~Hijano and D.~Neuenfeld, \emph{{Soft photon theorems from CFT Ward identites
  in the flat limit of AdS/CFT}},
  \href{http://dx.doi.org/10.1007/JHEP11(2020)009}{\emph{JHEP} {\bfseries 11}
  (2020) 009}, [\href{https://arxiv.org/abs/2005.03667}{{\ttfamily
  2005.03667}}].

\bibitem{Li:2021snj}
Y.-Z. Li, \emph{{Notes on flat-space limit of AdS/CFT}},
  \href{http://dx.doi.org/10.1007/JHEP09(2021)027}{\emph{JHEP} {\bfseries 09}
  (2021) 027}, [\href{https://arxiv.org/abs/2106.04606}{{\ttfamily
  2106.04606}}].

\bibitem{Costa:2014kfa}
M.~S. Costa, V.~Gon\c{c}alves and J.~a. Penedones, \emph{{Spinning AdS
  Propagators}}, \href{http://dx.doi.org/10.1007/JHEP09(2014)064}{\emph{JHEP}
  {\bfseries 09} (2014) 064},
  [\href{https://arxiv.org/abs/1404.5625}{{\ttfamily 1404.5625}}].

\bibitem{Donnay:2018neh}
L.~Donnay, A.~Puhm and A.~Strominger, \emph{{Conformally Soft Photons and
  Gravitons}}, \href{http://dx.doi.org/10.1007/JHEP01(2019)184}{\emph{JHEP}
  {\bfseries 01} (2019) 184},
  [\href{https://arxiv.org/abs/1810.05219}{{\ttfamily 1810.05219}}].

\bibitem{Banerjee:2017jeg}
N.~Banerjee, S.~Banerjee, S.~Atul~Bhatkar and S.~Jain, \emph{{Conformal
  Structure of Massless Scalar Amplitudes Beyond Tree level}},
  \href{http://dx.doi.org/10.1007/JHEP04(2018)039}{\emph{JHEP} {\bfseries 04}
  (2018) 039}, [\href{https://arxiv.org/abs/1711.06690}{{\ttfamily
  1711.06690}}].

\bibitem{Witten:1998qj}
E.~Witten, \emph{{Anti-de Sitter space and holography}},
  \href{http://dx.doi.org/10.4310/ATMP.1998.v2.n2.a2}{\emph{Adv. Theor. Math.
  Phys.} {\bfseries 2} (1998) 253--291},
  [\href{https://arxiv.org/abs/hep-th/9802150}{{\ttfamily hep-th/9802150}}].

\bibitem{schwartz_2013}
M.~D. Schwartz, \emph{Quantum Field Theory and the Standard Model}.
\newblock Cambridge University Press, 2013,
  \href{http://dx.doi.org/10.1017/9781139540940}{10.1017/9781139540940}.

\bibitem{Nguyen:2009jk}
D.~Nguyen, M.~Spradlin, A.~Volovich and C.~Wen, \emph{{The Tree Formula for MHV
  Graviton Amplitudes}},
  \href{http://dx.doi.org/10.1007/JHEP07(2010)045}{\emph{JHEP} {\bfseries 07}
  (2010) 045}, [\href{https://arxiv.org/abs/0907.2276}{{\ttfamily 0907.2276}}].

\bibitem{Hodges:2011wm}
A.~Hodges, \emph{{New expressions for gravitational scattering amplitudes}},
  \href{http://dx.doi.org/10.1007/JHEP07(2013)075}{\emph{JHEP} {\bfseries 07}
  (2013) 075}, [\href{https://arxiv.org/abs/1108.2227}{{\ttfamily 1108.2227}}].

\bibitem{Hodges:2012ym}
A.~Hodges, \emph{{A simple formula for gravitational MHV amplitudes}},
  \href{https://arxiv.org/abs/1204.1930}{{\ttfamily 1204.1930}}.

\bibitem{Parke:1986gb}
S.~J. Parke and T.~R. Taylor, \emph{{An Amplitude for $n$ Gluon Scattering}},
  \href{http://dx.doi.org/10.1103/PhysRevLett.56.2459}{\emph{Phys. Rev. Lett.}
  {\bfseries 56} (1986) 2459}.

\bibitem{Puhm:2019zbl}
A.~Puhm, \emph{{Conformally Soft Theorem in Gravity}},
  \href{http://dx.doi.org/10.1007/JHEP09(2020)130}{\emph{JHEP} {\bfseries 09}
  (2020) 130}, [\href{https://arxiv.org/abs/1905.09799}{{\ttfamily
  1905.09799}}].

\bibitem{DeWitt:1967uc}
B.~S. DeWitt, \emph{{Quantum Theory of Gravity. 3. Applications of the
  Covariant Theory}},
  \href{http://dx.doi.org/10.1103/PhysRev.162.1239}{\emph{Phys. Rev.}
  {\bfseries 162} (1967) 1239--1256}.

\bibitem{PhysRevD.34.1749}
S.~Sannan, \emph{Gravity as the limit of the type-ii superstring theory},
  \href{http://dx.doi.org/10.1103/PhysRevD.34.1749}{\emph{Phys. Rev. D}
  {\bfseries 34} (Sep, 1986) 1749--1758}.

\bibitem{Brandhuber:2022qbk}
A.~Brandhuber, J.~Plefka and G.~Travaglini, \emph{{The SAGEX Review on
  Scattering Amplitudes Chapter 1: Modern Fundamentals of Amplitudes}},
  \href{http://dx.doi.org/10.1088/1751-8121/ac8254}{\emph{J. Phys. A}
  {\bfseries 55} (2022) 443002},
  [\href{https://arxiv.org/abs/2203.13012}{{\ttfamily 2203.13012}}].

\bibitem{PhysRevLett.84.3531}
Z.~Bern, A.~De~Freitas and H.~L. Wong, \emph{Coupling gravitons to matter},
  \href{http://dx.doi.org/10.1103/PhysRevLett.84.3531}{\emph{Phys. Rev. Lett.}
  {\bfseries 84} (Apr, 2000) 3531--3534}.

\bibitem{Fu:2017uzt}
C.-H. Fu, Y.-J. Du, R.~Huang and B.~Feng, \emph{{Expansion of
  Einstein-Yang-Mills Amplitude}},
  \href{http://dx.doi.org/10.1007/JHEP09(2017)021}{\emph{JHEP} {\bfseries 09}
  (2017) 021}, [\href{https://arxiv.org/abs/1702.08158}{{\ttfamily
  1702.08158}}].

\bibitem{Plefka:2018zwm}
J.~Plefka and W.~Wormsbecher, \emph{{New relations for graviton-matter
  amplitudes}}, \href{http://dx.doi.org/10.1103/PhysRevD.98.026011}{\emph{Phys.
  Rev. D} {\bfseries 98} (2018) 026011},
  [\href{https://arxiv.org/abs/1804.09651}{{\ttfamily 1804.09651}}].

\bibitem{Kawai:1985xq}
H.~Kawai, D.~C. Lewellen and S.~H.~H. Tye, \emph{{A Relation Between Tree
  Amplitudes of Closed and Open Strings}},
  \href{http://dx.doi.org/10.1016/0550-3213(86)90362-7}{\emph{Nucl. Phys. B}
  {\bfseries 269} (1986) 1--23}.

\bibitem{Banerjee:2020zlg}
S.~Banerjee, S.~Ghosh and P.~Paul, \emph{{MHV graviton scattering amplitudes
  and current algebra on the celestial sphere}},
  \href{http://dx.doi.org/10.1007/JHEP02(2021)176}{\emph{JHEP} {\bfseries 02}
  (2021) 176}, [\href{https://arxiv.org/abs/2008.04330}{{\ttfamily
  2008.04330}}].

\end{thebibliography}\endgroup

\end{document}